%% file: ms.tex
\documentclass[fleqn,useAMS,usenatbib]{mn2e}

\usepackage{mathptmx}
\usepackage{graphicx}
\usepackage{amsmath}
\usepackage{amssymb}
\usepackage{booktabs}
\usepackage{color}
\usepackage{units}

\input{commands}

\title[The Lyman-$\alpha$ forest in optically-thin hydrodynamical simulations]
{The Lyman-$\alpha$ forest in optically-thin hydrodynamical simulations}

\author[Luki\'c et al.]{
    Zarija Luki\'c$^{1}$, Casey W.~Stark$^{2}$, Peter Nugent$^{1, 2}$,
    Martin White$^{1, 2}$, Avery A.~Meiksin$^{3}$,
\newauthor
    and Ann Almgren$^{1}$ \\
$^1$ Lawrence Berkeley National Laboratory, CA 94720-8139, USA \\
$^2$ Department of Astronomy, University of California Berkeley, CA
       94720-3411, USA \\
$^3$ Institute for Astronomy, University of Edinburgh,
       Edinburgh EH9 3HJ, Scotland
}

\begin{document}

\date{Submitted to MNRAS: June 25, 2014; accepted: November 7, 2014}
\pagerange{\pageref{firstpage}--\pageref{lastpage}} \pubyear{2014}

\maketitle

\label{firstpage}

\begin{abstract}

  We study the statistics of the \Lyaf in a flat $\Lambda$CDM cosmology with the
  $N$-body + Eulerian hydrodynamics code Nyx. We produce a suite of simulations,
  covering the observationally relevant redshift range $2 \leq z \leq 4$. We
  find that a grid resolution of 20 $\hinv$kpc is required to produce one
  percent convergence of \Lyaf flux statistics, up to $k = 10$ $\hinv$Mpc. In
  addition to establishing resolution requirements, we study the effects of
  missing modes in these simulations, and find that box sizes of $L > 40$
  $\hinv$Mpc are needed to suppress numerical errors to a sub-percent level. Our
  optically-thin simulations with the ionizing background prescription of
  \citet{haardt_and_madau_2012} reproduce an IGM density-temperature relation with $T_0
  \approx 10^4$ K and $\gamma \approx 1.55$ at $z = 2$, with a mean transmitted
  flux close to the observed values. When using the ionizing background
  prescription of \citet{faucher_giguere_et_al_2009}, the mean flux is 10-15 per
  cent below observed values at $z = 2$, and a factor of 2 too small at $z = 4$.
  We show the effects of the common practice of rescaling optical depths to the
  observed mean flux and how it affects convergence rates. We also investigate
  the practice of `splicing' results from a number of different
  simulations to estimate the 1D flux power spectrum and show it is accurate at
  the 10 per cent level. Finally, we find that collisional heating of the gas
  from dark matter particles is negligible in modern cosmological simulations.

\end{abstract}


% Body
% ----

% Intro
\input{sec1}

% Modeling the Lya forest
\input{sec2}

% Physics of the Lya forest
\input{sec3}

% Resolution
\input{sec4}

% Box size
\input{sec5}

% Splicing
\input{sec6}

% UVB rescaling
\input{sec7}

% Line statistics
\input{sec8}

% Collisionality
\input{sec9}

% Conclusion
\input{sec10}

\section*{Acknowledgements}

The authors would like to thank An\v{z}e Slosar, Pat McDonald, and
Matt McQuinn for many useful discussions, and are grateful to Francesco
Haardt, Piero Madau, and Claude-Andre\`e Faucher-Gigu\'ere for making
their UVB rates publicly available. 
We acknowledge the helpful review 
of an original version of the manuscript by an anonimous referee.
ZL and CWS acknowledge the hospitality of Triple Rock where many concepts
were refined.
This work was in part supported by the Scientific Discovery through Advanced
Computing (SciDAC) program funded by U.S. Department of Energy Office of
Advanced Scientific Computing Research and the Office of High Energy Physics.
Calculations presented in this paper used resources of the National Energy
Research Scientific Computing Center (NERSC), which is supported by the
Office of Science of the U.S.~Department of Energy under Contract
No.~DE-AC02-05CH11231.
This work made extensive use of the NASA Astrophysics Data System and of
the astro-ph preprint archive at arXiv.org.

% References
% ----------
\footnotesize{
  \bibliographystyle{mn2e}
  \bibliography{ms}
}

\appendix
\input{appen_a}
\input{appen_b}
\input{appen_c}

\label{lastpage}

\end{document}

%% file: commands.tex
\usepackage{xspace}

\newcommand{\Lya}{Ly$\alpha$\xspace}
\newcommand{\Lyaf}{Ly$\alpha$ forest\xspace}
\newcommand{\LCDM}{$\Lambda$CDM\xspace}

\newcommand{\comment}[1]

\newcommand{\Msun}{\ensuremath{\mathrm{M_{\odot}}}}

\newcommand{\HI}{\ensuremath{\textrm{H} \, \textsc{i}}}
\newcommand{\HII}{\ensuremath{\textrm{H} \, \textsc{ii}}}
\newcommand{\HeI}{\ensuremath{\textrm{He} \, \textsc{i}}}
\newcommand{\HeII}{\ensuremath{\textrm{He} \, \textsc{ii}}}
\newcommand{\HeIII}{\ensuremath{\textrm{He} \, \textsc{iii}}}

\def\ltsima{$\; \buildrel < \over \sim \;$}
\def\lsim{\lower.5ex\hbox{\ltsima}}
\def\gtsima{$\; \buildrel > \over \sim \;$}
\def\gsim{\lower.5ex\hbox{\gtsima}}

\newcommand{\rhob}{\rho_{\rm b}}
\newcommand{\rhodm}{\rho_{\rm dm}}

\newcommand{\hinv}{h^{-1}}

\newcommand{\NHI}{N_{\HI}}

\newcommand{\mf}{\langle F \rangle}

\newcommand{\expd}[1]{\times 10^{#1}}
\newcommand{\mean}[1]{\langle #1 \rangle}

%% file: sec1.tex
\section{Introduction}
\label{sec:intro}

The cold dark matter (CDM) model of structure formation has proven impressively
successful at describing observations on large scales and at early times.
Perhaps the most successful confrontation of theory and observation is with the
study of anisotropies in the cosmic microwave background, where such a
comparison has yielded tight constraints on many of the basic cosmological
parameters and an increased confidence in our understanding of the Universe in
the linear regime.  The next simplest structure is the intergalactic medium
(IGM), the beaded filamentary network of structures between galaxies, which
contains most of the mass in the Universe \citep[for a recent review,
see][]{Mei09}. The IGM is best probed by the \Lyaf, the collection of
intervening absorption systems detected in the spectra of distant quasars.

\citet{1986Ap&SS.118..509I} and \citet{1986MNRAS.218P..25R} were the first to
suggest the \Lyaf originated from partially ionized hydrogen, in their case
confined gravitationally by haloes of collisionless or cold dark matter.
Using numerical simulations, \citet{Cen94} demonstrated that \Lyaf systems
arise naturally within the framework of theories of structure formation
through gravitational instability in CDM dominated cosmologies.
The installation of HIRES at Keck
\citep{1994SPIE.2198..362V} made it possible to make precision comparisons with
the models, confirming the success of the gravitational instability scenario for
the origin of the \Lyaf.

Modern hydrodynamic simulations recover many of the measured statistical
properties of the \Lyaf, such as the \HI\ column density
distribution, the pixel flux distribution function and the flux power spectrum
at a level capable of distinguishing between plausible variants of the CDM
model, with the $\Lambda$CDM model the most successful \citep{ZhaAnnNor95,
HKWM96,   1997ApJ...489....7R, Zha97, croft_et_al_1998, mcdonald_et_al_2000,
MBM01, 2002ApJ...581...20C,viel_et_al_2004}.
However, some differences are found. Most notable is the distribution of the
absorption line widths (Doppler widths) characterized by the $b$-parameter
(typically about 30 km/s). While the line widths are consistent with the amount
of broadening characteristic of photoionized gas, the measured distributions
show too many broadened lines compared with the predictions of the original
simulations. This is likely an indication that \HeII\ was reionized late, at $z
\lesssim 4$
\citep{2000ApJ...534...57B,2000ApJ...534...41R,Schaye00,MBM01,Worseck2014}.
Allowing for a late \HeII\ reionization and including radiative transfer during
reionization a range of Doppler widths may be achieved consistent with
the data \citep{2007MNRAS.380.1369T,   2012MNRAS.423....7M}. Since the line
widths control the scale and number of features, the distribution of wavelet
coefficients is also strongly affected and, to a lesser extent, the column
density distribution and the pixel flux distribution \citep{MBM01}. The flux
power spectrum is most affected at high wavenumbers.

The last decade has seen increasing use of \Lya absorption to investigate large-
scale structure and cosmology. The Sloan Digital Sky Survey (SDSS)
\citep{2000AJ....120.1579Y} provided an enormous increase in the amount of \Lyaf
data with thousands of quasars suitable for 1D analysis, but at the cost of the
spectra being low-resolution and fairly noisy. Still, this volume of data
allowed a much-improved measurement of the 1D flux power spectrum
\citep{mcdonald_et_al_2006}, placing constraints on the large scale spectral
index $n_s$ and the amplitude of fluctuations $\sigma_8$. The BOSS experiment of
SDSS-III \citep{dawson_et_al_2012} further increased the sky density of suitable
quasar lines of sight. The close proximity of large numbers of lines of sight
has enabled 3D correlations in the forest to be measured over large scales for
the first time using a sample of some 14000 QSOs
\citep{slosar_et_al_2011,slosar_et_al_2013,busca_et_al_2013}.  The 3D flux information
has also been cross-correlated with other high redshift tracers
\citep{fontribera_et_al_2012,fontribera_et_al_2013}.  The 1D flux power spectrum
has been measured to unprecedented precision
\citep{palanquedelabrouille_et_al_2013}. The 3D \Lya absorption correlations
are a promising means of constraining the nature of dark energy through the
measurements of the angular diameter distance and the Hubble constant at high
redshifts by detecting the large-scale Baryon Acoustic Oscillation (BAO) peak
\citep{slosar_et_al_2011,busca_et_al_2013,font-ribera_et_al_2014}.
At the same time, the measured signal provides a novel test of the
gravitational instability origin of the \Lyaf\ and the large-scale power
in the meta-galactic ionizing background
\citep[e.g.][]{mcquinn_and_white_2011,mcquinn_et_al_2011}.
In the near future, tomographic reconstruction with closely-separated 
\Lya forest sightlines from star-forming galaxies will enable direct 
3D mapping of the IGM on $\sim$Mpc scales \citep{KGLee2014}.

The \Lyaf may also be used to constrain galaxy formation models. Galactic winds
driven by feedback effects from galaxy formation models can impact statistics of
the \Lyaf flux \citep{viel_et_al_2013} and the circumgalactic medium. Searches
for the impact on the circumgalactic medium are underway around Lyman-break
galaxies \citep{2011MNRAS.414...28C, 2012ApJ...750...67R}. \Lyaf simulations are
required in this context to provide a quantitative interpretation of the
results.

To fully compare numerical simulations with the growing body of
increasingly accurate measurements of the \Lyaf, it is critical that
the simulations have converged to a level of precision comparable to
that of the data. Various studies have addressed convergence
issues. These have tended to be heterogeneous, with conclusions
dependent on the statistic for which convergence is sought, redshift,
and cosmological model. We summarize the principal findings in the literature.

It is clear that the Jeans length of the gas ($\sim 500 \, h^{-1}$kpc depending on
the redshift) must be resolved to recover the correct absorption line widths and
small-scale wavelet coefficients, with a suggested resolution of at least 40
$\hinv$kpc (comoving) at $z = 2 - 3$ \citep{1999ApJ...517...13B, Schaye00,
MeiWhi01, Tyt09, lidz_et_al_2010}. Resolution on this scale is also adequate for
converging to better than 5 percent on the hydrogen ionization rate required to
match the measured effective optical depth of the IGM \citep{MeiWhi04,
2005MNRAS.357.1178B, Tyt09}, as well as the effective optical depth itself
\citep{2009MNRAS.398L..26B}. At $z > 4$, however, convergence on the effective
optical depth diminishes to poorer than 15 percent at this mean resolution
\citep{2009MNRAS.398L..26B}. A resolution of 40 $\hinv$kpc is also
inadequate for converging on the Doppler parameter and wavelet coefficient
distributions at $z = 4$. This is particularly the case for the narrowest
features, for which a comoving mean resolution of better than 20 $\hinv$kpc
appears necessary, with results still not clearly well converged
\citep{1999ApJ...517...13B, lidz_et_al_2010}.

The results are also sensitive to box size. The inferred mean ionizing
background is converged to a few percent for comoving box sizes of 30
$\hinv$Mpc for $z > 2$ \citep{MeiWhi04, 2005MNRAS.357.1178B, Tyt09}. The
linewidths increase with box size, possibly not converged to better than 5
percent at $z = 2$ for a comoving box size as large as 54 $\hinv$Mpc
\citep{Tyt09}, although the distribution of smoothed wavelet coefficients
appears well converged at this redshift for the smaller box size of 25
$\hinv$Mpc \citep{lidz_et_al_2010}. At $z > 3$, the wavelet coefficients are not
well converged for box sizes as large as 50 $\hinv$Mpc \citep{lidz_et_al_2010}.

Convergence requirements on the 1D flux power spectrum are also demanding.
\citet{2005ApJ...635..761M} found better than 5 percent convergence from the
fundamental mode up to $k < 0.025$ km$^{-1}$s for $2 < z < 4$, and up
to $k < 0.1$ km$^{-1}$s at $z < 3$, for a resolution of 39 $\hinv$kpc,
but in a comoving box size of only 5 $\hinv$Mpc. In larger boxes
(30 $\hinv$Mpc), \citet{2006MNRAS.365..231V} found 5 percent convergence at
$k < 0.01$ km$^{-1}$s for a mean resolution of 150 $\hinv$kpc, but
of only 12 per cent for $k = 0.02$ km$^{-1}$s at $z = 4$. Other work found
convergence of up to 10 per cent may be achieved at $k < 0.03$ km$^{-1}$s in
20 - 40 $\hinv$Mpc boxes at $z = 2 - 5$ (although possibly as poor as 20 per cent at
$z = 5$), with resolutions of 60 - 200 $\hinv$kpc, although requiring better
than 50 $\hinv$kpc resolution for 5 percent convergence at $k = 0.1$
km$^{-1}$s at $z = 2$, with even this inadequate at $z = 5$
\citep{MeiWhi04, 2009MNRAS.398L..26B, Tyt09}. Even at this level, the spatial
flux correlation function converges to better than 10 per cent over only 3 percent of
the box size \citep{MeiWhi04}. The convergence of absorber pair and higher
multiple statistics along neighboring lines of sight are not expected to fare
better, which is perhaps why they have been largely ignored in simulation
comparisons with data.

The primary goal of this paper is to establish the box size and resolution
requirements to produce converged statistics of the \Lyaf flux over the
redshifts of observational interest, with the minimum set of physical processes.
While this is not the final word, doing so already presents several numerical
challenges.
The gaseous nature of the IGM requires simulating hydrodynamics along with the
gravitating dark matter particles to obtain accurate results.
Since the signals derive from almost the entire volume, and from the gas close
to the mean density, Lagrangian methods which naturally increase resolution in
the high-density regions are not favored in terms of "time-to-solution".
In fact, they tend to be prohibitively expensive unless the hydrodynamic
calculations are somehow truncated in the high density regions (e.g.~by invoking
artificial star formation as done in Gadget with the `Quick Lyman-alpha' flag). 
Lagrangian methods will also have slightly worse resolution in
underdense regions, which become important for the $z > 3$ forest.
In our code Nyx, we use an Eulerian (comoving) grid.
Capturing the Jeans scale while covering a cosmologically representative
volume requires a large dynamic range.
Even with the modern supercomputers and a scalable cosmological hydrodynamics
code, we are not able to directly simulate the volumes probed by observations
while simultaneously resolving the \Lyaf. However, we are now able to simulate
boxes large enough to be cosmologically representative and not suffer --- to a
desired accuracy --- from the missing larger-scale modes.

The outline of the paper is as follows. In Section \ref{sec:sim} we describe the
physics and numerical capabilities required to simulate the IGM. Here we
introduce our simulation code and the runs we performed for this study. We also
describe how we model the \Lyaf using the simulation data. In Section
\ref{sec:properties} we discuss the physical properties of the IGM we find, and
compare with previous results. In Section \ref{sec:resolution} we demonstrate
the convergence of flux statistics with respect to physical resolution. We also
explore the possibility of increasing the resolution via Richardson
extrapolation. We demonstrate the convergence of flux statistics with respect to
the domain size in Section \ref{sec:box_size}. For the first time, we have
produced simulations sufficiently large to capture large-scale effects while
also resolving the Jeans scale. In Section \ref{sec:splicing} we compare our
full-range simulation to a common technique for splicing together simulations
with smaller dynamic ranges. It is commonly the case that simulations do not
recover the observed mean flux, but that simulated fluxes are rescaled to match
the observed mean. In Section \ref{sec:uvb} we explore effects of such
rescaling. Statistics of \Lya lines and wavelets are presented in Section
\ref{sec:small_scale}. In Section \ref{sec:collisionality} we quantify the
effects of collisionality between dark matter particles and gas which can
artificially increase the temperature of the gas. Finally, we present our
conclusions in Section \ref{sec:conclusion}.

Throughout this work, we assume the flat $\Lambda$CDM cosmological model.
All distances are comoving and are quoted in ``h units'', i.e.~$\hinv$Mpc
or $\hinv$kpc. Masses are in $\Msun$. When discussing the gas velocity, we use
the peculiar velocity, $v = a \dot{x}$, where $x$ is the comoving scale. When
discussing the velocity coordinate of a pixel, we use the Hubble flow velocity
$v = \dot{a} x$, evaluated at that redshift.

%% file: sec2.tex
\section{Simulating the Ly-$\alpha$ Forest}
\label{sec:sim}

\begin{figure*}
  \begin{center}
    \includegraphics[width=\textwidth]{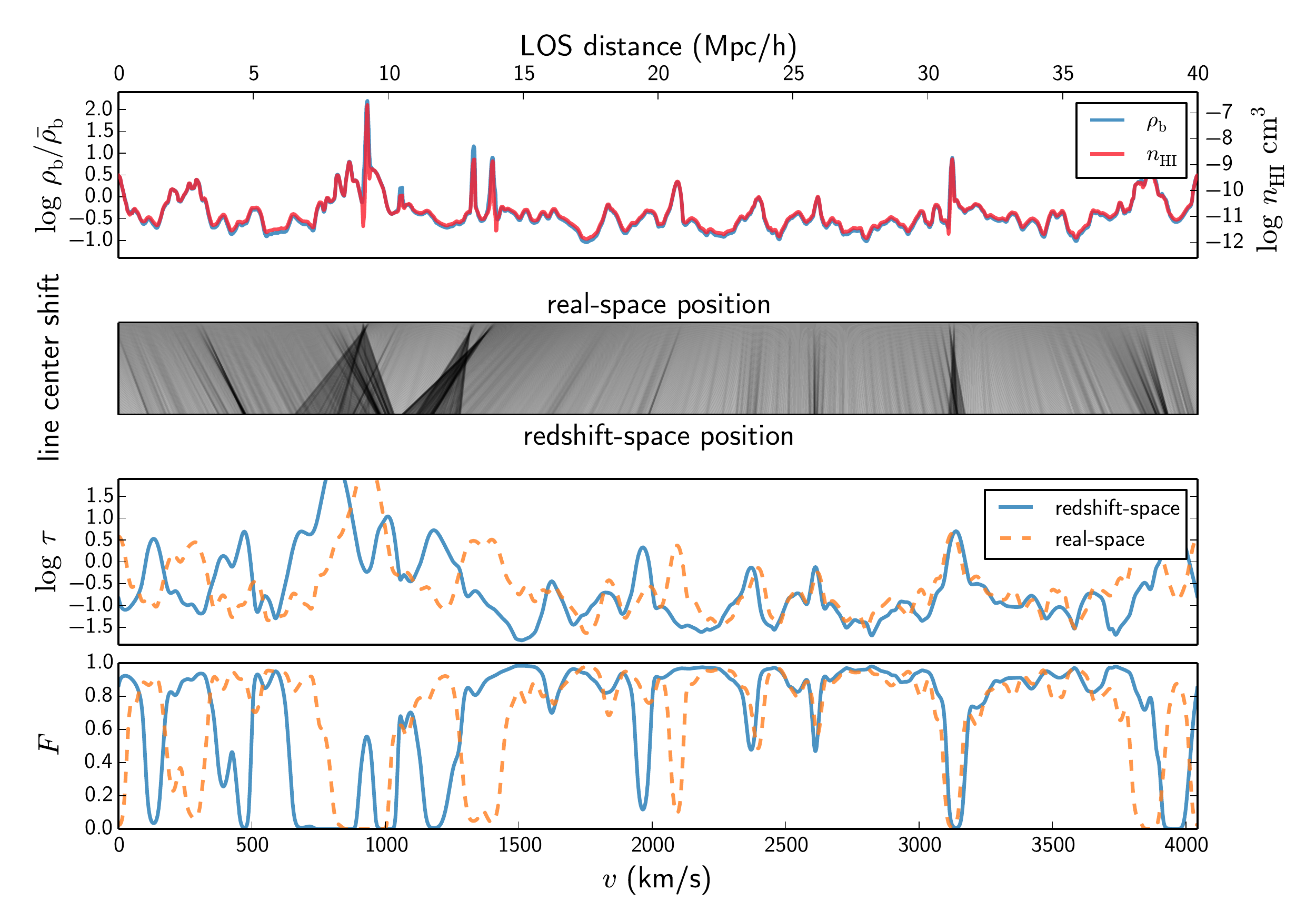}
  \end{center}
  \caption{A sample skewer, from a $2048^3$ simulation in a periodic box of side
    length $40 \, \hinv$Mpc at $z = 2.5$, showing the ingredients in the flux
    calculation. The top horizontal axis gives the (comoving) distance in the
    line-of-sight direction through the box. The lower horizontal axis gives the
    same distance in velocity units (see Table \ref{tab:conversion} for
    conversion factors). The top panel shows the baryon density, on a log scale.
    The middle panel illustrates how the velocity component along the line of
    sight $v_\parallel$ shifts the line center going from real- to
    redshift-space. The lower two panels show the optical depth and flux,
    respectively. The differences between the real- and redshift-space flux
    show that the redshift-space distortions do not just shift lines, but also
    change the blending of the lines.}
  \label{fig:sample_skewer}
\end{figure*}

The \Lyaf arises primarily from four physical effects:\ the
gravitational collapse of baryons and cold dark matter, ideal
gas pressure support, radiative heating and cooling, and
photoionization. The standard picture \citep[see, e.g.][]{Mei09} is that at
moderate overdensities and scales larger than $1 \, \hinv$Mpc, the
baryons simply trace the dark matter. The IGM at redshifts $z < 6$ has
a typical temperature $T \sim 10^4$ K, so that at scales of $\sim 100$
$\hinv$kpc the gas is pressure supported, and the density fluctuations
are thus suppressed relative to the dark matter, or as often put, the
baryons in the IGM are {\it filtered} on this scale. \citet{GneHui98}
first provided a detailed description of this process in the context of
linear theory. Briefly, the scale at which the gravitational and
pressure forces are equal is the Jeans length, and in comoving units
is equal to
\begin{equation}
  \lambda_{\rm J} = (1 + z) c_s \sqrt{ \frac{\pi}{G \rho_{\rm m}} }
    = 0.783 \sqrt{ \frac{T / (10^4 \, \mathrm{K})}
                        {\Omega_{\rm m} (1 + \delta) (1 + z)} }
      \; \hinv \mathrm{Mpc}
\end{equation}

In the case of adiabatic expansion (before reionization), $T \propto (1 + z)^2$,
the Jeans length decreases with time. In the case of constant temperature (after
reionization) it increases with time.
For the cosmology we assume here, with $\Omega_{\rm m} = 0.275$, the Jeans
length for mean density gas at $T = 10^4$ K changes from 0.67 to
0.86 $\hinv$Mpc from $z = 4$ to $z = 2$.
Observations suggest that the temperature of the IGM does not change much
over the redshift range $2 < z < 4$
\citep{becker_et_al_2011, bolton_et_al_2014}, so we can immediately see that
higher redshifts will require higher resolution to resolve \Lyaf absorbers
to a similar accuracy.

As defined, the Jeans scale $k_{\rm J} = 2 \pi / \lambda_{\rm J}$ is an
instantaneous measure that does not take into account the evolution of
density or sound speed. Since the amount of filtering at a given
epoch also depends on the thermal history of the gas, a more
interesting dynamical quantity is the filtering scale $k_{\rm f}$, the
scale at which baryon fluctuations are suppressed relative to cold
dark matter. In linear theory, the filtering scale is
\begin{equation}
  \frac{1}{k_{\rm f}^2(t)} = \frac{1}{D_+(t)}
    \int_0^t dt^\prime a^2(t^\prime)
      \frac{\ddot{D}_+(t^\prime) + 2 H(t^\prime) \dot{D}_+(t^\prime)}
          {k_{\rm J}^2(t^\prime)}
       \int_{t^\prime}^t \frac{dt^{\prime\prime}}{a^2(t^{\prime\prime})}
\end{equation}
The filtering scale in linear theory is always equal to the Jeans
scale at an earlier time. This means that before reionization the
filtering scale is larger than the Jeans scale, and after reionization
the filtering scale is smaller than the Jeans scale. The key point
here is that after reionization, in the case of roughly constant
temperature history, $k_{\rm f}$ is smaller than the Jeans scale. A rule of
thumb is that for typical growth factors and thermal histories, the filtering
scale is roughly half the Jeans scale for $2 < z < 4$.  In the linear regime
the filtering of baryon power is roughly exponential:
\begin{equation}
  P_b(k) \approx P_{\rm dm}(k) e^{-2 k^2 / k_{\rm f}^2} \; .
\end{equation}

The chemical composition of the IGM is close to primordial, thus the
dominant radiative processes involve only hydrogen and helium. The
competition between photoionization heating and adiabatic cooling
drives the gas to a tight power-law relation between density and temperature
\citep{katz_et_al_1996, hui_and_gnedin_1997}. As explained in
\citet{hui_and_gnedin_1997}, the slope of the power-law steepens in time,
rapidly close to the reionization epoch, and more slowly in later times.
After reionization,
around $z = 10$, the IGM is in photoionization equilibrium, with the
ionization maintained by a UV metagalactic radiation field. The \HI\
optical depth to \Lya scattering is proportional to the \HI\ density,
approximately described by a simple relation between the optical depth
and baryon density:
\begin{equation}
  \tau = A \left( \rhob / \bar{\rho}_{\rm b} \right)^\beta
\end{equation}
Finally, the transmitted flux fraction is defined as $F = e^{-\tau}$, which we
refer to as just flux. The normalization constant $A$ may be fixed by comparing
with measurements of the mean \Lya transmission through the IGM, $\mf \equiv
\langle e^{-\tau} \rangle$.

Improvements on this method may be achieved by using a polytropic
equation of state and approximating the density fluctuation
distribution as lognormal \citep{1997ApJ...479..523B} or using
$N$-body simulations under the assumption that the baryons trace the
dark matter, possibly with some pressure support
\citep{1995A&A...295L...9P, croft_et_al_1998, GneHui98, MeiWhi01}.
While such simplified descriptions are sufficient for qualitatively
characterizing the properties of the \Lyaf, they are unfortunately not
adequate for detailed statistical predictions. In reality, we expect
scatter in the $\rhob$-$T$ relation as a result of shock
heating (as the peculiar velocities of most of the baryons are
supersonic), other radiative processes, and possibly even \HeII\
reionization at $2 < z < 4$. In addition, the relation between the
local \HI\ density and the flux is complicated by peculiar velocities
along the line of sight.

As a result, it is likely that the most accurate description requires full
hydrodynamical simulations coupled to an $N$-body code. Hydrodynamic
simulations directly model the heating and cooling processes in the
IGM, along with capturing the effects of shock
heating. Pseudo-hydrodynamics methods like Hydro Particle Mesh (HPM)
can capture the coarse properties of pressure support, but only
shock-capturing methods can produce the correct $\rhob$-$T$ phases. As
shown in \citet{viel_et_al_2013}, capturing the hot gas that makes up
the WHIM creates 5 per cent differences in the flux probability distribution
function (PDF) and flux 1D power. In addition, accretion shocks on the
outskirts of halos and filaments can clearly alter the density and
temperature profiles of those regions and therefore the \Lya
transmission through them.

The biggest computational challenge for accurately capturing the state of the
IGM is the required dynamic range. In order to appropriately model bulk flows,
simulations must cover linear scales of $\mathcal{O}(100 \; \hinv
\mathrm{Mpc})$. The bulk flows play an important role in determining the
temperature distribution of the IGM via shock heating. As shown in
\citet{Tyt09}, using too small a box will result in an underestimate of the mean
temperature $T_0$. At the same time, simulations must resolve the filtering
scale, which is $\mathcal{O}(100 \; \hinv \mathrm{kpc})$ for the densities of
interest. The required dynamic range ends up being closer to $10^4$ than $10^3$
however, because adequately resolving a given scale in a simulation means
covering it with several resolution elements, so the required minimum scale
turns out to be 10 to 20 $\hinv$kpc.

With modern numerical techniques and supercomputers, resolving a
dynamic range of $\sim 10^4$ in full volume of a 3D simulation is now
practical.  Using Lagrangian techniques like Smoothed Particle
Hydrodynamics (SPH) or the Eulerian adaptive resolution technique
Adaptive Mesh Refinement (AMR), it is straightforward to achieve still
larger dynamic ranges. However, these techniques only help simulations
focused on resolving small fractions of the total domain volume.  The
difficulty in simulating the IGM is that it covers almost all of the
volume of the domain.  The gas responsible for the \Lya
  forest is close to the cosmic mean density rendering Lagrangian
  methods computationally non-optimal as they spend a majority of the
  compute cycles evolving dense regions. This is especially true at
  higher redshifts where most of the signal comes from the underdense
  regions (Fig.~\ref{fig:F_rho}).

The \Lyaf arises from relatively low column density regions, 
with the contribution from higher column densities decreasing as a power law.
Lines with low hydrogen column density 
are difficult to observe in practice, thus observations recover well lines with 
$\NHI > 10^{13}$ cm$^{-2}$, especially at high redshifts, $z > 2$ 
\citep[see, e.g.~][]{janknecht2006, haardt_and_madau_2012}. 
In these environments the $\HI$ ionizing photon mean free path is much larger than
the boxes simulated here for $z < 6$. Therefore we model radiative processes
under the assumption of an optically-thin medium and assume a uniform ionizing
background radiation field. We note that accurate modeling of high column
density systems, like Lyman limit systems with $\NHI > 10^{17}$ cm$^{-2}$,
or Damped \Lya systems characterized by $\NHI > 10^{20}$ cm$^{-2}$, should
include explicit radiative transfer. While these systems do form in our
simulations, the relevant physics is not present in our optically thin
simulations and we shall not investigate them in detail.

\begin{table}
\begin{center}
  \caption{Conversion factors versus redshift}
  \begin{tabular}{ r r r r r r }
    \toprule
    $z$  & $\lambda_\alpha$ & $H(z)$ & $d\lambda / d\chi$ & $dv / d\chi$ & $10^3 dz / d\chi$ \\
    \midrule
    2.00 &            3645  &   285  &              1.16  &          95  &             0.95  \\
    2.25 &            3949  &   319  &              1.29  &          98  &             1.06  \\
    2.50 &            4253  &   354  &              1.43  &         101  &             1.18  \\
    3.00 &            4860  &   428  &              1.73  &         107  &             1.42  \\
    3.50 &            5468  &   508  &              2.06  &         113  &             1.69  \\
    4.00 &            6075  &   592  &              2.40  &         118  &             1.98  \\
    \bottomrule
  \end{tabular}
  \medskip

    Conversion factors for the flat $\Lambda$CDM cosmology considered here with
    $h = 0.702$ and $\Omega_{\rm m} = 0.275$. Wavelengths are given in \AA,
    comoving distances in $\hinv$Mpc and velocities in km s$^{-1}$.

  \label{tab:conversion}
 \end{center}
\end{table}

Hydrogen (and HeI) reionization takes place at high redshift, $z > 10$.
Therefore the details of this epoch are unimportant for the thermodynamical
properties of the gas at redshifts relevant for \Lyaf observations ($z < 4$).
In contrast, \HeII\ reionization takes place at an observationally relevant
epoch ($3 < z < 5$) although the observational picture is not yet resolved
\citep[e.g.][]{Worseck2014}.
In addition, the size of fluctuations in the \HeII\ ionizing background are very
poorly constrained, varying by an order of magnitude in recent studies
\citep{Shull2010, Syphers2014, McQuinn2014}.
However, the main effect of \HeII\ reionization on the IGM is that the
additional photoheating increases the temperature of the IGM.
The UV background prescriptions we employ in this study model this increase
in the temperature via an increase in photoheating rates and ionize \HeII\ by
$z = 3$. We note, however, that \HeII\ reionization could result in higher
temperatures with explicit radiative transfer and significant (spatial)
fluctuations of the ionizing background.
Thus, including \HeII\ reionization correctly in simulations requires
incorporating radiative transfer \citep{2007MNRAS.380.1369T}, which
remains an active area of current research in cosmological
hydrodynamics codes 
\citep[][]{2007MNRAS.380.1369T, mcquinn_et_al_2011, 
2012MNRAS.423....7M, Compostella2013}.

\subsection{Simulations}

The simulations we present here are performed with the Nyx code
\citep{almgren_et_al_2013}. Nyx follows the evolution of dark matter
simulated as self-gravitating Lagrangian particles, and baryons
modeled as an ideal gas on a uniform Cartesian grid. Nyx includes Adaptive
Mesh Refinement (AMR) capabilities, which we can use to extend the
simulated dynamic range. We do not make use of AMR in the current
work, as the \Lyaf signal spans nearly the entire simulation domain
rather than isolated concentrations of matter where AMR is more
effective. The Eulerian gas dynamics equations are solved using a
second-order accurate piecewise parabolic method (PPM) to accurately
capture shock waves. Our implementation uses a dimensionally unsplit
scheme with full corner coupling \citep{colella_1990} to better
reconstruct the 3D fluid flow.
The same mesh structure that is used to update fluid quantities is
also used to compute the gravitational field and to evolve the
particles via a particle-mesh (PM) method, using Cloud-In-Cell (CIC)
interpolation to switch between particle- and mesh-based quantities.
The gravitational source terms in the momentum and energy equations are
discretized in time using a predictor-corrector approach.
The additional physics of radiative
heating and cooling is included via source terms in the 
equations for internal and total energy. 
As the relevant time scale for heating and cooling can be
significantly different from the stability criterion 
required by the explicit discretization of gas dynamics equations 
(the Courant-Friedrichs-Lewy or CFL condition), the heating and cooling source
terms are integrated in time using VODE \citep{vode} and coupled to the 
hydrodynamics using a Strang splitting \citep{strang:1968} approach. 
For more details of our numerical methods, see \citet{almgren_et_al_2013}.

\begin{table}
  \begin{center}
    \caption{List of simulations}
    \begin{tabular}{l c c c r}
      \toprule
      Name       & Box size     & Elements & Resolution    & $m_{\rm dm}$       \\
                 & [$\hinv$Mpc] &          & [$\hinv$kpc] & [\Msun]             \\
      \midrule
      L10\_N128  &           10  &  128$^3$ &           78  & $4.3 \times 10^7$ \\
      L10\_N256  &           10  &  256$^3$ &           39  & $5.4 \times 10^6$ \\
      L10\_N512  &           10  &  512$^3$ &           20  & $6.7 \times 10^5$ \\
      L10\_N1024 &           10  & 1024$^3$ &           10  & $8.4 \times 10^4$ \\
      \cmidrule(r){1-2}
      L20\_N256  &           20  &  256$^3$ &           78  & $4.3 \times 10^7$ \\
      L20\_N512  &           20  &  512$^3$ &           39  & $5.4 \times 10^6$ \\
      L20\_N1024 &           20  & 1024$^3$ &           20  & $6.7 \times 10^5$ \\
      L20\_N2048 &           20  & 2048$^3$ &           10  & $8.4 \times 10^4$ \\
      \cmidrule(r){1-2}
      L40\_N512  &           40  &  512$^3$ &           78  & $4.3 \times 10^7$ \\
      L40\_N1024 &           40  & 1024$^3$ &           39  & $5.4 \times 10^6$ \\
      L40\_N2048 &           40  & 2048$^3$ &           20  & $6.7 \times 10^5$ \\
      \cmidrule(r){1-2}
      L80\_N1024 &           80  & 1024$^3$ &           78  & $4.3 \times 10^7$ \\
      L80\_N2048 &           80  & 2048$^3$ &           39  & $5.4 \times 10^6$ \\
      L80\_N4096 &           80  & 4096$^3$ &           20  & $6.7 \times 10^5$ \\
      \bottomrule
    \end{tabular}
    \medskip

      The simulations used in this work. Resolution refers to the cell size, and to
      ease comparison with SPH simulations we list the mass of dark matter particles
      in each simulation. See text for details.

    \label{tab:sims}
  \end{center}
\end{table}

We simulate the \textit{WMAP} 7-yr data constrained \LCDM cosmology, with
parameters: $\Omega_{\rm m} = 0.275$, $\Omega_\Lambda = 1 - \Omega_{\rm m} =
0.725$, $\Omega_{\rm b} = 0.046$, $h = 0.702$, $\sigma_8 = 0.816$, and $n_s =
0.96$ \citep{komatsu_et_al_2011}. We provide Table \ref{tab:conversion} to help
convert between scale, wavelength, and velocity coordinates at redshifts used in
this work. The latest \textit{Planck} constraints \citep{ade_et_al_2014} differ
somewhat from WMAP-7 values, most notably in the values for the Hubble constant
$h$ and total matter content $\Omega_{\rm m}$. These differences will not play
an important role in this paper, as we aim to explore numerical prescriptions
for achieving 1 per cent accurate \Lyaf statistics. The conclusions here will inform
future work for running many \textit{viable} cosmologies and understanding their
numerical limitations.

The full set of simulations is listed in Table \ref{tab:sims}. We designed the
set of simulations to cover the expected maximum box size and minimum resolution
needed to show convergence. 
All simulations are initialized at $z = 159$, starting 
from a grid distribution of particles and Zel'dovich approximation 
\citep{Zeldovich1970}.  Transfer functions were generated using both 
analytical approximation \citet{EisensteinHu1999} and Boltzmann solver code
CLASS \citep{CLASS}.  The conclusions presented here have no sensitivity 
on the particular transfer function used, but it is of course important to 
maintain the same transfer function accross a series of runs one is
comparing to each other.
We focus on snapshots in the range $2 \leq z \leq 4$, relevant for most
observations. To simplify the comparison, simulations performed in the same box
size share the same large-scale modes, the only difference being that higher
resolution runs have more modes sampled on small scales.

\begin{figure}
  \centering
  \includegraphics[width=\columnwidth]{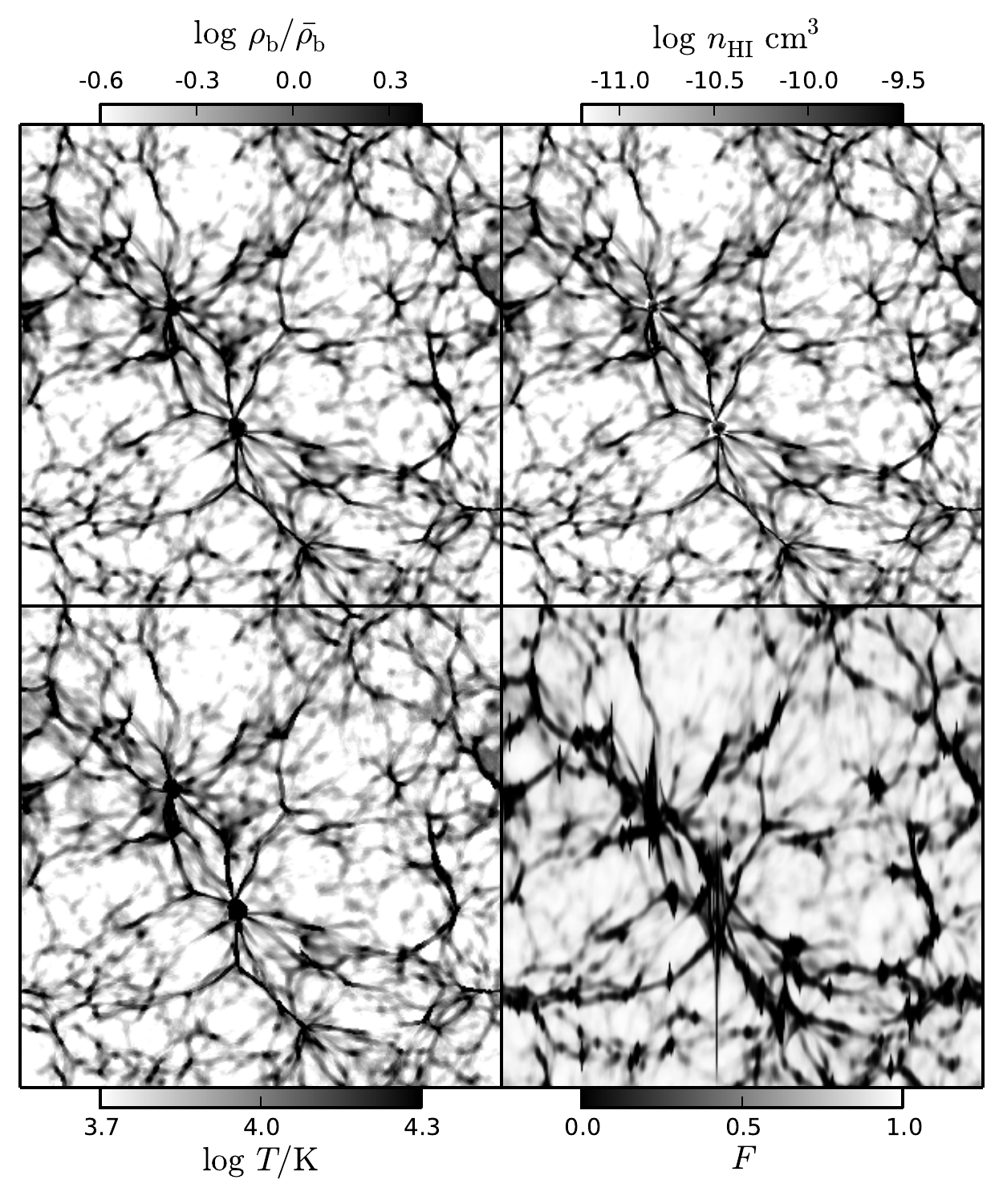}
  \caption{A slice of the baryon density, temperature, \HI\ number density, and
    flux from the L20\_N2048 simulation at $z = 2.5$. The slice covers the
    domain of 20 x 20 $\hinv$Mpc, with a thickness of about 100 $\hinv$kpc (10
    cells). Note that the $F$ line of sight is the y-axis direction, so that
    broadened lines show up as vertical black streaks.}
  \label{fig:web}
\end{figure}

\subsection{Included Physics}
\label{sec:nyx}

Besides solving for gravity and the Euler equations,
we model the chemistry of the gas as having a primordial composition
with hydrogen and helium mass abundances of $X = 0.75$, and $Y =
0.25$, respectively. The choice of values is in agreement with the recent
CMB observations and Big Bang
nucleosynthesis \citep{Coc2013}. The resulting reaction network includes 6 atomic species:
\HI, \HII, \HeI, \HeII, \HeIII\ and e$^-$, which we evolve under the
assumption of ionization equilibrium. The resulting system of
algebraic equations is:
\begin{equation}
  \begin{aligned}
    & \left( \Gamma_{e, \HI} n_e + \Gamma_{\gamma, \HI} \right) n_{\HI}
      = \alpha_{\rm r, \HII} n_e n_{\HII} \\[1.5mm]
    & \left( \Gamma_{e, \HeI} n_e + \Gamma_{\gamma, \HeI} \right) n_{\HeI}
      = \left( \alpha_{\rm r, \HeII} + \alpha_{\rm d, \HeII} \right)
      n_e n_{\HeII} \\[1.5mm]
    & \left[ \Gamma_{\gamma, \HeII} + \left(\Gamma_{e, \HeII}
      + \alpha_{\rm r, \HeII} + \alpha_{\rm d, \HeII} \right)
      n_e \right] n_{\HeII} \\[1.5mm]
    & \qquad = \alpha_{\rm r, \HeIII} n_e n_{\HeIII}
      + \left( \Gamma_{e, \HeI} n_e + \Gamma_{\gamma, \HeI} \right)
      n_{\HeI}
  \end{aligned}
  \label{eq:equil_species}
\end{equation}
in addition, there are three closure equations for the conservation of charge
and hydrogen and helium abundances. Radiative recombination
($\alpha_{\rm r, X}$), dielectronic recombination ($\alpha_{\rm d, X}$), and
collisional ionization ($\Gamma_{e, {\rm X}}$) rates are strongly dependent on
the temperature, which itself depends on the ionization state through the mean
mass per particle $\mu$
\begin{equation}
  T = \frac{2}{3} \frac{m_p}{k_{\rm B}} \mu\ e_{\rm int}
\end{equation}
where $m_p$ is the mass of a proton, $k_{\rm B}$ is the Boltzmann
constant, and $e_{\rm int}$ is the internal thermal energy per mass of
the gas. Here we assume adiabatic index for monoatomic ideal gas.
For a gas composed of only hydrogen and helium, $\mu$ is related to the
number density of free electrons relative to hydrogen by $\mu = 1 /
\left[ 1 - (3 / 4) Y + (1 - Y) n_e / n_{\rm H} \right]$.
We iteratively solve the reaction network equations together with the
ideal gas equation of state, $p = 2/3 \rho e_{\rm int}$, to
determine the temperature and equilibrium distribution of species.

We compute radiative cooling as in \citet{katz_et_al_1996}, and assume
a spatially uniform, but time-varying ultraviolet background (UVB)
radiation field from either \citet{faucher_giguere_et_al_2009} or
\citet{haardt_and_madau_2012}. We do not follow radiation transport
through the box, nor do we explicitly account for the effects of
thermal feedback of stars, quasars, or active galactic nuclei; all
cells are assumed to be optically thin, and radiative feedback is
accounted for via the UVB model. In addition, we include inverse
Compton cooling off the microwave background. For the
exact rates used in the Nyx code and comparison of two UV
backgrounds we refer the reader to Appendix \ref{app:rates}.

\subsection{Simulated Spectra}
\label{ssec:spectra}

The optical depth $\tau$ for \Lya\ photon scattering is
\begin{equation}
  \tau_\nu = \int n_{\rm X} \sigma_\nu dr
\end{equation}
where $\nu$ is the frequency, $n_{\rm X}$ is the number density of
species X, $\sigma_\nu$ is the cross section of the interaction, and
$dr$ is the proper path length element. For our current work, we
assume a Doppler line profile, so the resulting optical depth is
\begin{equation}
  \tau_\nu = \frac{\pi e^2}{m_e c} f_{12}
    \int \frac{ n_{\rm X} }{ \Delta \nu_{\rm D} }
    \frac{ \exp \left[ -\left(
        \frac{ \nu - \nu_0 }{ \Delta \nu_{\rm D} } \right)^2 \right] }
         {\sqrt{\pi}} dr,
  \label{eq:taunu}
\end{equation}
where $\Delta \nu_{\rm D} = (b / c) \nu_0$ is the Doppler width with the
Doppler parameter $b = b_{\rm thermal} = \sqrt{2 k_{\rm B} T / m_{\rm H}}$, and
$f_{12}$ is the upward oscillator strength of the \Lya resonance transition of
frequency $\nu_0$. See Appendix \ref{app:tau} for a more detailed discussion of
our optical depth calculation, including the discretization of Equation
(\ref{eq:taunu}).

We choose sightlines, or ``skewers'', crossing the domain parallel to one of the
axes of the simulation grid and piercing the cell centers. Computationally, this
is the most efficient approach. This choice of rays avoids explicit ray-casting
and any interpolation of the cell-centered data, which introduce other numerical
and periodicity issues. We cover the entire $N^3$ grid with skewers, which provides
the equivalent of $N^2$ spectra.
Although large-scale modes along different spatial dimensions are statistically
independent allowing some gain in statistics from multiple viewing directions,
in this work we use a single line-of-sight axis rather than combining together
skewers using all 3 axes.
The process of going from simulated baryon values to flux $F$ is illustrated
in Figure \ref{fig:sample_skewer}.

%% file: sec3.tex
\section{Physical Properties of the Ly$\alpha$ Forest}
\label{sec:properties}

\begin{figure}
  \begin{center}
    \includegraphics[width=\columnwidth]{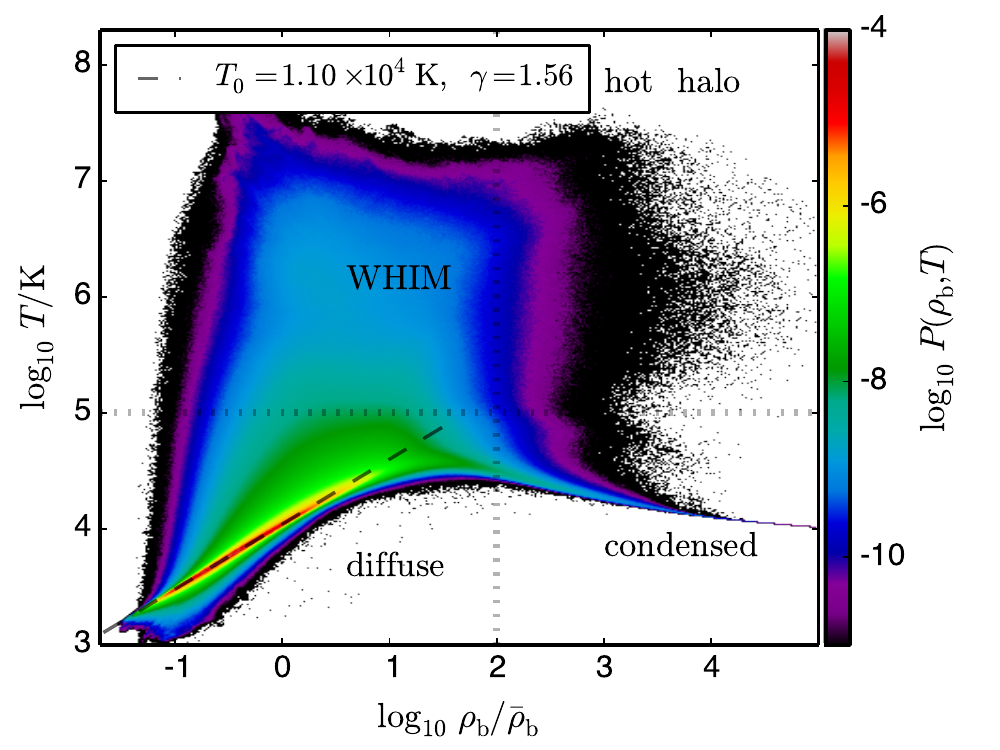}
  \end{center}
  \caption{
    The density--temperature distribution of gas (volume-weighted histogram) at
    $z = 2.5$ in the L40\_N2048 simulation, showing broadly four thermodynamical
    regimes for baryons in cosmological simulations.  Simulation was done with
    \citet{haardt_and_madau_2012} UV background.  The \Lyaf signal primarily
    comes from the diffuse region in the lower-left, which includes most of the
    baryons in the universe. Note the tight temperature-density relation
    in this regime.
  }
  \label{fig:rho-T}
\end{figure}

\citet{Zha98} discuss the physical properties of the \Lyaf in
hierarchical models such as CDM. The discussion in this section can
largely be considered as an update of that work.

As described above, the state of the IGM is relatively simple with a few power
laws approximately tying together the spatial distribution of baryon density,
temperature, proper \HI\ number density, and optical depth to \HI\ \Lya\ photon
scattering. Figure \ref{fig:web} shows a slice of these quantities in one of our
high-resolution simulations, except with the optical depth replaced by the
transmitted flux. We choose to show flux because it highlights the range of each
quantity that is relevant for observations. That is, we want to highlight
differences between an optical depth of 1 or 2, which changes the flux
drastically, but not between 10 and 100, which is essentially opaque. We
adjusted the gray-scale intensity ranges of density, temperature, and \HI\
number density to roughly match the morphology of the flux, which provides a
good guide to what ranges of each quantity affects the \Lyaf. We note that over
the relevant redshift range, the comoving density and temperature evolve slowly,
so that these ranges roughly apply to all redshifts. However, the physical \HI\
density changes drastically primarily due to expansion. The striking
morphological similarity between the fields demonstrates how well the usual
approximations work. The flux field is clearly the least like the other fields due
to two effects: the optical depth is in redshift-space and is therefore
distorted by peculiar velocities; in addition it is also thermally broadened,
smearing high temperature regions across the line-of-sight axis.

A key component of the robustness of \Lyaf predictions is the
$\rhob$-$T$ relation in the diffuse IGM. The relation is approximated
by the power law
\begin{equation}
  T = T_0 \left( \frac{\rhob}{\bar{\rho}_{\rm b}} \right)^{\gamma - 1}
\end{equation}
where $T_0$ is the temperature at mean density, and $\gamma$ is the
slope of density-temperature relation, both of which are set by the
metagalactic ionizing background. Typical values are $T_0$ between
10,000 and 20,000 K and $\gamma$ between 1 and $1.6$.
Figure \ref{fig:rho-T} shows the joint
PDF of density and temperature (volume-weighted) in the L40\_N2048 as an image
and the power law relation over plotted with a dashed line. We tried fitting the
density-temperature relation line several ways and found that a linear
least-squares fit is sufficient. The number of points in the diffuse IGM phase
is very large even for small simulations, so there is very little uncertainty
in the fit parameters.
However, we noticed a small but systematic difference in the best-fit $\gamma$
depending on the density range fit.
Fitting underdense regions, i.e.~points with $-1 < \log_{10}\rhob < 0$
yields $\gamma$ values a few percent higher than fitting near the mean
density, $-0.5 < \log_{10} \rhob < 0.5$.
Thus even if we neglect the scatter in the $\rhob$-$T$ relation a single
power-law approximation breaks down at a few percent accuracy.

\begin{figure}
  \begin{center}
    \includegraphics[width=\columnwidth]{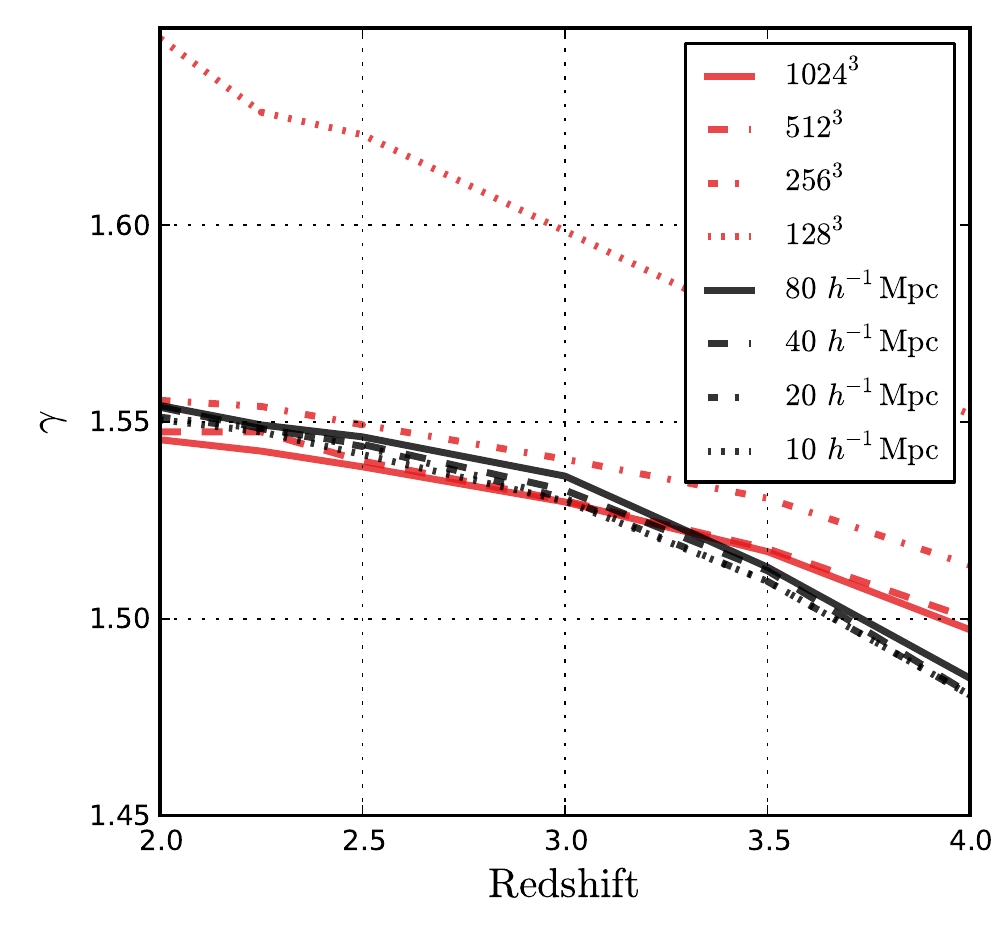}
  \end{center}
  \caption{
    Redshift evolution of $\gamma$, the slope of density-temperature relation.
    Lines in black show the weak dependence on the box size (the resolution is
    kept fixed at 20 $\hinv$kpc), while the
    the red lines show a rapid convergence with respect to the spatial
    resolution (in 10 $\hinv$Mpc box).  Black and red lines are simulations with
    \citet{haardt_and_madau_2012} and
    \citet{faucher_giguere_et_al_2009} UV
    backgrounds, respectively.
  }
  \label{fig:gamma_z}
\end{figure}

Figure \ref{fig:gamma_z} shows the evolution of our best-fit values for $\gamma$
in the resolution series of simulations and the box size series of simulations.
We fit the $\rhob$-$T$ relation with linear least squares in $\log \rhob$ and
$\log T$, fitting the range $-0.5 < \log_{10} \rhob < 0.5$ and $\log_{10}
T/\mathrm{K} < 4$. We see that convergence with spatial resolution is rather
fast, and that box size does not affect recovered value of $\gamma$. In
addition, we see that UV background as given by \citet{haardt_and_madau_2012},
shown in black, exhibits marginally more redshift evolution than that of \citet
{faucher_giguere_et_al_2009} (the red lines in Figure \ref{fig:gamma_z}). We
find similar results in the fit $T_0$ values, where there is a small resolution
effect for poor resolution, but the fit value remains the same between the
L10\_N256, 512, and 1024 runs. Box size appears to have no effect on the
resulting $\rhob$-$T$ line, as expected.

A large fraction of the gas lies on the $\rhob$-$T$ relation line --- about 90
per cent by volume and 50 per cent by mass in this case.
The significant scatter above the power-law relation line is due to shock 
heating, whereas the small scatter just below
the line is due to a subtlety in the discretization of the gravitational
source term in the total energy equation.  As discussed in \cite{almgren_et_al_2013},
the most obvious discretization is to compute the product of the
momentum and the gravitational vector.
While this is spatially and temporally second-order accurate, it allows
gravitational work to change the internal energy since
the update to the total energy is no longer numerically equivalent
to the update to the kinetic energy calculated using the updates to the momenta.
An alternative discretization defines the update to total energy 
only through the update to kinetic energy as calculated from the momentum equation. 
This maintains the analytically expected behavior 
of gravitational work contributing to the kinetic energy only.  
Through numerical testing we have determined that the latter formulation greatly
reduces the number of cells scattered below the line in void regions, 
while having a negligible effect on
the results otherwise.  Due to the small number of cells affected, 
the difference to the flux mean, pdf, and power spectrum 
at $k \leq 10$ $\hinv$Mpc is only 0.1\%.
The fraction of gravitational work in a timestep that directly
contributes to the internal energy (thereby increasing the temperature)
rather than kinetic energy ranges from $5 \times 10^{-3}$ at early times to
$5 \times 10^{-2}$ at late times for a run with CFL number of 0.5. 
These numbers are quite independent of the spatial resolution employed.
However, while the two discretizations of the gravitational source 
produce this difference in $\rhob$-$T$ regardless of the spatial resolution, 
they do converge to the same answer when refining the time-step:
in simulations run with a CFL number of 0.05,
the two formulations yield indistinguishable $\rhob$-$T$ plots, and
the fraction of gravitational work that contributes to the internal energy
stays below $5 \times 10^{-3}$ throughout the run.

\begin{figure}
  \begin{center}
    \includegraphics[width=\columnwidth]{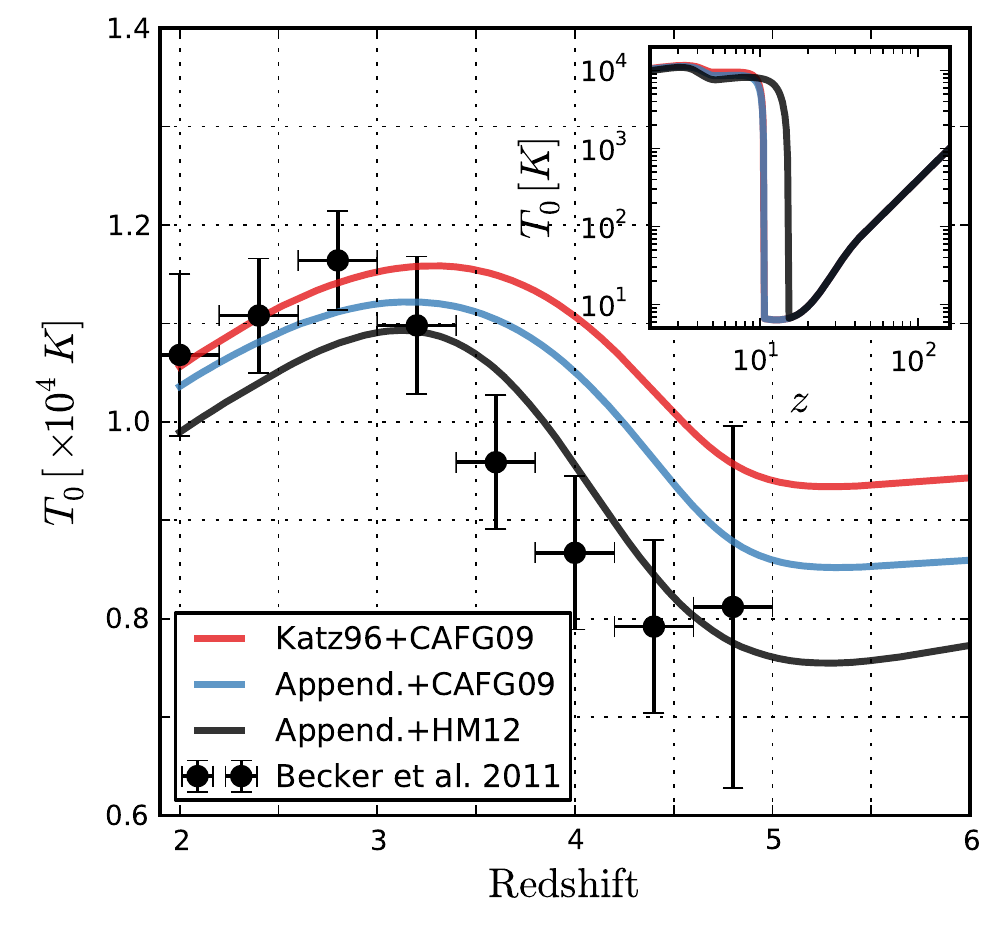}
  \end{center}
  \caption{The mean temperature of the IGM as a function of redshift
    in our simulations compared to the observations presented in
    \citet{becker_et_al_2011} (error bars are 2-$\sigma$). The red
    line shows a simulation using \citet{katz_et_al_1996} atomic rates
    and the \citet{faucher_giguere_et_al_2009} UVB. The blue line is
    obtained using the rates presented in the Appendix \ref{app:rates}
    of this paper and \citet{faucher_giguere_et_al_2009} UVB. The
    black line shows a simulation with the rates presented in Appendix
    \ref{app:rates} and the \citet{haardt_and_madau_2012} UVB.
    While the main figure shows the $T_0$ evolution over
    the observationally relevant redshifts, the inset figure
    shows the full simulation range starting at $z=159$, on a
    logarithmic scale.
    }
  \label{fig:T0_evolution}
\end{figure}

In Figure \ref{fig:rho-T}, we have also roughly marked
the four phases of the IGM: the diffuse IGM giving rise to the \Lyaf, the Warm
Hot IGM (WHIM) --- rarefied, shocked gas falling onto the filaments and halos,
the hot halo gas in the process of virialization, and the cooling and collapsing
condensed phase.
The overall shape of the $\rhob$-$T$ diagram is reproduced in almost any
cosmological simulation, even with low-resolution, as long as it includes
primordial gas heating and cooling.
However, we do find that larger box size simulations
produce more shocked gas around filaments (a more significant WHIM).
We do not see a significant resolution dependence on the fraction of gas in
the WHIM, but we see both that larger boxes have more gas in the WHIM,
and that the WHIM is shocked to higher temperatures. This is expected behavior,
as small-box simulations miss large-scale velocity components.
For the most interesting, diffuse gas region, the $\rhob$-$T$ relation and the
amount of scatter around it can also be affected by \HeII\ reionization.
For instance, \citet{mcquinn_et_al_2009} found that in their post-processed
radiative transfer simulations most of the reionization models increased $T_0$
and decreased $\gamma$ while significantly broadening the $\rhob$-$T$ relation,
mostly due to spatial variations in the $\rhob$-$T$ relation from RT effects
like shadows. Understanding the full effects of \HeII\ reionization on IGM is
beyond the scope of this work.

\begin{figure}
  \begin{center}
    \includegraphics[width=\columnwidth]{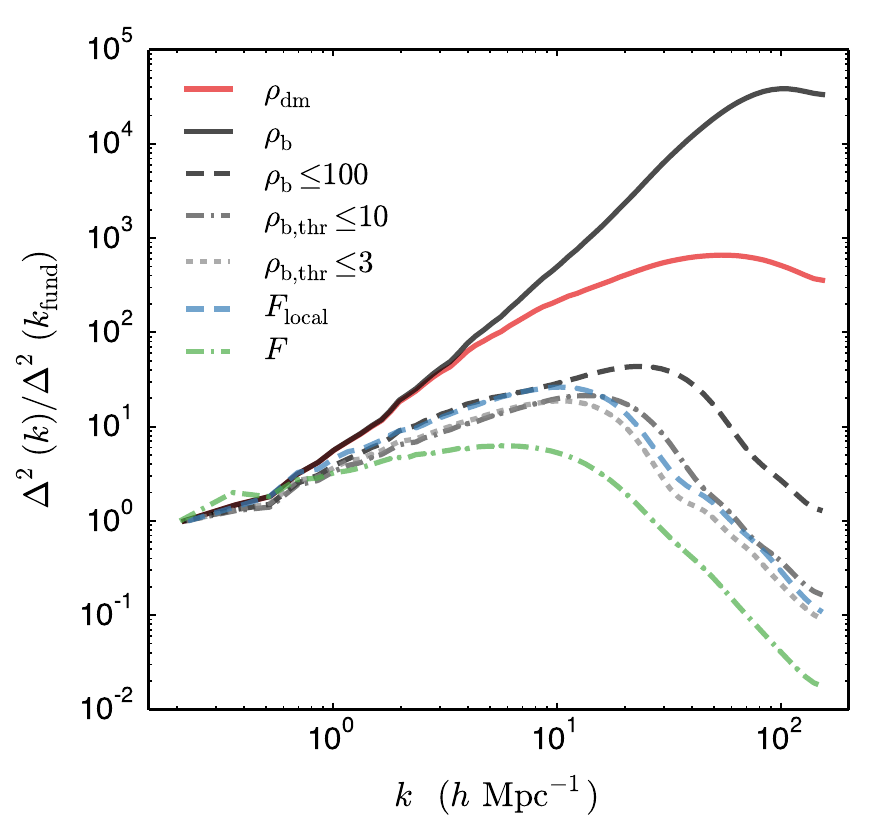}
  \end{center}
  \caption{Illustration of the effect of the filtering scale on power
    spectra. Here we show the power spectra of baryon density $\rhob$,
    dark matter density $\rhodm$, the local flux $F_{\rm local}$, the monopole
    of the redshift-space (observable) flux $F$, and the thresholded baryon
    density $\rho_{\rm b, thr}$ limited to 100, 10, and 3 times the
    mean density, all from the L40\_N2048 simulation at redshift
    $z = 2$.}
  \label{fig:filtering}
\end{figure}

In Figure~\ref{fig:T0_evolution}, we show the evolution of the
temperature at mean density. This is calculated as an average (in log
space) of the gas at mean density for temperatures $T < 10^5$ K at each time
step. We also show the effects of different UV
backgrounds, from \citet{faucher_giguere_et_al_2009} and
\citet{haardt_and_madau_2012}, and differing atomic rates (see
Appendix \ref{app:rates}). Qualitatively the temperature of the IGM
decreases at high redshifts due to the expansion and inverse Coulomb
cooling, then rises sharply during hydrogen reionization at
$z \sim 10$. We carried out a study spanning several orders of
magnitude in initial temperature for our simulations and have
determined that, due to adiabatic expansion, Compton cooling and
hydrogen reionization, no memory of the initial temperature is
retained at $z \lesssim 10$. In Figure~\ref{fig:T0_evolution}, we also
show recent observational results from \citet{becker_et_al_2011},
which is in good agreement with the $z = 2.4$ measurement recently
carried out by \citet{bolton_et_al_2014} but lower than the temperatures
inferred by \citet{lidz_et_al_2010}. It is interesting to point
out that the differences in temperature evolution that different
modern UV backgrounds produce, roughly 10 per cent, are less than
observational uncertainties. We also note that both of the UV
backgrounds we consider show two visible jumps in the temperature of
the IGM, corresponding to \HI\ and \HeII\ reionizations.

\begin{figure*}
  \begin{center}
    \includegraphics[width=\textwidth]{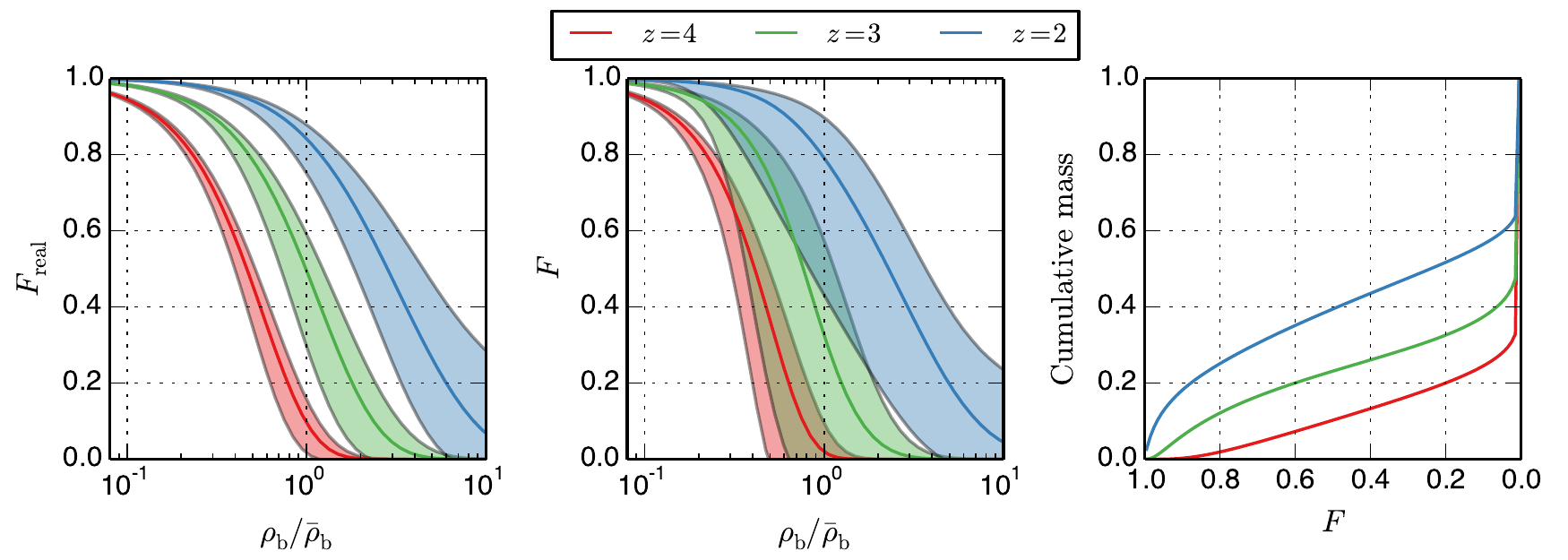}
  \end{center}
  \caption{
    Contribution of different density regions to the \Lyaf flux at $z = 2$, 3,
    4. Meax flux was rescaled to match Becker et al.~2011.
    \textit{Left:} the real-space flux vs.~gas density. The lines show
    the medians and the filled regions show the normalized median absolute
    deviation (normalized to match one standard deviation for a normal
    distribution). \textit{Middle:} the redshift-space flux vs.~gas density.
    \textit{Right:} the cumulative distribution of mass vs.~redshift-space
    flux.}
  \label{fig:F_rho}
\end{figure*}

Due to the direct influence of pressure forces, baryon fluctuations
are suppressed compared to dark matter (which is affected by the gas
pressure only because the gravitational acceleration has a component
due to the gas).  Our simulations do not account for the details of
star formation, feedback from stars or Active Galactic Nuclei; the
regions that should be galaxies are only blobs of overcooled gas. Due
to this overcooling inside halos, small-scale fluctuations in the
baryonic component are artificially enhanced as shown in the solid
black line of Figure~\ref{fig:filtering} (the red line shows the dark
matter power spectrum for reference).  Since we know that our
simulations do not realistically represent the gas quantities in high
density regions, we can exclude them from our analysis at which point
the filtering scale becomes clear.  To highlight this, in
Figure~\ref{fig:filtering}, we show the baryon power spectrum with
several density thresholds.  These are obtained by ``clipping'' the
original baryon density field, i.e.~resetting the densities higher
than the threshold down to the selected threshold value.
The clipping is done here only for illustrative
  purposes. This is qualitatively similar to what happens with the
  \Lyaf signal --- where the flux drops to zero at a certain density,
  and any higher density has no additional effect.  Clipping of the
small-scale fluctuations also introduce a linear bias on large
scales. For clarity, we have normalized all power spectra to be 1 at
the fundamental (box-scale) mode.  The different threshold value power
spectra also illustrate that there is a density dependence of the
filtering scale. As the threshold density decreases, the filtering
scale increases. We also show two flux power spectra to see how they
probe the filtering and thermal broadening scale. We computed the
\Lyaf flux without redshift-space distortions or thermal broadening,
which we call the local flux $F_{\rm local}$. In this case, the
optical depth for the local flux is just the appropriate rescaling of
the \HI\ number density,
\begin{equation}
  \tau_{\rm local} = \frac{\pi e^2 f_{lu} \lambda_0}{m_e c H(z)} n_{\HI}  \; ;
  \; F_{\rm local} = e^{-\tau_{\rm local}} \; .
  \label{eq:local_f}
\end{equation}
The dashed blue line is the local flux spectrum, which shows
pressure support smoothing at a scale roughly matching the $\rho_{\rm b, thr}
\leq 10$ result. Note the little difference between thresholding at 10 times
the mean baryon overdensity and 3 times the mean.
We also show the monopole of the 3D flux power spectrum as the
dashed green line, which includes smoothing not just from pressure support but
also contributions from thermal broadening and redshift-space distortions,
giving rise to even more filtering on small scales.

In Figure \ref{fig:F_rho}, we show relations between the \Lyaf\ flux and the gas
density. In the left panel we plot the real-space flux of cells as a function of gas
density. For each density bin we plot the median and normalized median absolute
deviation (normalized to equal the standard deviation for a normal distribution)
independently above and below. This serves as a qualitative estimate of what density
regimes contribute to the \Lyaf signal at different redshifts. For instance, we
immediately see that a majority of the signal at high redshift originates in
under-dense regions, while at $z = 2$, it lies in the mild overdensities. In the
middle panel, we show similar info, but this time we use the redshift-space
flux. As redshift-space distortions couple regions several Mpc away and can map
different cells to the same redshift-space position (see Figure \ref{fig:sample_skewer}),
the redshift-space flux is less correlated with density and thus
exhibits more scatter than in real space. However, the median lines are similar
at all redshifts. In the right panel, we show the cumulative mass of cells with
fluxes above some value. The sharp
rise in the cumulative mass at $F = 0$ shows the mass fraction in the
saturated regions of the forest, filaments and halos. This figure also shows the difficulty
of simulating \Lyaf signal at high redshifts, $z \gtrsim 4$: we immediately
see that small fluctuations in density produce significant difference in flux.
Arguably, this effect is more critical for numerical convergence than the
decrease in filtering scale described in Section \ref{sec:sim}.

Historically, the \Lyaf was studied in the context of absorption line
systems. However, the process of \Lyaf line finding and fitting is
not well-defined and results can vary between implementations.
For this reason we will explore line statistics separately in
Section \ref{sec:small_scale}, and in the following sections we will focus
on the flux $N$-point correlation statistics.

%% file: sec4.tex
\section{Resolution study}
\label{sec:resolution}

\begin{figure*}
  \begin{center}
    \includegraphics[width=\textwidth]{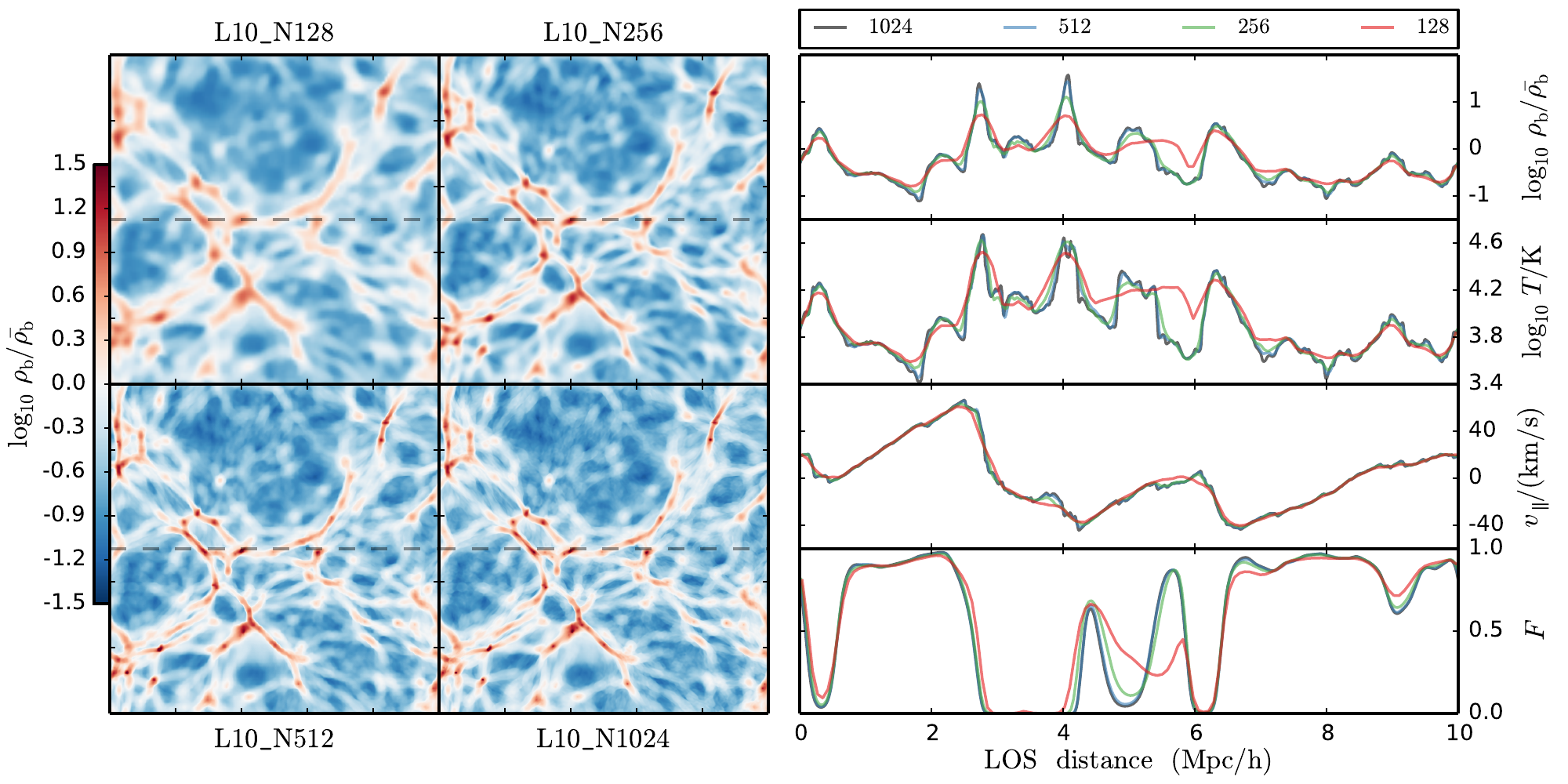}
  \end{center}
  \caption{Illustration of the effects of resolution in the 10 $\hinv$Mpc
    simulations at $z = 2.5$. On the left, we plot slices $\sim 150$ $\hinv$kpc
    deep (2 cells in the 128$^3$ simulation) of the baryon density. On the right
    we show values along the skewers marked as dashed lines in the slices. The
    skewers were selected in the same position in each simulation.}
  \label{fig:res_series}
\end{figure*}

The physical resolution required to model \Lyaf flux statistics varies
significantly with redshift, with higher redshifts requiring higher
resolution for the same relative error.  There are
  several physical effects which contribute to this behavior.  One is
  the change in the comoving filtering scale, which decreases with
  increasing $z$.  We further demonstrate the increasing steepness of
  the flux--density correlation as a function of redshift in Figure
  \ref{fig:F_rho}, which means that for the same relavite error in
  $\rho_b$ or $\NHI$ the relative error in flux will be larger at
  $z=4$ than at $z=2$.  Finally, the gas in voids is $\sim$2 times
  colder than the gas at the mean density (and a factor of $\sim$4
  colder than in mild overdensity regions), therefore thermal
  broadening of lines is less at high redshifts than at low ones.

The average
transmission sharply increases going to lower redshifts as the physical density
of neutral hydrogen decreases, primarily due to the expansion of the universe,
with secondary changes due to the ionizing background radiation. Here we explore the
accuracy of simulated \Lyaf flux statistics at $2 \leq z \leq 4$, the relevant
range for most of current and near future observations. We focus on results in
our 10 $\hinv$Mpc boxes, as they offer the easiest path to an increase in
resolution, but we have also explicitly checked that the conclusions presented
here are valid in the case of the larger-box simulation series as well
(see Table \ref{tab:sims}). In other words, we do not observe that numerical
errors due to missing modes (explored in the next section of this paper) couple
with resolution error at more than the percent level.
The same behavior was observed in Gadget simulations presented in
\citet{stark_et_al_2014a}. The simulations we present in this section were
done with \citet{faucher_giguere_et_al_2009} UV backround.

The results of this section are applicable to grid/Eulerian codes,
where the effective resolution of pressure forces is commonly better than the
effective resolution of gravitational forces. For example, many tests show that
hydrodynamic quantities are already very accurate at 1 - 2 cells away from
discontinuities --- see e.g. \citet{almgren_et_al_2010} for the case of the hydro
algorithm implemented in Nyx but the same is true for schemes used in virtually
any other cosmological Eulerian hydro code to date. On the other hand, the
gravitational force resolution is much worse. Grid codes use a Particle-Mesh
(PM) scheme to compute gravitational forces, which is very fast but suffers from
smoothing the density field at small scales. Generally, two particles must be
separated by at least 5 cell sizes for the gravitational force to match
$1/r^2$ (for example, see \citealt{rru2009}). The opposite is true in most SPH
\Lyaf simulations presented in the literature. In this case, gravitational
resolution is much higher for the same grid size/number of particles, with
the gravitational forces computed with a TreePM or particle-particle PM hybrid
scheme.  This provides a much (roughly 10 times) higher gravitational
resolution than the grid codes for the same grid configuration. At the
same time, the SPH kernel smooths the hydrodynamic quantities on scales of
$\sim 2\times$ the mean inter-particle spacing for gas around mean density.
In this regard, the resolution study presented here is not directly applicable
to \textit{all} codes.
However, we believe that the results of other studies we conduct in this
paper are largely code independent.

In Figure~\ref{fig:res_series}, we provide an illustration of how the grid
resolution affects relevant IGM structures and the \Lyaf flux. Here we use our
four 10 $\hinv$Mpc simulations and plot a slice and skewer in the same position
from each simulation. The baryon density slice is on the left, while on the
right we show baryon density, temperature, velocity parallel to the line of
sight, and transmitted flux along the skewer. In both slices and skewers, we see
a clear pattern of converging values. Overall, the L10\_N256, L10\_N512, and
L10\_N1024 results agree very well, and the L10\_N128 results are similar,
but have structural differences.
In the baryon density slices, we see that structures in L10\_N128 are severely
under-resolved. The large cell size prevents the collapse
of dense regions, and the solution contains puffy filaments and halos, and less
depleted voids. The relatively small number of resolution elements also means
that the simulation misses the rare, extremely low and high density regions. In
the L10\_N256 slice we can see structure that resembles the highest-resolution
case much more closely, although the filaments and halos are still a bit
puffier. Finally, the differences between the L10\_N512 and L10\_N1024 slices
are minor. The filament widths are essentially the same and the differences
noticeable by eye are restricted to the very dense galaxy-like regions. This is
fortunate for modeling the \Lyaf signal, since the dense regions are saturated
in absorption, rendering those differences undetectable in flux. In the baryon
density and temperature skewer values we see the same patterns. The L10\_N128
simulation reproduces the broad shapes, but fails catastrophically at the extremes.
The other simulations match each other much better, and the L10\_N512 and
L10\_N1024 values are very close at all positions. One difference is in the
dense structure near the LOS distance of 4 $\hinv$Mpc, where the L10\_N1024
simulation resolves two temperature peaks, almost certainly accretion shocks.
The L10\_N512 simulation just barely reproduces the two peaks while this feature
is smeared out as one bump in the two lower resolution simulations. The flux
field proves to be unaffected by those kinds of details as can be seen in the
lowest panel. Interestingly, the parallel velocity values show much less
difference between runs. This reinforces the common knowledge that bulk flows
are not as sensitive to resolution as they are to the box size. Finally, the
most important differences lie in the flux values. Here, we see that L10\_N512
and L10\_N1024 runs look virtually identical. Further, many of the small
differences in baryon quantities between the L10\_N256 and higher resolution
runs are washed out in the optical depth calculation --- the similar velocity
shifts and significant broadening provide a fortunate ``fudge factor'' when only
considering the flux. The same cannot be said for the L10\_N128 simulation
fluxes though, which show significant differences, especially at the LOS
distance of 5 $\hinv$Mpc. We have checked this for several random skewers and
with all other redshifts available and note that the overall conclusions remain
the same, although differences very mildly increase with redshift. In the
following sections, we quantify the above differences in resolution.

\subsection{The Mean Flux}

\begin{figure}
  \begin{center}
    \includegraphics[width=\columnwidth]{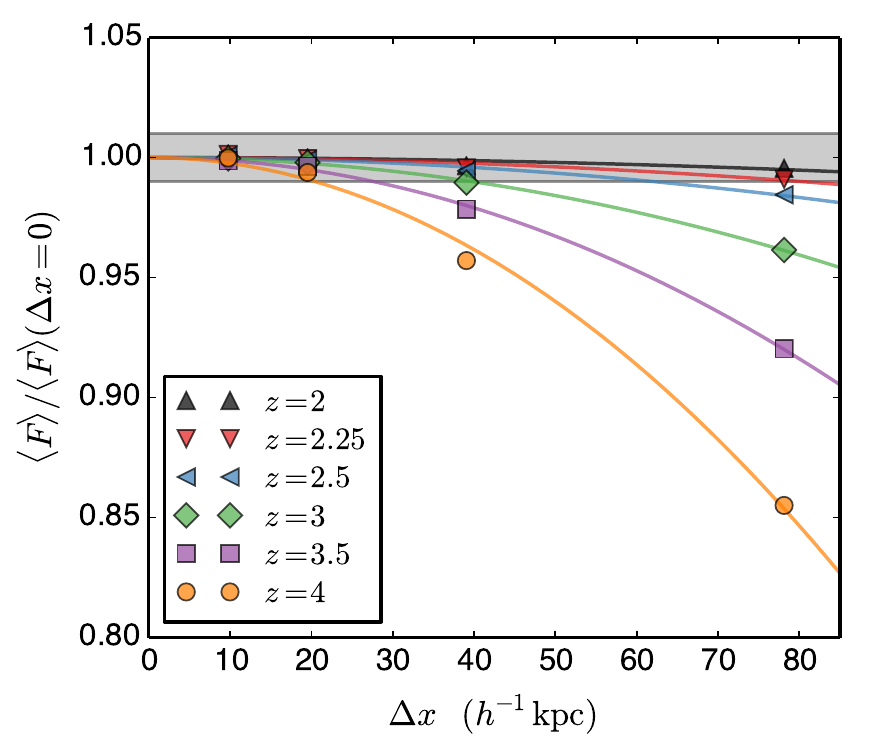}
  \end{center}
  \caption{Dependence of the mean flux on resolution for 6 redshifts. The lines
    are the best fits of the form $F(\Delta x) = F(0) - k \Delta x^2$, and the
    data plotted is normalized to the case $\Delta x \rightarrow 0$, i.e.
    $F(0)$. The gray shaded region shows a $\pm$1 per cent interval.}
  \label{fig:mf_conv}
\end{figure}

The simplest possible flux statistic is the mean transmitted flux
$\mf$, or equivalently, the effective optical depth $\tau_{\rm eff} =
-\log \mf$.  Observations show that the mean flux smoothly evolves
from about $0.4$ at $z = 4$, to about $0.9$ at $z = 2$, as expansion
gradually lowers the (proper) \HI\ density and the UVB intensity
slowly increases \citep{becker_et_al_2013}. Figure~\ref{fig:mf_conv}
shows the mean flux in four of our 10 $\hinv$Mpc simulations at
the snapshot redshifts. Here we immediately see that higher redshifts
need higher resolution to maintain the same accuracy. The coarsest
resolution run is within 1 per cent of the highest resolution run at $z = 2$,
but $\sim$15 per cent different at $z = 4$. Nyx is second-order accurate in
both the gas dynamics solver and gravity. Although the \Lyaf flux is a
heavily processed quantity derived from the density, velocity, and
internal energy of the gas, its mean clearly exhibits quadratic
convergence, as shown in Figure~\ref{fig:mf_conv}. The resolution
series allows us to determine $F(0)$ --- the simulated mean flux in
the limit $\Delta x \rightarrow 0$. Understanding the effect of
resolution on the mean flux is important for simulation results that
rescale optical depths to match an observed mean flux, as we explore
later in Section \ref{sec:uvb}.

\subsection{The Flux PDF}

The pixel flux probability distribution function (FPDF) is the probability
density function of the pixel fluxes
\citep{1997ApJ...489....7R, mcdonald_et_al_2000, becker_et_al_2007}.
The probability density function $P(F)$ is defined such that the integral
of $P$ over the full $F$ range is equal to 1, $\int P(F) \, dF = 1$.
In the case of equally-spaced $F$ bins, the $P(F)$ values are just the
appropriately rescaled histogram.  We compute the FPDF with 50 equally
spaced bins of $F \in [0, 1)$.
The FPDF is a relatively smooth function, with a shape typically peaked
at $F = 0$ and 1, and rising at intermediate fluxes (although the slope will
be negative and there will be no $F \sim 1$ peak at a high enough redshift).
In principle this one-point statistic is a good probe of the thermal history
of the IGM and the amplitude of density fluctuations, however, the FPDF is
very sensitive to systematic effects such as the resolution of the
spectrograph, determination of the quasar continuum level and/or pixel noise.
Recently, the FPDF was measured using a sample of 3,393 high signal-to-noise
BOSS quasar spectra \citep{Lee2014}, where they found a good fit to the data with a
temperature-density slope of $\gamma = 1.6$ and strongly disfavoring inverted
$\rho$-$T$ models ($\gamma < 1$).

\begin{figure}
  \begin{center}
    \includegraphics[width=\columnwidth]{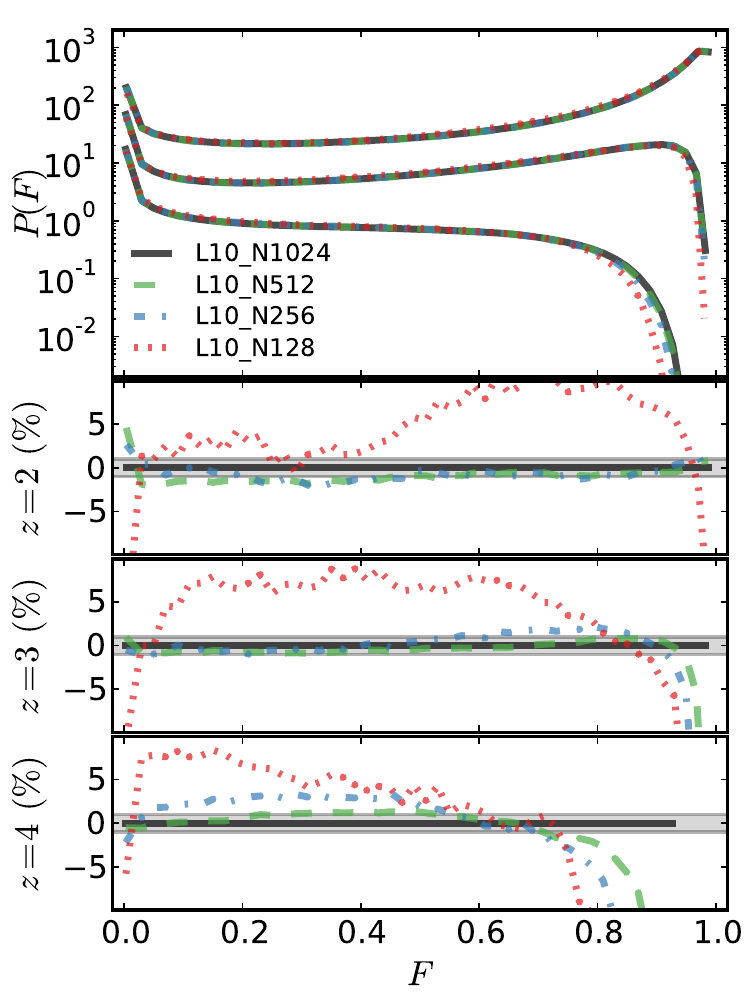}
  \end{center}
  \caption{Convergence of the flux PDF with respect to grid resolution. For
    clarity, in the upper panel we have multiplied the $z = 2$ data by a factor
    of 100, and the $z = 3$ data by 10.}
  \label{fig:res_fpdf}
\end{figure}

We consider the convergence of the flux PDF $P(F)$ at redshifts 2, 3, and 4,
which we show in Figure \ref{fig:res_fpdf}. Again, we note that the resolution
requirements increase with redshift. It appears that this is mostly due to the
rarity of transmissive regions at high redshift. In the $z = 4$ ratio panel, we
can see that even the L10\_N1024 simulation does not capture pixels with $F
\approx 1$, as the black line cuts short near 0.9. It is instructive to look
again at Figure \ref{fig:F_rho}, which clearly shows how difficult it is to obtain
$F = 1$ cells at high redshifts, even in very underdense regions.
The L10\_N512 simulation does match
the highest resolution to a few percent up to $F < 0.6$. At lower redshifts
however, L10\_N512 is in percent agreement with the 1024$^3$ run, while the
L10\_N256 is within a few percent. As expected from the qualitative inspection
at the beginning of this section, the L10\_N128 simulation results are
qualitatively different at all redshifts. At high redshift, it severely
underestimates transmissive pixels, pushing the low-flux end above the other
simulations. At low redshift, it misses the extreme fluxes at both ends, raising
the probability at moderate fluxes above other simulations.

\subsection{The 1D Flux Power Spectrum}

\begin{figure}
  \begin{center}
    \includegraphics[width=\columnwidth]{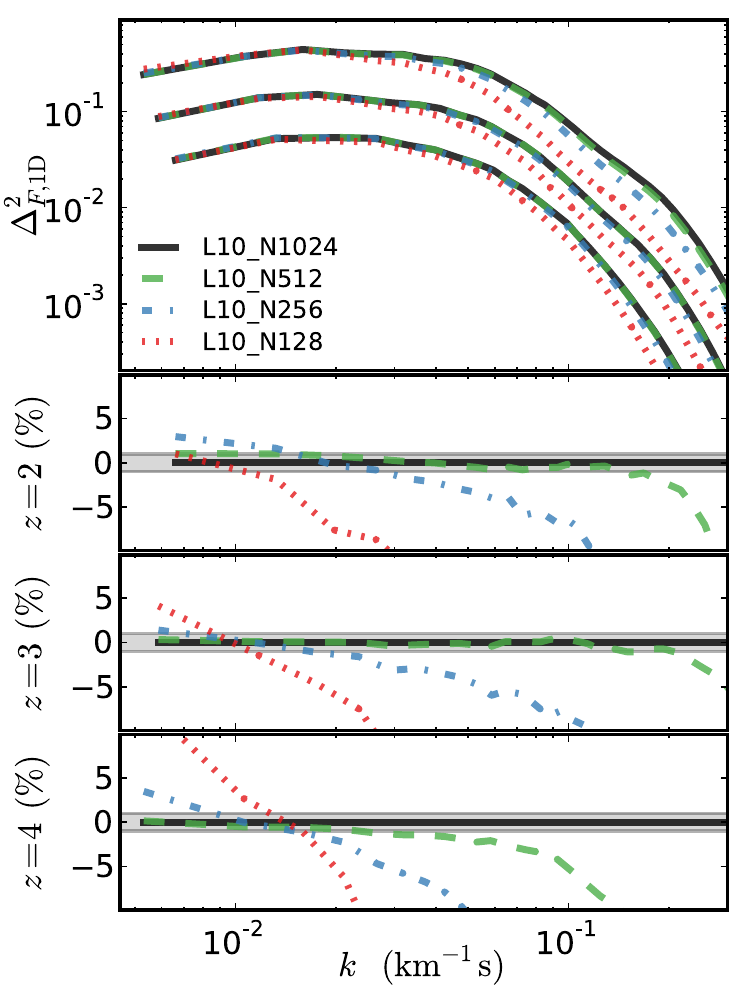}
  \end{center}
  \caption{Convergence of the 1D power spectrum at redshifts 2, 3,
    and 4. Here we do not modify the values in the upper panel --- the
    flux power increases with increasing redshift, so from top to
    bottom are redshifts 4, 3, and 2.}
  \label{fig:res_pf1d}
\end{figure}

Spatial correlations of the \Lyaf offer a promising route to measuring the
density fluctuations at high redshifts and small scales.
Analogously to density fluctuations, one can define the \Lyaf flux contrast
with respect to the mean as $\delta_F \equiv (F - \mf) / \mf$.
Here, $\mf$ is the global average for a given redshift, but we caution that
some older publications use the per skewer mean flux as $\mf$.
In 3D, the Fourier transform $\hat{\delta}$ and dimensionless power spectrum
$\Delta^2$ are
\begin{equation}
  \begin{aligned}
    & \hat{\delta}(\mathbf{k})
      = V^{-1} \int \delta(\mathbf{x}) e^{i \mathbf{k} \cdot \mathbf{x}} d^3x \\
    & \Delta^2(k) = \frac{k^3 P(k)}{2 \pi^2}
      = \frac{k^3}{2 \pi^2} V \langle \hat{\delta} \hat{\delta}^* \rangle
  \end{aligned}
  \label{eq:pk_def}
\end{equation}
where $V$ is the domain volume, $P$ is the power spectrum (in units of volume),
and the average $\langle \rangle$ is over shells in $k$-space. In 1D, along the
line of sight, the Fourier transform $\hat{\delta}_{\rm 1D}$ and dimensionless
power spectrum $\Delta_{\rm 1D}^2$ are
\begin{equation}
  \begin{aligned}
    & \hat{\delta}_{\rm 1D}(k_\parallel)
      = L^{-1} \int \delta(x_\parallel)
        e^{i k_\parallel x_\parallel} dx_\parallel \\
    & \Delta^2_{\rm 1D}(k_\parallel)
      = \frac{k_\parallel}{\pi} P_{\rm 1D}(k_\parallel)
      = \frac{k_\parallel}{\pi} L
        \langle \hat{\delta}_{\rm 1D} \hat{\delta}_{\rm 1D}^* \rangle \\
  \end{aligned}
\end{equation}
where $L$ is the domain side length, $P_{\rm 1D}$ is the 1D power spectrum (in
units of length), and the average $\langle \rangle$ is over modes with magnitude
$k_\parallel$. The expressions above are written in comoving coordinates $x$
because this is most convenient in simulations. Observationally, however, one
measures the flux in velocity units rather than comoving scale, so we also
present some results in these units. This creates a small redshift-dependence in
the transform between $k$ in comoving coordinates ($\hinv$Mpc) and $k$ in
velocity coordinates (km$^{-1}$ s).

\begin{figure*}
  \begin{center}
    \includegraphics[width=\textwidth]{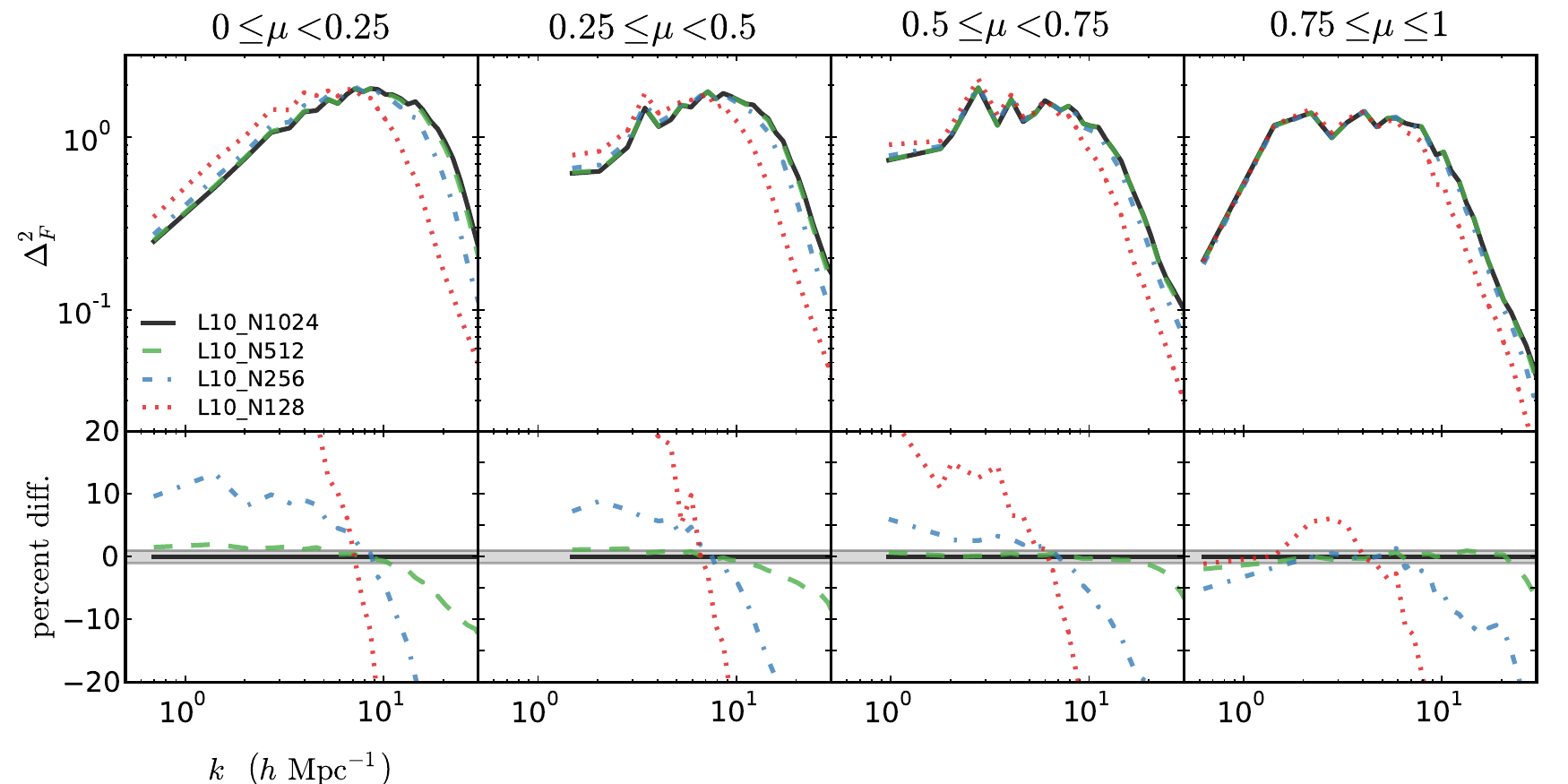}
  \end{center}
  \caption{Convergence of the 3D flux power spectrum in 4 $\mu$ bins
    at redshift $z = 4$ (the agreement is better at lower redshifts).
    The leftmost panel shows the power spectrum mostly perpendicular to the
    line of sight, the rightmost is mostly parallel to the line of sight.
    The agreement is better along the line of sight due to thermal
    broadening which erases some of the differences.}
  \label{fig:res_pf3d}
\end{figure*}

The flux power spectrum is redshift-space distorted and thus anisotropic. But before
going into the full anisotropic power spectrum, we will explore its 1D
counterpart, obtained by Fourier transforming $\delta_F$ along each line of
sight and averaging in $k_\parallel$ bins. Historically, there were only few a
high signal-to-noise quasar spectra which were located far apart from each
other. In this limit, individual spectra can be assumed to be independent from
each other, and the 1D flux power spectrum is the only relevant measure of flux
clustering. Even with SDSS increasing the number of quasar spectra to $\sim
10^4$, estimating 3D correlations was still too inaccurate. Today, the main
strength of the 1D flux power spectrum is that it can probe relatively small
scales, down to $\sim 0.1$ $\hinv$Mpc. Therefore, it is a good test of the
nature of dark matter and the mass of neutrinos. The observed 1D flux power
spectrum has been studied in
\citet{croft_et_al_1998, croft_et_al_1999, mcdonald_et_al_2000,
2002ApJ...581...20C, kim_et_al_2004, mcdonald_et_al_2006,
palanquedelabrouille_et_al_2013}.

Here we consider the resolution convergence of the dimensionless 1D
flux power spectrum $\Delta_{F, \mathrm{1D}}^2$ at redshifts 2, 3, 4,
as shown in Figure \ref{fig:res_pf1d}, leaving other effects for the
subsequent sections. We want to emphasize again that while we show
results for the 10 $\hinv$Mpc box, we have checked that conclusions
are the same in other convergence series in larger box sizes.
Figure \ref{fig:res_pf1d} shows that 20 $\hinv$kpc resolution
run (L10\_N512) agrees with 10 $\hinv$kpc run
(L10\_N1024) to better than 1 per cent at
redshifts $z = 2$ and $z = 3$ even beyond $k = 0.1$ km$^{-1}$s. Those are much
smaller scales than what is usable for making cosmological constrains, as
interpreting observations becomes difficult at such small scales due to
metal lines and other contaminents. Those scales are also not correctly
modeled with the physics included in our simulations. The $z = 4$ ratio panel
again shows how hard it is to get flux statistics accurately at very high
redshifts: the 20 $\hinv$kpc run departs from the 10 $\hinv$kpc run by 1 per
cent around $k = 0.03$ km$^{-1}$s.
This is still sufficiently good for cosmological purposes, especially since
the number of observed quasars at such high redshifts is rather small.

One important difference between density and flux fields, is that density
is manifestly conserved in our simulations, and its mean value is an
input parameter. In contrast, the mean flux will differ --- even when the
cosmology and physical models for cooling and heating processes are kept
constant --- due to numerical resolution, box size, and the random realization
of the initial density field.
Another characteristic feature of the flux field is that it is bounded in
value: $0 \le F \le 1$.  The maximum possible fluctuations around the mean
value are therefore also limited.  This is again in contrast to the density
fluctuations, as density contrast can in principle go to infinity.
As a result strong suppression of flux fluctuations on small scales --- for
example due to numerical effects like lack of resolution --- results in
increased fluctuations on large scales.
This effect is also clearly visible in the Figure \ref{fig:res_pf1d}, and
is more noticeable when the fluxes in simulations of different resolutions
are not rescaled to the same mean (as done here).
This effect is the biggest issue to getting the 1D flux power spectrum correct
in numerical simulations.
Whereas the density power spectrum can, to some extent, be simulated with low
resolution simulations using a series of nested-size boxes, each box
recovering accurately only a small portion of $P_m(k)$, the flux power spectrum
in an under-resolved \Lyaf simulation will be inaccurate on all scales.

\subsection{The 3D Flux Power Spectrum}

Because of redshift space distortions along the line of sight, the 3D flux field
is not isotropic, and therefore it is inappropriate to simply average it over
$k$ shells as indicated in Equation \ref{eq:pk_def}. A common way to describe
the anisotropic power spectrum is in terms of the cosine of the angle between
the wavevector $\bf k$, and its line of sight component $k_\parallel$, $\mu =
k_\parallel / |\mathbf{k}|$. In other words, the 3D power spectrum is binned in
$k$ and $\mu$, rather than $k$ alone, resulting in $P(k, \mu)$. On large scales,
the shape of flux power spectrum is very similar to the matter power spectrum
(see e.g.~\citealt{slosar_et_al_2009}).
As such, the \Lyaf is a tracer of the large-scale structure
which can measure the characteristics scale of baryon acoustic oscillations (BAO)
and use it as a standard ruler to measure the distances and the expansion rate
of the Universe.  Recently, the first measurements of the
cosmological parameters via the location of the BAO peak
in the \Lyaf was made with BOSS data \citep{slosar_et_al_2011,
slosar_et_al_2013, busca_et_al_2013}.

The 3D flux power spectra for 4 $\mu$ bins are shown in
Figure \ref{fig:res_pf3d}. The leftmost panel shows the power spectrum
mostly perpendicular to the line of sight, and the rightmost is mostly
parallel to the line of sight. Here we show only redshift $z = 4$
data, since this is where the agreement is worst. The agreement is better
at lower redshifts, as expected from previous 
considerations presented in Sections \ref{sec:sim}, and \ref{sec:properties}.
We immediately notice that
different resolutions agree more along the line of sight than across
it. This is a result of thermal broadening which erases much of the
small scale differences, bringing the results of lower resolution runs
closer to the high resolution solution. We nevertheless see that the
20 $\hinv$kpc resolution is good enough for most practical
purposes, typically 1 per cent away from the 10 $\hinv$kpc result at
all redshifts for $k < 10$ $h$ Mpc$^{-1}$.  From the
observed rate of convergence we expect that the difference between a
10 $\hinv$kpc and a (hypothetical) 5 $\hinv$kpc run would be
sub-percent.

\subsection{Richardson Extrapolation}

For a convergent numerical method, it is in principle possible to increase
the accuracy of a measured quantity by carefully combining
the results from a sequence of simulations where the only difference is the spatial
resolution. Here, we discuss the case of combining results via Richardson
extrapolation. A numerical method which is $p$-th order accurate in space
(meaning the error term is proportional to $h^p$ where $h$ is discretization
element), produces a numerical approximation $Q(h)$ as
\begin{equation}
  Q(h) = Q + Ah^p + O(h^{p+1}) \ .
\end{equation}
The first term on the right-hand side $Q$ is the exact value, the second term is
the leading error, and the third term is the higher-order error. The leading
error can be removed with simulations using two different values of $h$, for
example $h$ and $rh$, where $r$ is the refinement ratio, giving an extrapolation
expression
\begin{equation}
  Q_R = \frac{r^p Q(h) - Q(rh)}{r^p - 1} \ .
  \label{eq:richarson}
\end{equation}
The order of accuracy, $p$, is theoretically known from the algorithm
implemented, but can also be determined from actual numerical results. This
requires at least 3 simulations, in which case $p$ can be calculated as:
\begin{equation}
  p = \frac{\ln \left( \frac{Q(r^2h) - Q(rh)}{Q(rh) - Q(h)} \right) }{ \ln r } \ .
\end{equation}

\begin{figure}
  \begin{center}
    \includegraphics[width=\columnwidth]{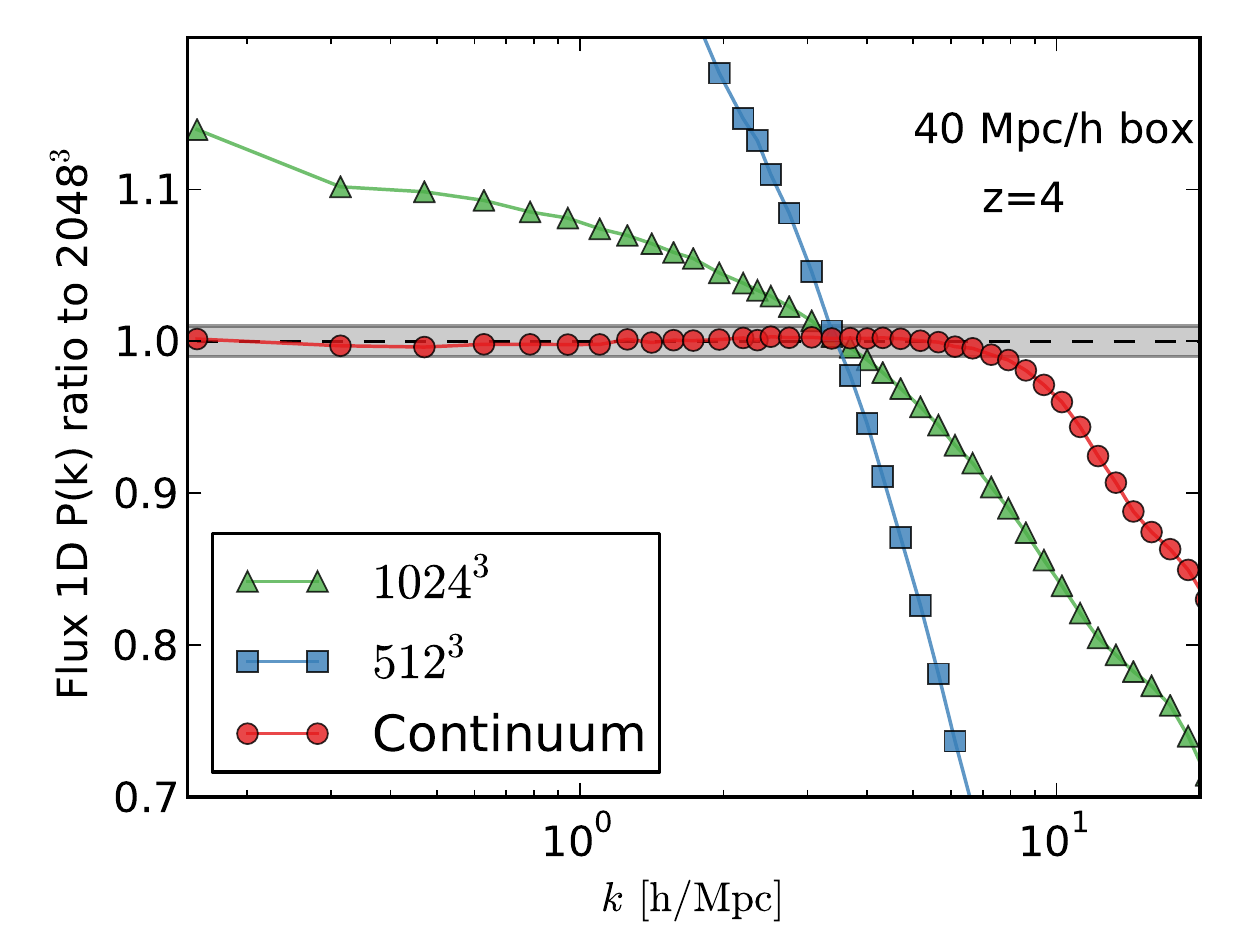}
  \end{center}
  \caption{ Convergence of 1D flux power spectrum at redshift $z=4$ in
    a 40 Mpc/h box.  We show the results for 512$^3$ and 1024$^3$
    simulations, together with the Richardson extrapolation from these
    two runs using the theoretically expected order of convergence,
    $p=2$. The shaded band shows $\pm$1 per cent range.  }
  \label{fig:rich_flux}
\end{figure}

In Figure \ref{fig:rich_flux} we show one such Richardson extrapolation, applied
to our 1D flux power spectrum results. Here we consider the results at $z = 4$,
since the differences are largest at this time, and we also use our 40
$\hinv$Mpc box size simulations. We see that the run with 40 $\hinv$kpc
resolution (L40\_1024) is not as accurate, differing from the 20 $\hinv$kpc
resolution run (L40\_2048) by up to 15 per cent in the range $k < 10$ $h$
Mpc$^{-1}$. The run shown in blue (79 $\hinv$kpc resolution) --- is even
worse. However, the ``continuum'' value deduced from these two runs via
Richardson extrapolation, using Equation \ref{eq:richarson}, shows remarkable
agreement with the highest resolution reference run. Here, we have used an order
$p = 2$, since Nyx is formally second-order. The fact that the theoretical value
works so well on the 1D flux power spectrum is very reassuring. For larger $k$
values the extrapolation fails. This is expected as the extrapolation procedure
cannot reproduce power that is not present in the underlying low resolution
simulations, nor can it work in the regime where the convergence breaks down due
to a dramatic loss of accuracy close to the resolution limit. Nevertheless, we
see that extrapolation can significantly improve accuracy from low-resolution
simulations on scales where convergence does hold. This improvement is a strong
evidence that numerical errors beyond the discretization scheme Nyx employs are
small to none, and a confirmation of the desired rate of convergence even in a
very processed quantity like the 1D flux power spectrum.

%% file: sec5.tex
\section{Box Size / Missing Modes}
\label{sec:box_size}

\begin{figure}
  \begin{center}
    \includegraphics[width=\columnwidth]{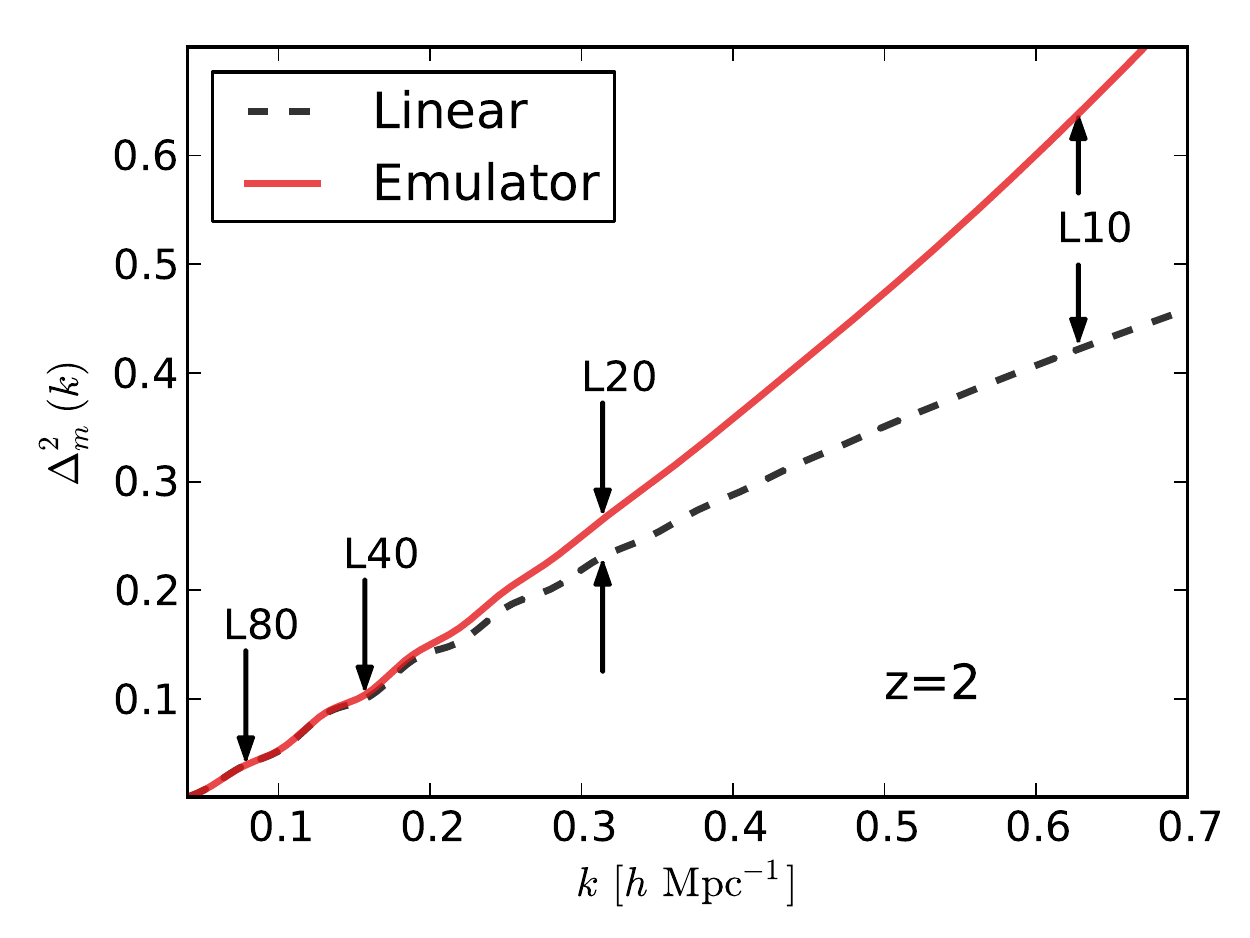}
  \end{center}
  \caption{
    Dimensionless linear and non-linear matter power spectrum at $z = 2$.
    Arrows show the scale of the fundamental mode in our boxes of 10, 20, 40,
    and 80 $\hinv$Mpc.
  }
  \label{fig:lin_emu}
\end{figure}

In cosmological simulations, we model a representative, but finite
volume of the universe using periodic boundary conditions in all 3
dimensions. As a result, perturbations on scales larger than the box
size are identically zeroed out while fluctuations which are smaller,
but comparable to the box size, are poorly sampled. A finite box size
can compromise \Lyaf results in at least two different ways. First, in
cosmological simulations in general, once the non-linear scale of
density fluctuations becomes similar to the box size the evolution of
modes is suppressed compared to what would be obtained with a larger
simulation box. Second, and relevant to the \Lyaf, a lack of
large-scale modes --- even linear ones --- can lead to an
underestimation of the bulk flow velocities of the gas. This, in turn,
leads to an underestimation of the heating from accretion shocks. This
has a significant impact on both the thermal broadening of lines (via
the amount of shock heating) and on the redshift-space distortions of
the optical depth. Thus if the simulated box is too small it cannot
produce representative \Lya flux statistics for the cosmology of
interest.

\begin{figure}
  \begin{center}
    \includegraphics[width=\columnwidth]{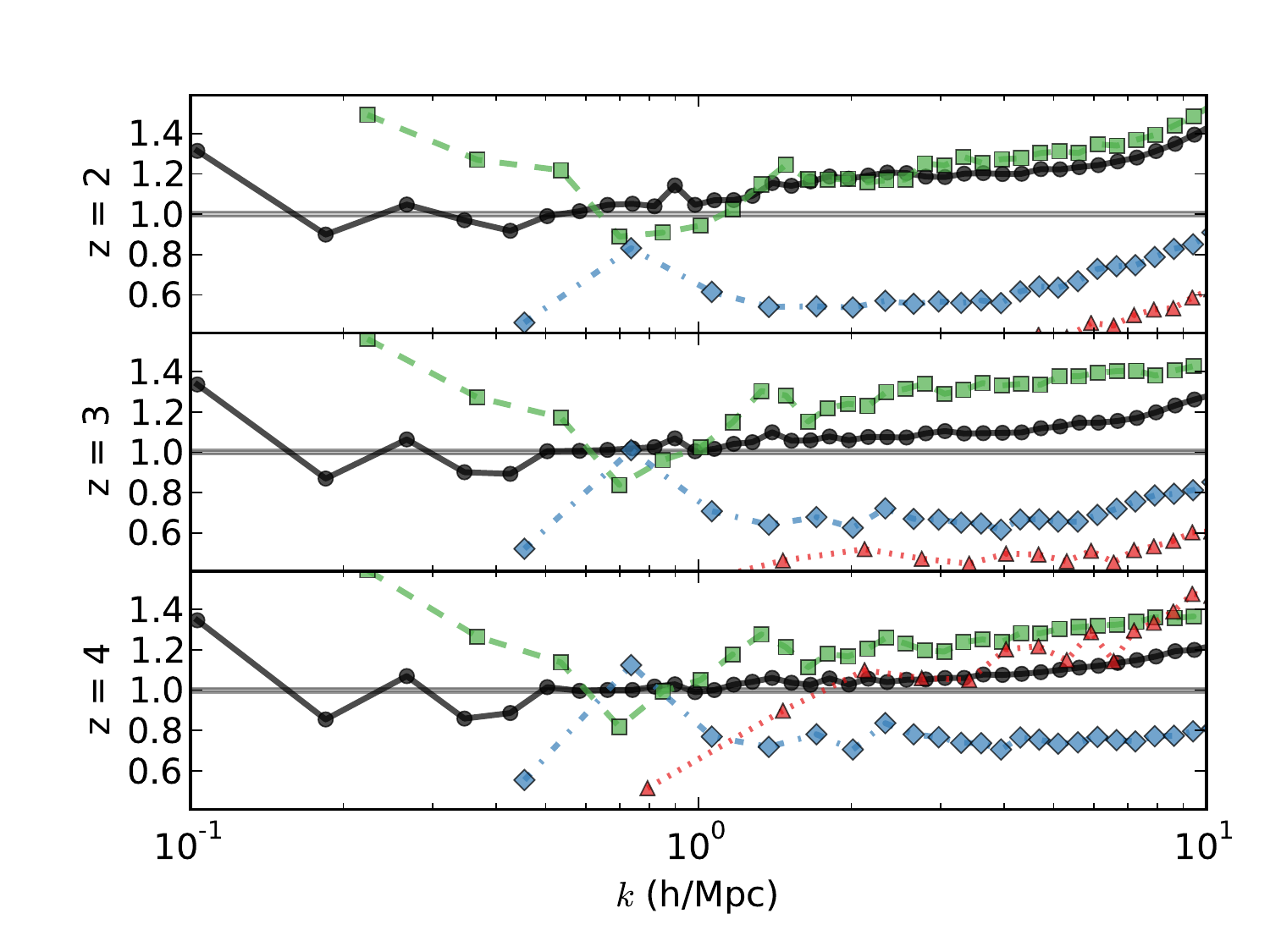}
  \end{center}
  \caption{ The matter power spectrum for different box-sizes.  We
    show ratios to the emulated predictions from a gravity-only
    simulation \citep{heitmann_et_al_2013} at redshifts 2, 3, and
    4. Red triangles are 10 $\hinv$Mpc box, blue diamonds are 20 $\hinv$Mpc,
    green squares 40 $\hinv$Mpc, and black circles 80 $\hinv$Mpc box. The
    resolution is constant in all runs: 19.5 $\hinv$kpc.}
  \label{fig:box_pk}
\end{figure}

To estimate the non-linear scale of density perturbations, we use the
power spectrum emulator FrankenEmu \citep{heitmann_et_al_2013}, shown
in Figure \ref{fig:lin_emu} for the cosmology considered here and at
redshift $z = 2$.  Since we end the simulations at $z = 2$, this is
the worst-case scenario in terms of the required box size. This alone
indicates that 40 $\hinv$Mpc is the bare minimum to avoid the box-size mode
becoming non-linear, with 80 $\hinv$Mpc being a more comfortable value. In
the context of ``missing modes'' in simulations of the \Lyaf, an
important and thorough recent work is that of \citet{Tyt09}. The range of box
sizes they consider is even larger than the one presented here: their
biggest box (54.5 $\hinv$Mpc) is similar to our largest, while they go
to box sizes as small as 1.7 $\hinv$Mpc. Thus most of their simulations are
over-evolved at $z = 2$, where the largest -- anchoring -- mode is
deeply in the non-linear regime according to
Fig.~\ref{fig:lin_emu}. Note that their choice of cosmology has
$\sigma_8 = 0.9$, therefore non-linearity starts at even larger scales
than the cosmology we consider here. In addition, due to the high
computational expense, they had to restrict their box size series
analysis to a spatial resolution of 53.3 $\hinv$kpc. This resolution is
significantly more coarse than the one we find necessary in this paper
(20 $\hinv$kpc, section \ref{sec:resolution}), but also coarser than what
Tytler et al.~would have likely run (13.3 $\hinv$kpc, see their section
11.3) if it were computationally feasible. Here we present a box size
convergence study extending to box sizes large enough to sample linear
modes even at the end of the simulations ($z = 2$), but also with the
desired spatial resolution to capture \Lya statistics to the percent
level.

Before turning to flux quantities, we will first look at the
convergence of the matter power spectrum in our runs as we increase
the box size while keeping the resolution constant. This is shown in
figure \ref{fig:box_pk}.  We clearly see the suppression in mode
growth in the small-box simulations with respect to 80 $\hinv$Mpc run.  The
differences in the matter power are rather significant, but as we will
show later --- and as shown in \citet{Tyt09} --- the differences in
the flux power are much less.

\subsection{N-Point Statistics}

\begin{figure}
  \begin{center}
    \includegraphics[width=\columnwidth]{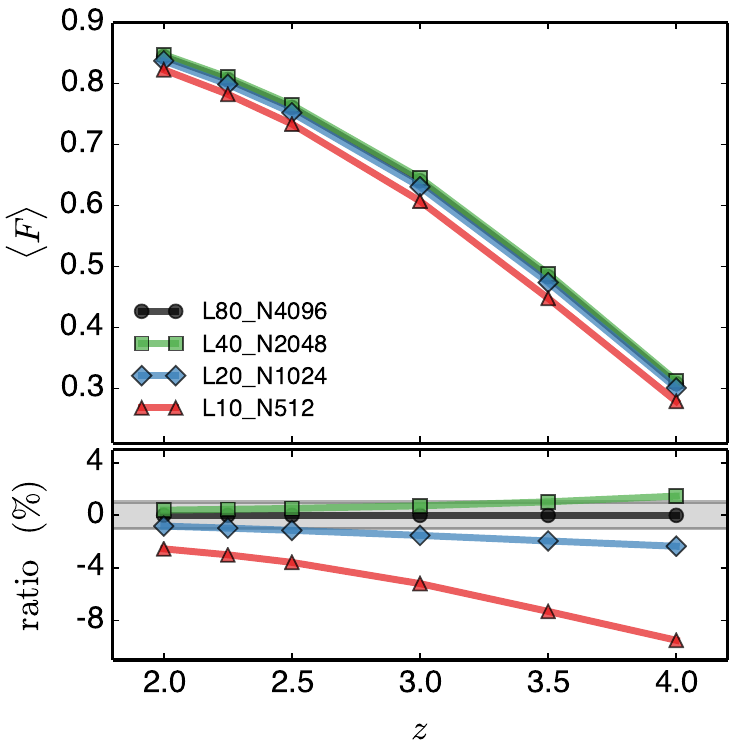}
  \end{center}
  \caption{Dependence of the mean flux on the box size, for 6
    different redshifts. Upper panel shows the mean flux, lower panel
    presents the ratio to the largest box-size run -- 80 $\hinv$Mpc.}
  \label{fig:box_taueff}
\end{figure}

In Figure \ref{fig:box_taueff}, we show the mean flux in different
box-sizes for a constant spatial resolution. As expected, the 10 $\hinv$Mpc
box is significantly inaccurate, while already in the 20 $\hinv$Mpc box we
obtain reasonable mean values. As in the resolution study in
Section \ref{sec:resolution}, we see the same trend of growing differences
as we move to higher redshift. This is not immediately intuitive
behavior, as one would expect small boxes to be less affected at $z = 4$
rather than at $z = 2$. As we do not have many independent realizations
of each box-size, we cannot state with certainty how much this effect
is due to runs having different realizations versus an actual physical
effect.  Another thing to note is that convergence is not one-sided
(e.g.~as the box size is increased the mean flux does not increase as
was the case in the resolution study).  Again, this could be just
statistical variance. Similar behavior is reported in \citet{Tyt09},
see their Table 5. Overall, we see the behavior one would expect
from Figure \ref{fig:lin_emu} --- namely, that there is only a small
difference between 40 $\hinv$Mpc and 80 $\hinv$Mpc boxes. The difference
increases with a further reduction in box-size, and becomes clearly
too large in the 10 $\hinv$Mpc box.

\begin{figure}
  \begin{center}
    \includegraphics[width=\columnwidth]{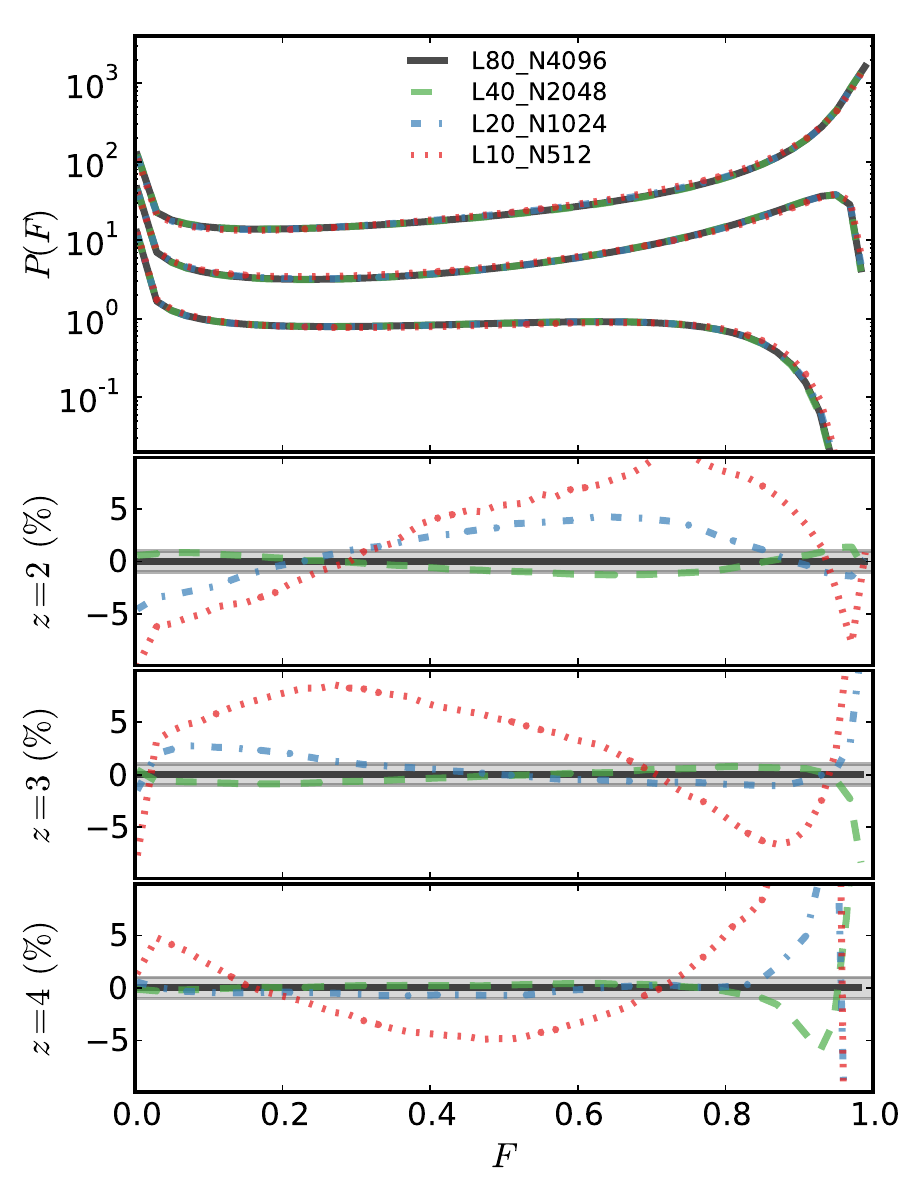}
  \end{center}
  \caption{Convergence of the flux PDF with respect to box size. For
    clarity, in the upper panel we have multiplied the $z = 2$ data by
    a factor of 100, and the $z = 3$ data by 10. The resolution is
    kept constant at 19.5 $\hinv$kpc, and the box-size increases from 10
    $\hinv$Mpc (dotted red line) to 80 $\hinv$Mpc (solid black line).}
  \label{fig:box_fpdf}
\end{figure}

As was done in Section \ref{sec:resolution}, we first remove the
differences in the mean flux value by rescaling the optical depth in
all boxes to the value in our ``best'' simulation, the 4096$^3$ run in
an 80 $\hinv$Mpc box. Since the rescaling is small for all but the 10
$\hinv$Mpc simulation this plays a rather minor effect, and our conclusions
would be the same if we presented unscaled results with the
differences being marginally higher. In Figure\ref{fig:box_fpdf} we
show the dependence of the flux PDF with respect to the box-size.  As
in the case of the flux mean, we see that the 40 $\hinv$Mpc and 80 $\hinv$Mpc
boxes are in percent agreement except at the very transmissive end, $F
\approx 1$, and at higher redshifts. As commented on in the resolution
study, the fully transmissive pixels become very rare at higher
redshifts due to the high physical density of neutral hydrogen and
therefore the error in determining the relative fraction of such
regions decreases. Qualitatively our results at $z = 2$ are in line
with those presented in \citet{Tyt09} (though note most of their boxes
are smaller than ours).

\begin{figure}
  \begin{center}
    \includegraphics[width=\columnwidth]{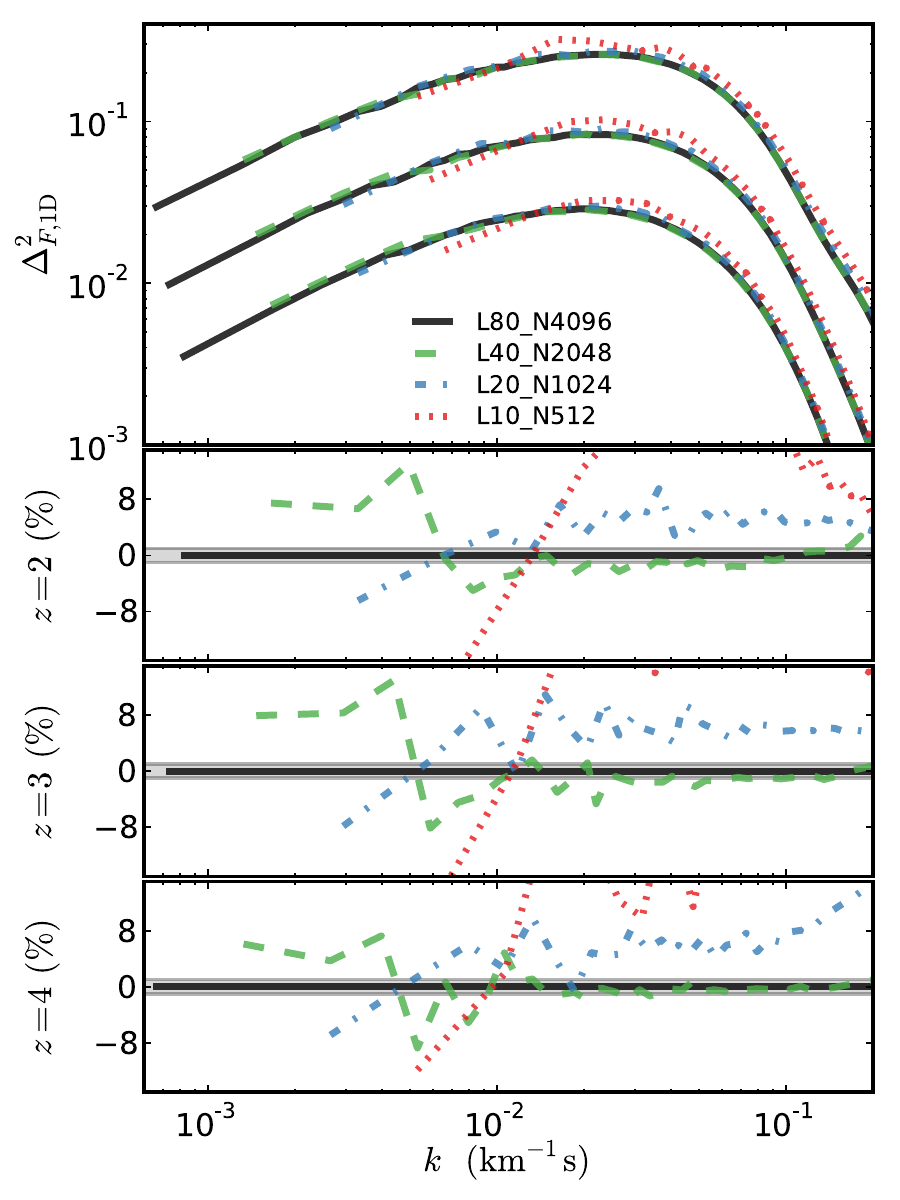}
  \end{center}
  \caption{Convergence of the flux 1D power spectrum with respect to
    box size at redshifts 2, 3, and 4. Dimensionless flux power is
    shown in the upper panel, while ratios to the 80 $\hinv$Mpc run at 3
    different redshifts are shown in the lower panels.  Colors and
    line styles follow that of Figure \ref{fig:box_fpdf}.}
  \label{fig:box_pf1d}
\end{figure}

Next, we turn to the 2-point correlation function of the \Lyaf flux
while first examining the 1D $P(k)$. In Figure \ref{fig:box_pf1d} we
immediately see that the differences in the flux power are much less
than in the matter power. This is not unexpected as the flux comes
from only moderate over densities, which are less affected by the sample
variance than halo regions. The convergence of the low-$k$ region is
difficult to assess due to different realizations of the initial conditions,
but overall we again see nice agreement between the 40 and 80 $\hinv$Mpc
boxes. Here, however, the 20 $\hinv$Mpc box is noticeably in error
(by 5-10 per cent) while the 10 $\hinv$Mpc box has no value for precision cosmology
measurements.
As before, our results are in good qualitative agreement with \citet{Tyt09}.
As we will show next, most of the differences in the 1D power originate from
the differences in power along the line of sight.  Despite those differences
we conclude that for 1D $P(k)$ constraints the 40 $\hinv$Mpc box size is a
reasonable one, while 80 $\hinv$Mpc is a safe choice.

Finally, we turn to the 3D P(k), looking at 4 $\mu$ bins, going from
across the line of sight $0 < \mu < 0.25$ to the power along the line
of sight $0.75 < \mu < 1$. This was investigated in
\citet{mcdonald2003}, where they ran simulations with box-size
spanning 20 to 80 $\hinv$Mpc, very much like the simulations presented
here. However, those were HPM simulations rather than full gas
dynamics, and the cell size was kept constant at a rather large value
of 156 $\hinv$kpc.

\begin{figure*}
  \begin{center}
    \includegraphics[width=\textwidth]{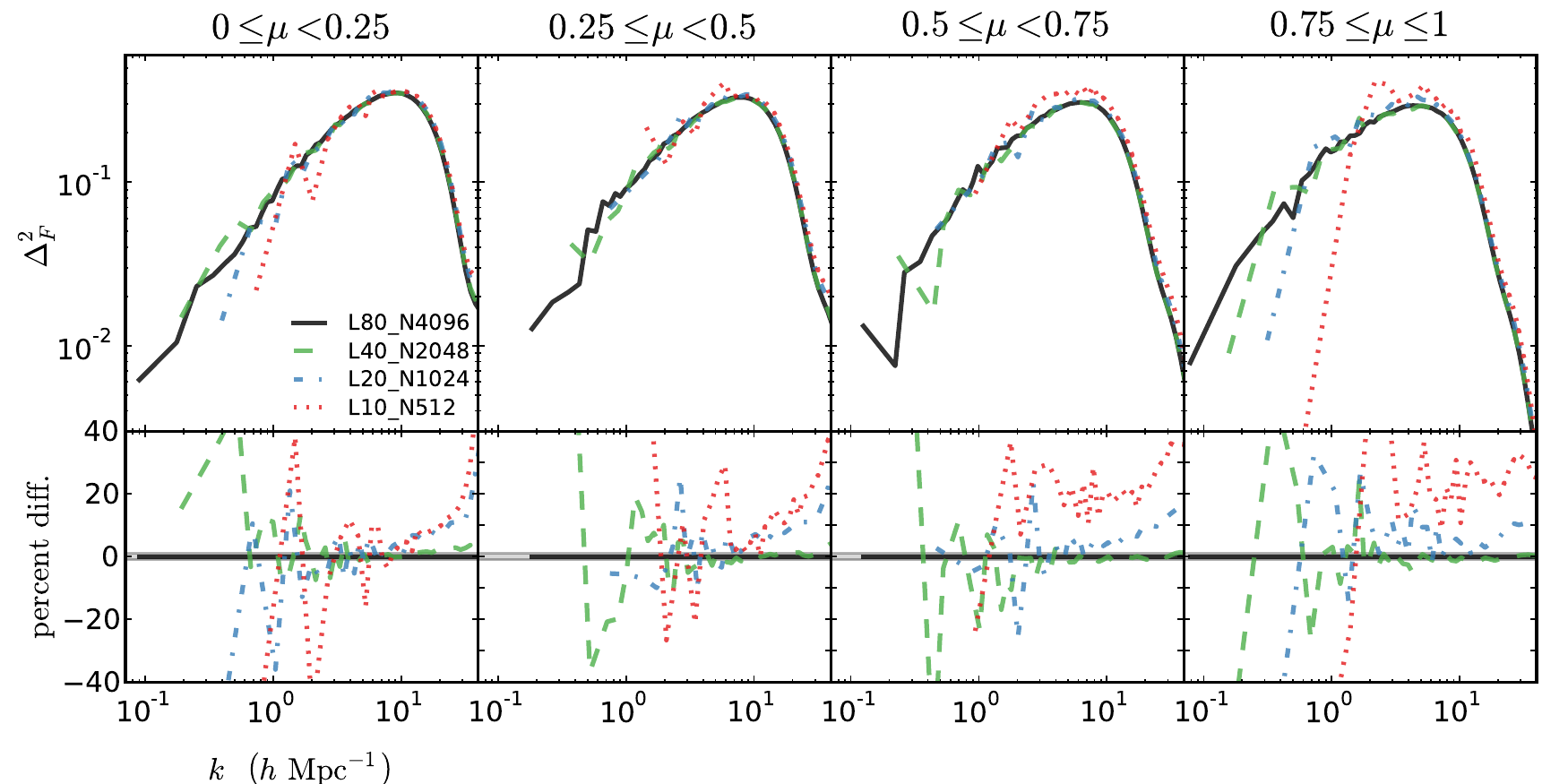}
  \end{center}
  \caption{Convergence of the flux 3D power spectrum with respect to
    box size in 4 $\mu$ bins. As in Figure\ref{fig:res_pf3d}, we
    show only $z = 4$ data, although here this is representative of the
    agreement at other redshifts as well. The modes at large scales
    are different due to sample variance.}
  \label{fig:box_pf3d}
\end{figure*}

In Figure \ref{fig:box_pf3d} we see good agreement between the 40 and
80 $\hinv$Mpc simulations when the spatial resolution is kept constant.  At
low $k$, there is a substantial scatter between simulations as random
phases in initial conditions differ in different box-sizes. As a
result, we cannot meaningfully compare our boxes at large scales.
However, at smaller scales, we see that a 40 $\hinv$Mpc simulation is in
percent-level agreement with the 80 $\hinv$Mpc run. Again, just by
observing the convergence rate for different box sizes we can be
confident that 80 $\hinv$Mpc is a very safe value to run at - most likely
sub-percent accurate.

Here, we also need to comment on the large-scale \Lyaf bias: on large
scales, the \Lya flux is a biased tracer of the matter field, and as
such is of great value for cosmology (see
\citet{slosar_et_al_2011, slosar_et_al_2013, busca_et_al_2013}).  At
present --- or in the near future --- running hydrodynamical
simulations with box sizes needed to sample the BAO peak, and at the
same time obtain the resolution necessary to resolve density
fluctuations in the IGM, is not a viable approach.  Still, one does
not necessarily need to have a BAO-regime simulation box to reach a
regime where the redshift-space \Lya flux power is related to a
real-space density power via a $k$-independent \citep{Kaiser1987}
formula:
\begin{equation}
 P_F(k, \mu) = b^2 (1 + \beta \mu ^2)^2 P_m(k)
 \label{eq:kaiser}
\end{equation}

We have examined bias in our simulations, and have found that they are
all too small for a reliable fit to $b$ and $\beta$ of Equation
\ref{eq:kaiser}. Our 80 $\hinv$Mpc box barely reaches a regime where
the parameters become scale-independent. Thus, while it is possible to
obtain those values using different approaches than directly fitting
Equation \ref{eq:kaiser}, that work --- and especially the comparison with
even smaller box-sizes presented here --- would be incongruent with
the accuracy carried out in the rest of this paper. For now, we will
leave it as a separate topic to be carried out in a future work.

%% file: sec6.tex
\section{Splicing}
\label{sec:splicing}

In the previous two sections we have confirmed and quantified both the
box-size and resolution requirements for achieving percent-level
accuracy for precision \Lyaf cosmology studies.  Although possible
with today's high-performance computing facilities, as demonstrated
with our L80\_N4096 simulation, currently performing a large number of
such simulations is impossible. In the past, even a single simulation
with that dynamical range was impossible.  One technique used to
compensate for the lack of dynamical range is splicing, first
introduced by \citet{mcdonald2003}, and most recently employed in
\citet{Borde2014}. Here we will assess the accuracy of splicing on a
1D flux power spectrum. For completeness, we will first briefly review
the method itself.

The mechanics behind splicing is to run three simulations, each
lacking sufficient dynamic range, and combine them into a result
that accurately represents a single full dynamic range simulation.
One runs
a simulation with enough large-scale power (i.e.~a big enough box),
but with too coarse a resolution, and another simulation where the box
is known to be too small but with good resolution. Finally a
simulation is carried out where both resolution and box size are
insufficient, the resolution set to the same as in first run, with the
box size the same as in the second. The idea is then to use two
small-box runs to capture the effect of coarse resolution on the power
spectrum, and two runs with coarse resolution to correct for the
missing modes in the small-box simulations.
Here we will attempt to
splice the result of our 4096$^3$ 80 $\hinv$Mpc run, which yields percent
accurate results as shown via the box-size and resolution convergence
tests in the two previous sections.  We will thus splice the
L80\_N1024, L20\_N1024 and L20\_N256 runs, and compare the result to
L80\_N4096.  Mathematically expressed, in the regime $k < k_{\rm min, 20}$,
where $k_{\rm min, 20} = 2 \pi / (20 \, \hinv \mathrm{Mpc})$ the flux power is
given as:
\begin{equation}
   P(k) = P_{L80\_N1024}(k) \frac{P_{L20\_N1024}(k_{\rm min,20})}{P_{L20\_N256}(k_{\rm min,20})} \, \, ,
\end{equation}
in the range $k_{\rm min,20} < k < k_{\rm Nyq,80} / 4$ where
$k_{\rm Nyq,80} = 1024\pi / 80\,h^{-1}{\rm Mpc}$ is
\begin{equation}
   P(k) = P_{L80\_N1024}(k) \frac{P_{L20\_N1024}(k)}{P_{L20\_N256}(k)} \, \, ,
\end{equation}
and for $k > k_{\rm Nyq,80} / 4$ it is
\begin{equation}
   P(k) = P_{L20\_N1024}(k) \frac{P_{L80\_N1024}(k_{\rm Nyq,80})}{P_{L20\_N256}(k_{\rm Nyq,80})} \, \, .
\end{equation}

\begin{figure}
  \centering
  \includegraphics[width=\columnwidth]{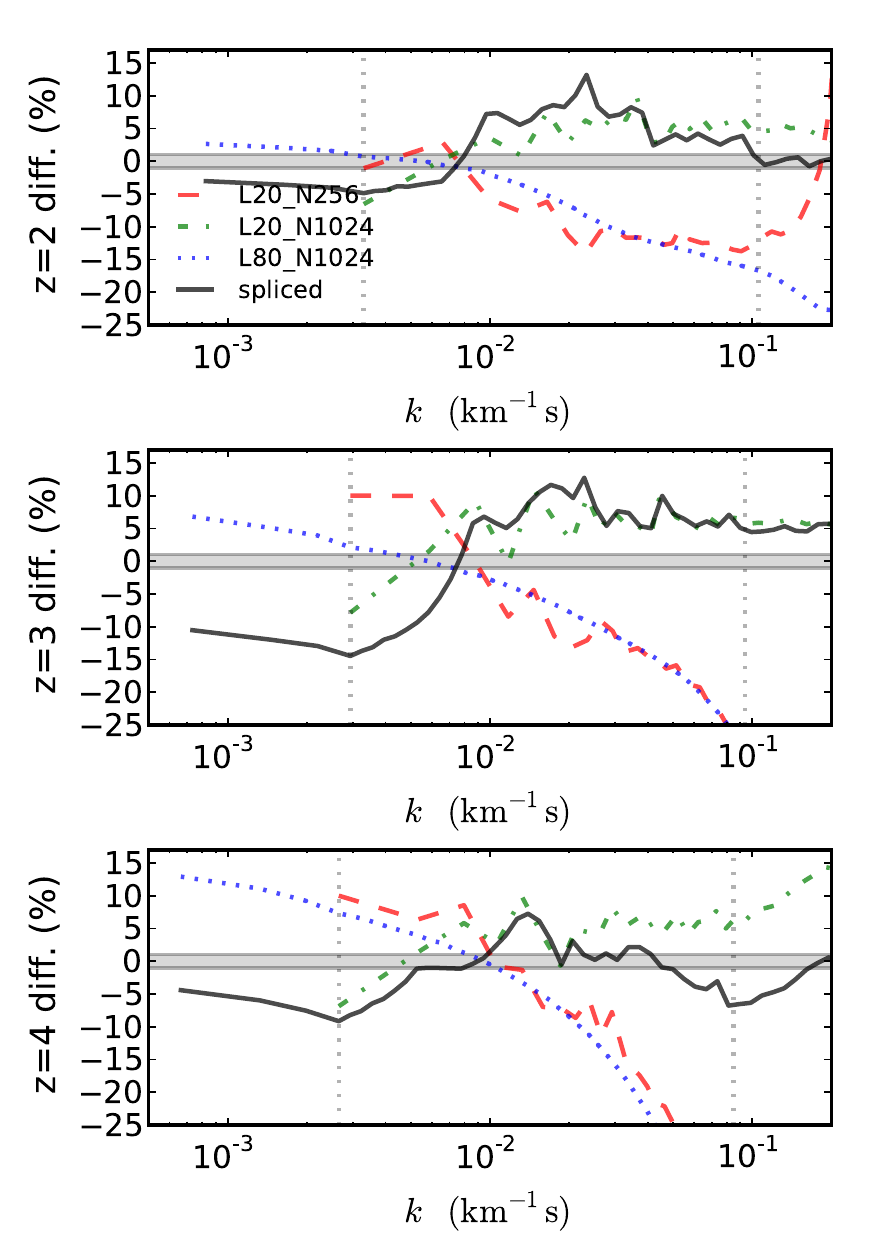}
  \caption{Comparison of a spliced 1D power spectrum and the actual
    one in a high-resolution, large-box simulation. From top to
    bottom we have redshift $z = 2$, $z = 3$, and $z = 4$. In addition to
    the spliced run, we show the ratios of power spectra for each of
    the three runs used for splicing. All ratios are taken with
    respect to L80\_N4096 run.}
  \label{fig:splicing}
\end{figure}

In Figure \ref{fig:splicing} we show the results of splicing the flux
power spectrum at 3 different redshifts.  The accuracy of splicing is
similar at all redshifts, and is mostly in the $\sim$10 per cent
range.  That is in agreement with the accuracy estimated by
\citet{mcdonald2003}, but noticeably above the 2 per cent accuracy
claimed by \citet{Borde2014}.  A possible reason for
  this discrepancy is the fact that \citet{Borde2014} tested the
  splicing method on a non-converged simulation (1024$^3$ particles in
  100 $\hinv$Mpc box).

%% file: sec7.tex
\section{Rescaling Optical Depths}
\label{sec:uvb}

Up to this point we have ignored the common practice of rescaling the
simulated optical depths. In most papers presenting the results of
\Lyaf\ predictions from optically-thin simulations, authors multiply
the optical depth in each pixel by some factor $A$ such that the
simulated mean flux matches the observed mean flux at the same
redshift. This is easily done with any root finding method on
$\mean{\exp(-A \tau)} - \mf_{\rm obs}$ and converges fairly
quickly. This fix is well-justified considering how poorly constrained
the amplitude, shape, and evolution of the ionizing background are.
When we rescale optical depths, it is understood that
  this is roughly equivalent to adjusting the specific intensity of
  the UV background used in the simulation.  Changes in the
  photoionization rate, $\Gamma$, in the simulation will most directly
  affect the \HI\ density while sub-dominant changes will come
  from photoheating rates.  The change in photoheating rate affects
  the temperature and pressure support of the gas at times when
  hydrogen or helium are not fully ionized.
%As $\tau \propto \Gamma^{-1}$, rescaling
%the optical depths $\tau^\prime = A \tau$ means roughly 
%$A \sim \Gamma / \Gamma^\prime$.

\begin{figure}
  \begin{center}
    \includegraphics[width=\columnwidth]{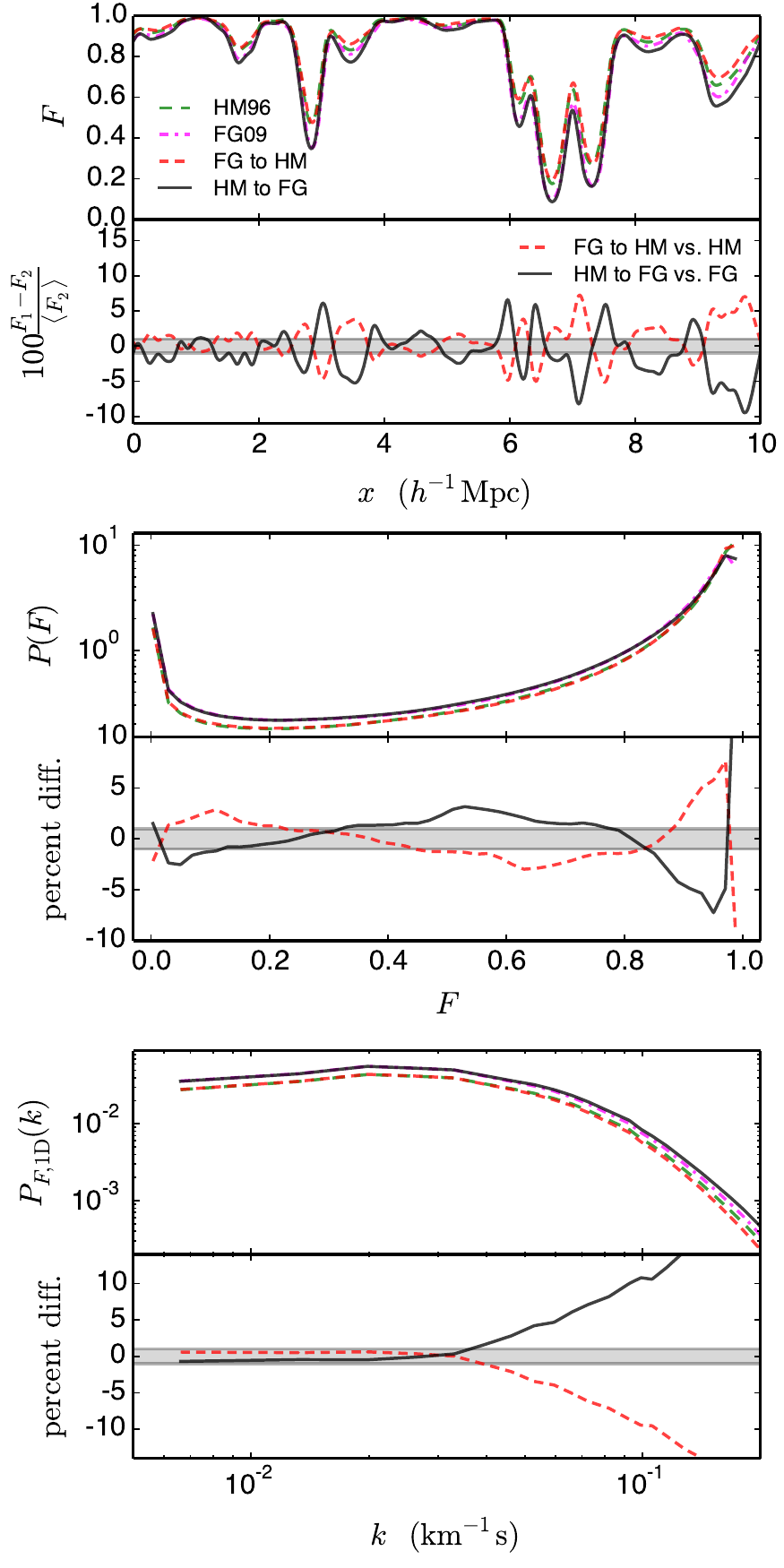}
  \end{center}
  \caption{
    {\it Top:} Flux statistics from the L10\_N1024\_HM96 and
    L10\_N1024\_FG09 runs at $z = 2$. At this redshift, the
    L10\_N1024\_HM96 run has $\mf = 0.8117$, and the L10\_N1024\_FG09
    run has $\mf = 0.7749$. Also shown are the flux statistics of
    these runs scaled to the mean flux of the other. {\it Middle:} the
    flux PDF and percent difference compared to the result with the
    same mean flux. {\it Bottom:} the flux 1D power and percent
    difference compared to the result with the same mean flux.}
  \label{fig:hm_vs_fg}
\end{figure}

In order to test the effect of rescaling optical depths, we first
tried taking two runs with different UV backgrounds (and otherwise the
same input parameters) and rescaling one to the mean flux of the
other. We used a run with the \citet{haardt_and_madau_1996} UVB
(labeled L10\_N1024\_HM96) and a run with the
\citet{faucher_giguere_et_al_2009} UVB (labeled L10\_N1024\_FG09). One
concern with starting from different UV backgrounds is that they can
result in different $\rhob$-$T$ relations, which would leave
differences in the flux statistics no matter how the rescaling is
done. The HM96 and FG09 UV backgrounds do result in slightly different
$\rhob$-$T$ relations, with very similar slopes but differing $T_0$
values. In the case of HM96, we fit $T_0 = 9.0 \times 10^3$ K and
$\gamma = 1.55$, and in the case of FG09, we fit $T_0 = 1.1 \times
10^4$ K and $\gamma = 1.55$ at redshift $z=2$. While this is not a
significant difference, a temperature difference like this should show
up in the flux power spectrum, for instance, as a different thermal
cutoff.  More importantly, while the two UV backgrounds
  result in similar instantenious $\rhob$-$T$ relation, the two have
  significantly different thermal histories:
  \citet{haardt_and_madau_1996} reionizes hydrogen at $z \approx 6$,
  while \citet{faucher_giguere_et_al_2009} has this occuring at $z
  \approx 12$.  This will result in two UVBs which have a different
  filtering scale, even if $T_0$ and $\gamma$ at a given redshift
  are the same.  At $z=2$, the HM96 run has a mean flux $\mf =
0.8117$ and the FG09 run has a mean flux $\mf = 0.7749$. Rescaling the
HM run to the FG mean flux requires $A = 1.403$ (or conversely
rescaling the FG run to the HM mean flux requires $A = 0.7138$). We
show an example skewer in the top panel of Figure
\ref{fig:hm_vs_fg}. The sample spectrum shows that between the HM96
and FG09 runs the flux in regions of high transmission is similar, but
the absorption features are generally deeper in FG09 primarily due to
the lower photoionization rate. It is difficult to tell if either run
has broader features by looking at just a few lines, but overall we
found that the HM96 run does have noticeably broader lines. We also
show the spectra after rescaling to the other mean flux. It is
reassuring to know that while the correction is an average over the
entire box, individual features agree well enough that the correction
also works well for individual lines.

The flux PDF and flux 1D power from these runs and their rescaled
versions are also shown in Figure \ref{fig:hm_vs_fg}. In both
statistics the HM96 and FG09 results differ by about 30 per cent, but the
process of rescaling to the other run's mean flux brings them to
within several percent. The remaining differences in the flux PDF are
not straightforward. The rescaled FG09 run has more pixels at low $F$,
fewer pixels at intermediate $F$, and more pixels at high $F$ compared
to the HM96 result. It appears that the rescaled FG09 rises faster
than the HM96 result at $F=0$ and $F=1$. In the 1D flux power spectra, the
rescaled versions agree very well at large scales. The rescaled FG09
result has slightly more power than the HM96 result, but it is within
1 per cent. On scales below $k > 0.4$ $h$ Mpc$^{-1}$, the slopes
of the rescaled versions start to diverge significantly. This is due
to the differing $T_0$ for each, resulting in a different thermal
cutoff. Overall, the rescaling process works remarkably well at
removing differences from different UV backgrounds, although one
should be careful with results that are sensitive to the $\rhob$-$T$
relation.

\begin{figure}
  \begin{center}
    \includegraphics[width=\columnwidth]{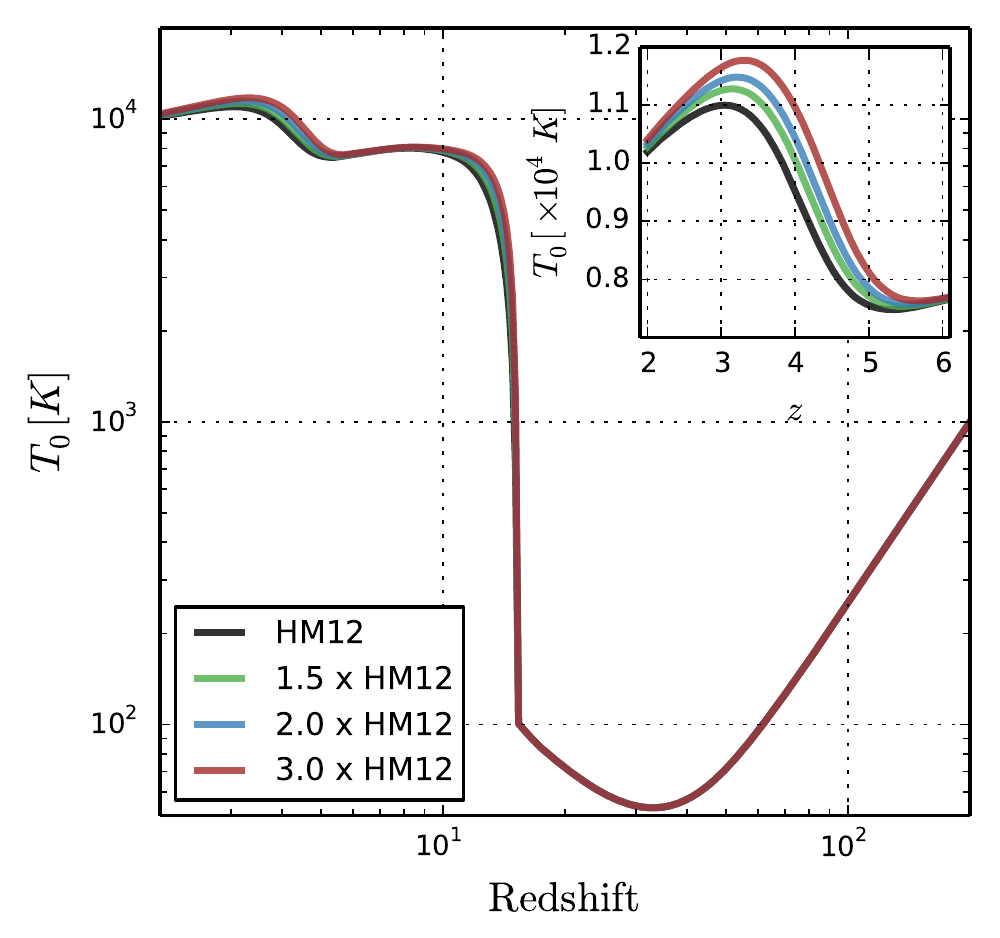}
  \end{center}
  \caption{
    The evolution of $T_0$ for a simulation with
    \citet{haardt_and_madau_2012} UV background (black line), and
    simulations where photoionization and photoheating
    rates for all ionic species have been multiplied by the same constant: 
    1.5 (green line), 2 (blue line), and 3 (red line).  Inset
    panel shows the evolution of $T_0$ over the $2 \leq z \leq 6$ redshift range.
    }
  \label{fig:scaled_T0}
\end{figure}

We also made another test of the UVB rescaling, by
  running simulations where the photoionization and photoheating
  rates for all ionic species have been multiplied by the same
  constant.  Here we use the \citet{haardt_and_madau_2012} rates and,
  since we multiply all of them by the same factor, the spectral shape
  of the original UVB is preserved and only the amplitude changes.  In
  figure \ref{fig:scaled_T0} we show the IGM temperature evolution. As
  expected modulating the amplitude of the UVB affects the temperature
  only when a species is not fully ionized.  Changes in the hydrogen
  leave no imprints on $T_0$ at observable redshifts, but
  the same change in helium photoheating rates do change the
  temperature due to its partial ionization.

\begin{figure}
  \begin{center}
    \includegraphics[width=\columnwidth]{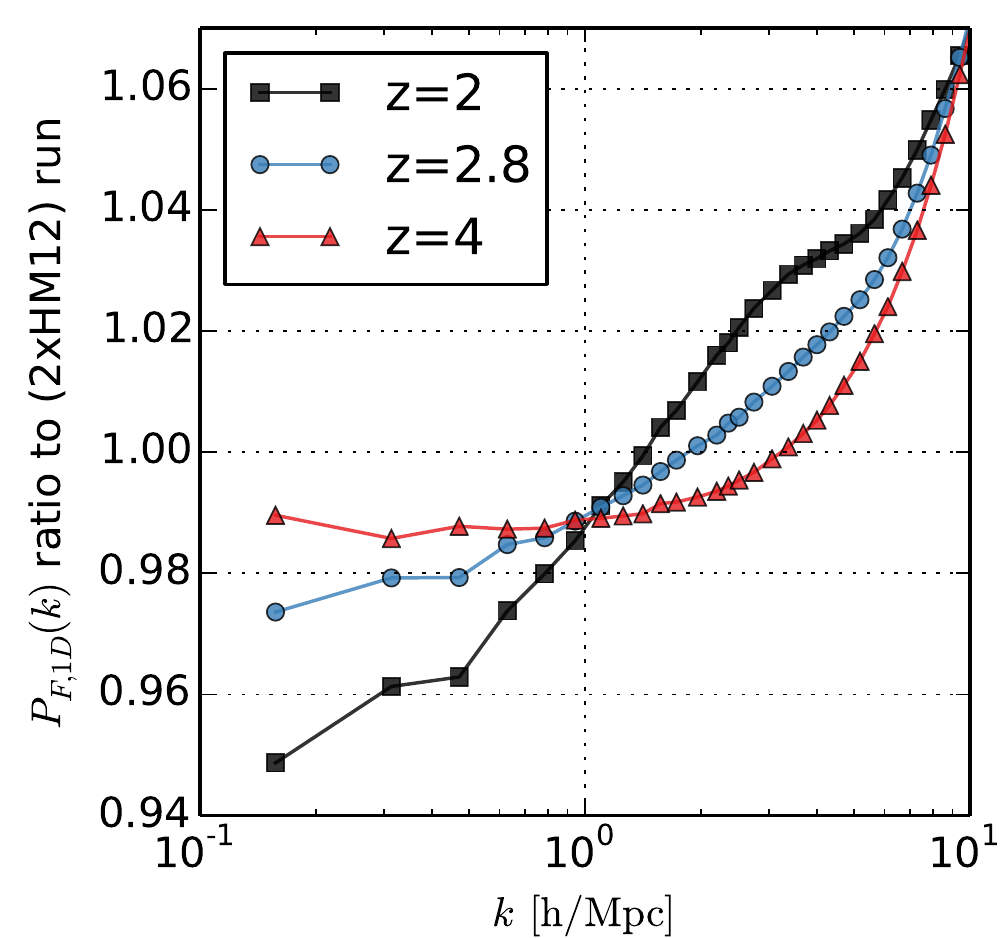}
  \end{center}
  \caption{
    The 1D power spectrum from a run using the original 
    \citet{haardt_and_madau_2012} UVB rates where the mean flux is rescaled 
    to match that of the run done with doubled UVB rates (2xHM12). We 
    show here the ratio to the actual 2xHM12 run, at 3 different redshifts: 
    $z=2$ (black, squares), $z=2.8$ (blue, circles), and $z=4$ (red, 
    triangles).
    }
  \label{fig:scaled_ps1d}
\end{figure}

We focus on two runs, one using the original
\citet{haardt_and_madau_2012} rates, and one where the multiplying
factor for all photo-rates is 2, approximately a value needed to
recover the observed mean flux in optically thin simulations with this
UVB. We compare the run done with doubled photo-rates, labeled 2xHM12,
with a flux-rescaled run performed with the original HM12 rates.  In
other words, the rescaling is done by simply finding the factor $A$
which will bring the mean flux of the original run to that of 2xHM12,
a procedure which ignores the differences in the instantenious gas
temperature and prior temperature history.  The rescaling factor $A$
we recover is close to, but not exactly equal to 0.5, and it shows the
tendency to decrease with increasing redshift. For example, it is 1\%
higher at $z=2$ ($A=0.505)$, than $A=0.5$ at redshift $z=2.4$, and
decreases to $A=0.479$ at $z=4$.  The results of these rescalings are
shown in solid lines in figure \ref{fig:scaled_ps1d} for 3 different
redshifts.  The difference on the 1D power spectrum between rescaling
the mean flux and actually running the full evolution with different
rates is a few percent, but it is present on all scales. We have also
checked the flux PDF, and found smaller differences of approximately
1\%.  One should note that the difference is smaller at higher
redshift, $z=4$, and is the greatest at $z=2$.  This means that the
effect of different temperature evolutions is more significant for the
IGM gas at mild overdensities, and not very significant for underdense
gas in void regions.

\begin{figure}
  \begin{center}
    \includegraphics[width=\columnwidth]{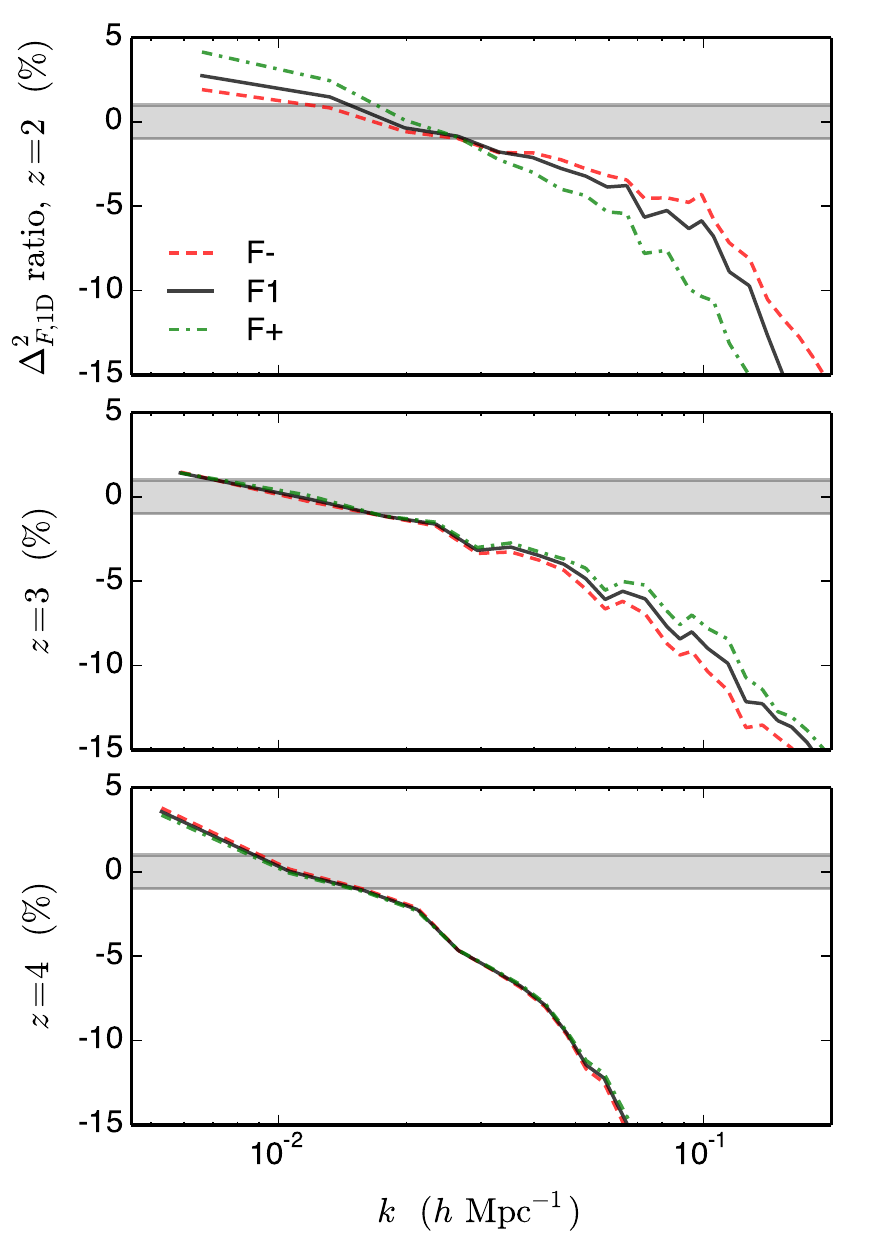}
  \end{center}
  \caption{The percent difference in the 1D flux power spectra
    between L10\_N256 and L10\_N1024 at $z =$ 2, 3, and 4, where the
    optical depths have been rescaled. F- is rescaled to 10 per cent smaller
    than the L10\_N1024 mean flux, F1 is rescaled to same mean flux,
    and F+ is rescaled to 10 per cent larger. This shows how the convergence
    rate depends on the mean flux. At $z = 2$, the F- power converges
    fastest, but at $z = 3$, the F+ power converges fastest. At $z =
    4$, the convergence rate is not noticeably affected by 10 per cent
    changes in the mean flux.}
  \label{fig:pf1d_rescale}
\end{figure}

Final test we performed was to take the 10 $\hinv$Mpc resolution
series runs and test the convergence rate of the flux statistics when
changing the mean flux. We compare the flux statistics computed from the same
optical depths, but with three rescalings
based on the mean flux of the L10\_N1024 simulation: one 10 per cent larger,
one equal, and one 10 per cent smaller, respectively labeled (F+, F1, and
F-). In Figure \ref{fig:pf1d_rescale}, we show the ratio of 1D flux power
spectra between the L10\_N256 and L10\_N1024 runs, computed from the rescaled
fluxes. From top to bottom, we show the result for $z = 2$, 3, and 4. At $z =
2$, we see that the 1D power converges faster for a lower mean flux. The
resolution error in the 1D flux power shows a characteristic slope difference
(the lower resolution result has a more negative slope), and at $z = 2$, the
slope difference is smallest for the lower mean flux rescaling, and largest for
the higher mean flux rescaling. Rescaling flux to a lower mean value means
shifting flux contributing regions to lower gas density, or equivalently less
nonlinear structures, and thus it is easier for a simulation to be converged.
At $z = 3$, we see the opposite trend, that
the results converge faster with a higher mean flux rescaling. This indicates
a ``sweet spot'' mean flux (or similarly, a redshift) where it is easiest to
resolve \Lyaf structures. When the mean flux is very high, the forest probes
higher densities closer to halos which are harder to converge.  Similarly,
when the mean flux is very low, the forest has significant sensitivity to
the very underdense regions. Although we are examining purely numerical
effects here, note that these conclusions also translate to the question
of the importance of additional physics in simulations. From what we have
presented in this section, one would clearly expect galactic outflows, AGN
feedback, and other processes originating within galaxies to matter much
more at e.g.~redshift $z = 2$ than $z = 3$.
At $z = 4$, we see no difference in the convergence rate with the different mean
flux rescalings. At this high of a redshift, the mean flux is low and
a rescaling of only $\pm10$ per cent does not significantly affect the density
range contributing to the forest (see the red bands in Figure \ref{fig:F_rho}).

Finally, another issue with the practice of rescaling to the observed mean flux
is that few current \Lyaf\ simulations are converged to the percent level
in $\mf$ at high redshift. As shown previously, even simulations with a
resolution of $\sim 40$ $\hinv$kpc are not converged to a percent in the mean
flux for $z > 3$. Taking a simulation of insufficient resolution and performing
a mean flux rescaling will result in the wrong
correction. Under-resolving IGM structures results in a mean flux
lower than it should be, so a rescaling to a higher mean flux, for
instance, will require a smaller rescaling factor $A$ than what would
be needed for a higher-resolution run.

%% file: sec8.tex
\section{Small Scale Statistics}
\label{sec:small_scale}

\subsection{Line Statistics}

The \Lyaf\ is classified as systems with $\NHI < 10^{17}$ cm$^{-2}$,
known for sitting in the linear portion of the
curve of growth. This makes it straightforward to fit the \Lyaf lines
with Doppler profiles. Each line fit provides the column density
$\NHI$ and Doppler parameter $b$ of the underlying system. However,
this neglects the issues of significant line blending in the forest,
as well as line broadening dominated by Hubble broadening rather than
thermal broadening. \citet{meiksin_et_al_2010}, for instance, mention
that line shapes in simulated spectra ignoring peculiar velocities are
qualitatively different than those in the full spectra. This indicates
that the $\NHI$ and $b$ derived from line fitting may not correspond
to the actual column density and temperature distributions of the gas
that makes up the forest.  However, line parameter distributions are a
sensitive measure of line shapes in the forest. Additionally,
interpreting line parameter statistics is fairly straightforward ---
the column density is a proxy for equivalent width and the Doppler
parameter is the line width.

\begin{figure}
  \begin{center}
    \includegraphics[width=\columnwidth]{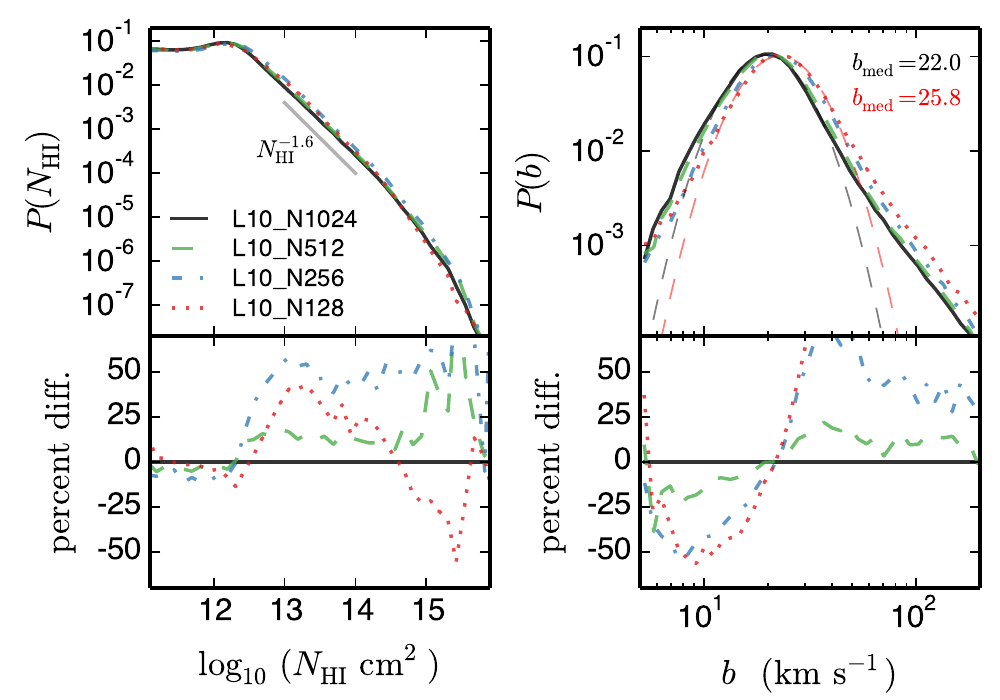}
  \end{center}
  \caption{
    The distributions of the line \HI\ column density (left) and the Doppler
    parameter (right) in the resolution series of the 10 $\hinv$Mpc boxes at $z
    = 2$. The percent difference relative to the L10\_N1024 simulation is shown
    in the bottom panels. In the $P(\NHI)$ panel, the gray line illustrates the
    power law slope $\propto \NHI^{-1.6}$. In the $P(b)$ panel, the thin dashed
    red and black lines show lognormal distribution fits to the L10\_N128 and
    L10\_N1024 results, and the red and black text gives the corresponding
    median $b$-parameter values (also the peak probability). The simulated
    $b$-parameter  distributions are more skewed than lognormal, but a lognormal
    fits the core of the simulated distributions well.
  }
  \label{fig:lines_res}
\end{figure}

In this work, we generate spectra with a fixed resolution of
$\Delta v_{\rm pix} = 1$ km/s to avoid possible issues of the fits depending
on spectral resolution and the assumed pixel SNR. We have chosen this resolution
because it is sufficiently smaller than what we expect for the minimum
line width of gas at $T = 10^3$ K, which corresponds
$b = 4$ km s$^{-1}$ for hydrogen. We want to reduce statistical uncertainty in
the distributions as much as possible, which means having many lines in
each bin. The line compilation in \citet{haardt_and_madau_2012} (their section
3), provides $dN / dz$ of systems above a given $\NHI$. At $z = 3$,
$dN/dz = 24.2$ for $\NHI > 10^{15}$ cm$^{-2}$. The
corresponding path length required to find $10^4$ absorbers (for 1 per cent
statistics) is $2.9 \times 10^5$ $\hinv$Mpc, or a total spectral
length of about $3.1 \times 10^{7}$ km s$^{-1}$.
The L10\_N128 simulation is actually smaller than this, so we use the
entire box in that case.
For all other runs, we evenly distribute the skewers throughout the
volume up to the required path length. We use the \textsc{specfit}
code \citep[first used in][]{MBM01} to perform the Voigt line fitting
of our simulated spectra.
For each spectrum, the code splits the spectrum into regions separated
by a threshold value $\tau > \tau_{\rm min}$.
In these absorption regions, \textsc{specfit} uses first and second
derivatives of the flux to identify line centers and then performs a
$\chi^2$ minimization of the line parameters.

We show the effect of simulation resolution on the line parameter distributions
in Figure \ref{fig:lines_res}. We show the probability distribution function of
the line column density in the range $11 < \log_{10}(\NHI \, \mathrm{cm}^2) <
16$ and the Doppler parameter in the range 5 km/s $< b <$ 200 km/s and the
percent difference to the L10\_N1024 results. The column density distribution is
relatively flat for $\NHI < 10^{12.5}$ cm$^{-2}$, and then turns over
to a power law $dN/d\NHI \propto \NHI^{-1.6}$, where the slope of the power
law depends on the UV background. The annotated gray line gives an example
of the power law slope.
Qualitatively, the different resolutions agree well. The L10\_N128 and L10\_N256
runs do not peak as much at $\NHI < 10^{12.5}$ cm$^{-2}$, and turn
over more slowly, resulting in an excess probability of lines in the $12.5 <
\log_{10}(\NHI \, \mathrm{cm}^2) < 14$ range. This same trend is present in the
L10\_N512, but it is less significant. The lowest resolution run, L10\_N128,
shows a deficit of high column density lines $\NHI > 10^{14}$ cm$^{-2}$
compared to the other runs. The Doppler parameter distribution
is close to lognormal with a peak around $b = 20$ km/s. The Doppler parameters
distributions show a much clearer convergence pattern. The overall shape of the
Doppler parameter distribution does not appear to change with resolution, but
the peak $b$ value decreases with increasing resolution. This holds together
with our qualitative picture of the resolution study -- the lower-resolution
runs are like artificially smoothed higher-resolution runs, so the resulting
absorption lines are broader as well. We also show two fits of the lognormal
distribution $P(b) = A / (b \sigma) \exp[-(\log b - \log b_{\rm med})^2 / (2
\sigma^2)]$, where $A$, $b_{\rm med}$, and $\sigma$ are the free parameters. The
red thin dashed line is the fit to the L10\_N128 and the black thin dashed line
is the fit to the L10\_N1024 result. From the lowest resolution to the highest
resolution result, the median $b$ value changes from 25.8 km/s to 22.0 km/s, and
the corresponding temperatures are $3 \times 10^4$ K and $4 \times 10^4$ K. This
is an important consideration for studies using small-scale statistics to infer
the temperature of the IGM. If we were to use the low-resolution runs to infer
the IGM temperature, we would be biased to lower temperatures than results based
on higher-resolution runs (the fit $T_0$ values are essentially the same in
these runs).

\begin{figure}
  \begin{center}
    \includegraphics[width=\columnwidth]{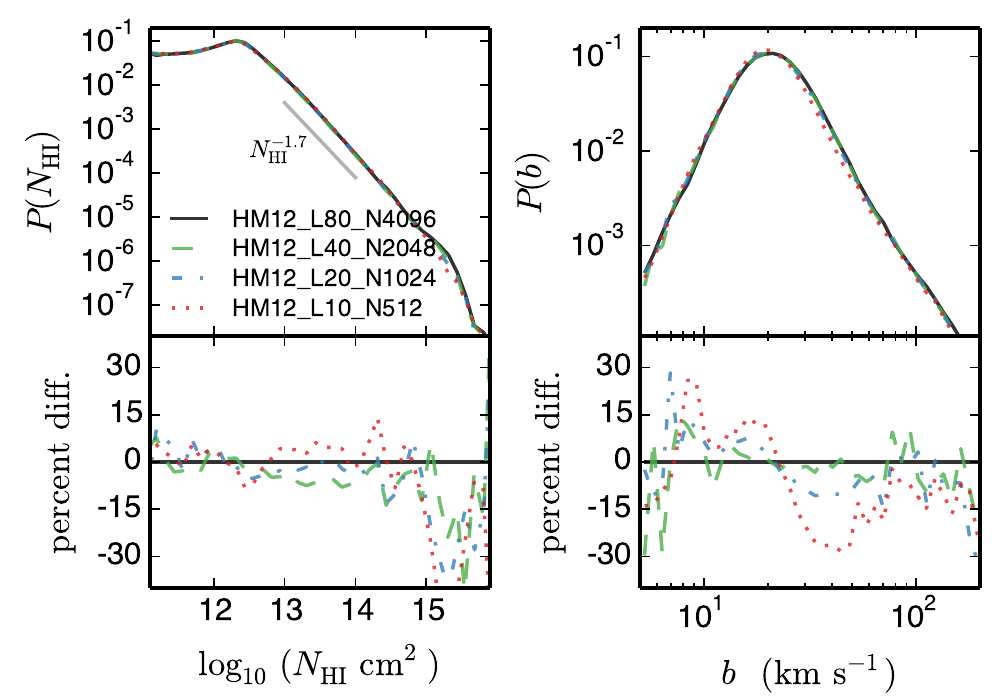}
  \end{center}
  \caption{
    The distributions of the line \HI\ column density (left) and the Doppler
    parameter (right) in the box size series with a grid scale of 20 $\hinv$kpc
    at $z = 2$. The percent difference relative to the HM12\_L80\_N4096
    simulation is shown in the bottom panels. In the $P(\NHI)$ panel, the gray
    line illustrates the power law slope $\propto \NHI^{-1.7}$.
  }
  \label{fig:lines_box}
\end{figure}

We show the effect of simulation box size on the line parameter
distributions in Figure \ref{fig:lines_box}. Compared to the
resolution series, the box size has little effect on the line
fits. The different UVB in the box series simulations results in a
slightly steeper $\NHI$ distribution power law, shown with the gray
line matching $\propto \NHI^{-1.7}$. The column density distributions
across box size are very close to each other, and differences appear
to be in the statistical noise. More significant differences appear
above $\NHI > 10^{15}$ cm$^{-2}$, however, this is also
mostly due to the rarity of well-fit high column density systems. The
box size has a clearer effect on the Doppler parameter distribution,
although it is still much smaller than the resolution effect. Again,
the distribution shape across box size is essentially the same, but
the peak position increases with increasing box size.  This is hard to
see in the top $P(b)$ panel, but in the bottom difference panel, we
see the curves flattening out around $b = 20$ km/s with increasing box
size.

\subsection{Wavelet Statistics}

\begin{figure}
  \begin{center}
    \includegraphics[width=\columnwidth]{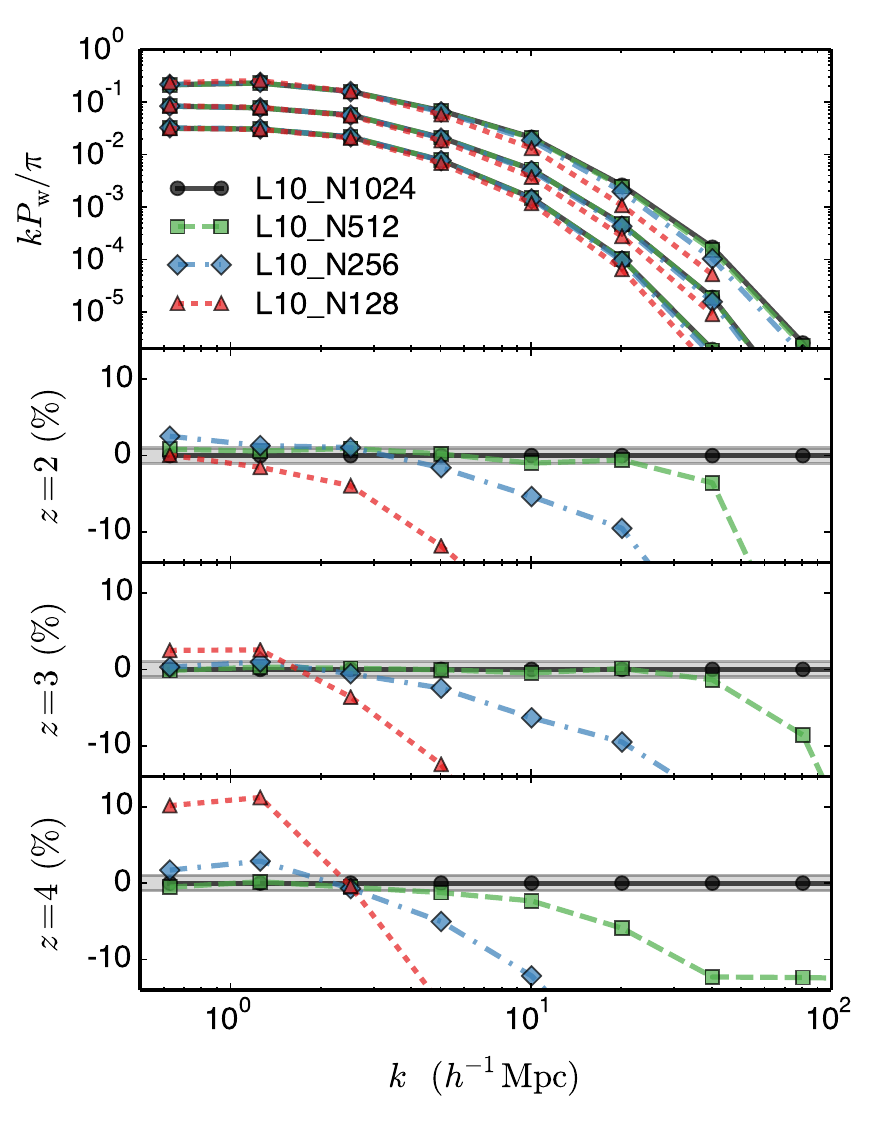}
  \end{center}
  \caption{Convergence of the flux discrete wavelet transform power
    with respect to physical resolution in 10 $\hinv$Mpc box.}
  \label{fig:pw}
\end{figure}

In addition to line fitting, we also performed a wavelet power
analysis of our spectra. Wavelets have previously been applied to the
\Lyaf as a means of objectively measuring the line widths, primarily
in order to probe the IGM temperature. \citet{meiksin_2000} introduced
wavelets as a tool for the \Lyaf as a means of data compression and a
measure of small-scale power. \citet{zaldarriaga_2002} extended the
use of wavelets to search for spatially localized line-width
(temperature) fluctuations, an analysis recently repeated with a
slightly different use of wavelets in \citet{lidz_et_al_2010} and 
\citet{Garzilli2012}.
\citet{theuns_et_al_2002} used wavelets to search for temperature
fluctuations in the IGM associated with \HeII\ reionization.

Wavelet basis functions are orthogonal and complete, providing
well-defined transforms into wavelet coefficients and back. More
importantly, wavelets are localized in real and Fourier space. The
discrete wavelet transform (DWT) is a decomposition of some signal $f$
into discretized wavelet bases $\psi_{jk}(x)$, where we use $j$ as the
level (sometimes called stretch) of the wavelet and $k$ as the shift
(or position). The DWT provides wavelet coefficients $w_{jk}$ such
that $f(x) = \sum_{j,k} w_{jk} \psi_{jk}(x)$, and in this case we are
transforming the flux fluctuations $\delta_F$
along the line of sight. We use the Daubechies
20 coefficient wavelet, which is the most common choice for wavelet
analysis. With the wavelet coefficients in hand, we can compute the
wavelet power spectrum $P_{\rm w}$ as average of the squares of the
coefficients, just as one would with Fourier coefficients.
\begin{equation}
  P_{\rm w}(k_j) = L \langle w_{jk}^2 \rangle
\end{equation}
where the average is taken over all of the shifts $k$ for the level
$j$. The $L$ factor is included to match our previous Fourier
convention and the dimensionless wavelet power spectrum is $k_j P_{\rm w} /
\pi$. We associate the level $j$ to the mode $k_j = 2 \pi / L 2^{j-1}$.

Figure \ref{fig:pw} shows the impact of spatial resolution on the wavelet power.
As expected, we see very similar behavior to the flux power spectrum,
namely the percent level agreement between the 20 $\hinv$kpc
(L10\_N512) and 10 $\hinv$kpc (L10\_N1024) runs.
As with the power spectrum we see that inadequate resolution to capture
the fluctuations responsible for the rise of the \Lyaf produces an error
on {\it all scales}, not only scales below the resolution limit.
We have checked that the box-size behavior is also very similar to that
seen in $P_F(k)$, Figure \ref{fig:box_pf1d}.

%% file: sec9.tex
\section{Collisionality}
\label{sec:collisionality}

In order to obtain a correct distribution of \Lya spectral lines it is
essential for simulations to correctly capture the temperature of the
IGM for a given physical and cosmological model. That is the reason we
opt here to use a highly accurate shock capturing scheme for gas
dynamics.  In this section we investigate the potential artificial
heating of the IGM gas due to the numerical collisionality of dark matter
particles. \citet{Steinmetz1997} have shown that the
discretization of dark matter particles gives rise to a steady flow of
energy from dark matter to the gas. In their paper, they both outlined
the analytic theory for the artificial heating of the gas, and
tested it in both the adiabatic case and in the presence of radiative
cooling -- a regime of interest here. The focus of their study was
dense regions, galactic-size halos, where this effect is expected to
be smaller than in the low-density IGM.  A
particular concern for us is that there is a non-negligible
contribution to the \Lyaf from regions which are quite underdense,
i.e.~where we have less than 1 particle per cell in our runs.  We are
further sensitive to this error due to our code's ability to capture
and propagate weak shocks indiscriminate of whether their origin is
physical or numerical. It is therefore important to consider and
quantify this numerical source of gas heating.

\begin{figure*}
  \centering
  \includegraphics[width=\textwidth]{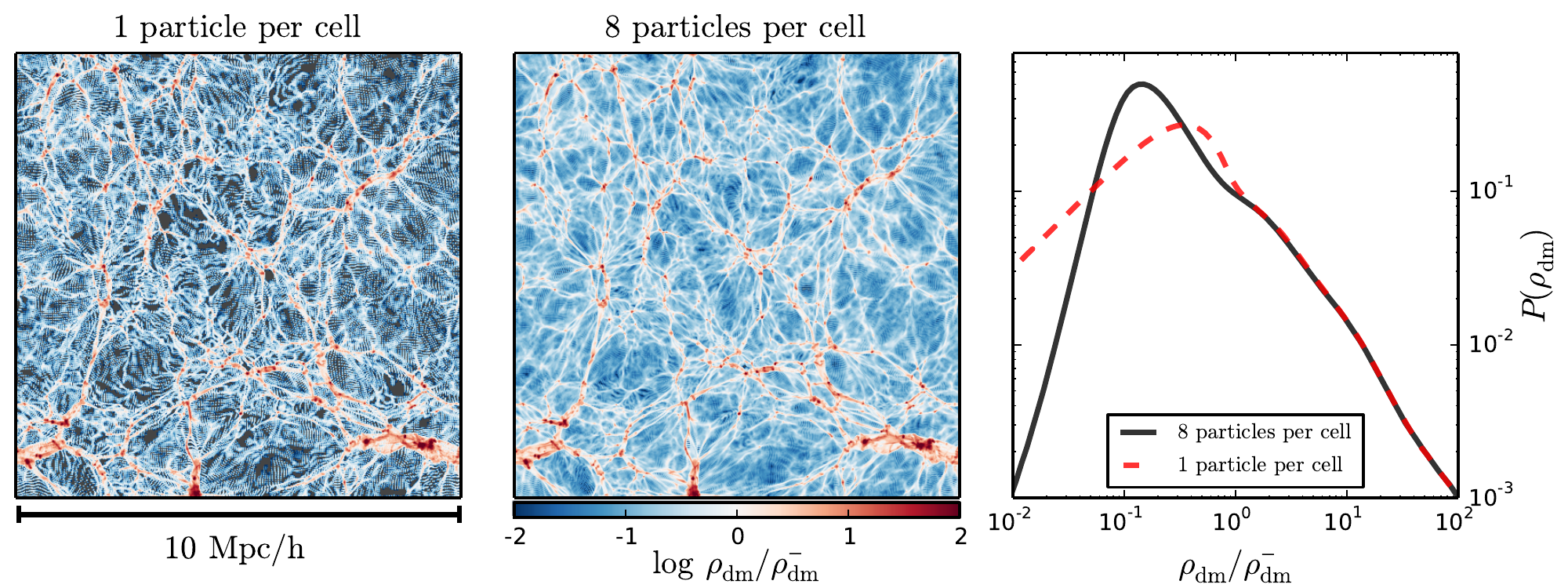}
  \caption{
    \textit{Left}: A slice of dark matter density from L10\_N512
    simulation (1 particle per cell). \textit{Middle}: A slice of dark
    matter density with 8 particles per cell (1024$^3$ particles on a
    512$^3$ grid). \textit{Right}: The PDF of dark matter from both
    runs. We have used a Cloud-In-Cell method to deposit particles on
    the grid, a common approach in cosmological simulations.}
  \label{fig:rhodm-overload}
\end{figure*}

Here we take our L10\_N512 run to be ``the fiducial'' one, as it has
one particle per cell, a common value encountered in \Lyaf
simulations. This roughly corresponds to runs with the same number of
particles for dark matter and gas particles in SPH simulations, which
is typical of almost all SPH simulations in the literature.  We
compare this run with 512$^3$ particles, with a run done on the same
grid, with the same code parameters and cosmological realization, but
with 1024$^3$ particles. Figure \ref{fig:rhodm-overload} shows a slice
of dark matter in both runs together with the PDF from the whole box.
We clearly see the discretization error the Cloud-in-Cell method
introduces to underdense regions. Here there is a clear transition at
one particle per cell with those at higher densities performing well
while those lower than this inaccurately estimating the dark matter
densities.
We have also checked the PDF of the baryon densities in the two runs,
and confirmed that differences are less than 1\%. In void regions,
where the dark matter density is least correctly represented on the grid,
the evolution of matter is driven by the large-scale gravitational field,
therefore the sampling of the dark matter on the grid scale does not
significantly alter the distribution of baryons directly, other
than via pressure forces due to the change in gas temperature.

Our numerical simulations show that the effect of collisional heating
in the gas is indeed present, but overall it is negligible. We do not
observe a significant change in the gas temperature as shown in
Figure \ref{fig:dt_overload}. Here, we show the median difference
between the two runs over the density range relevant for the \Lyaf,
along with the median absolute deviation scatter (defined as the median
of the absolute deviations in temperature of each cell from the full-box
temperature median, scaled to give 1$\sigma$ for a Gaussian distribution).
The differences in the median are within 1 per cent throughout this density range
while the MAD scatter increases with density. The latter is at least
partially an artifact of sampling, as the number of cells decreases
with increasing density (recall Figure \ref{fig:rho-T}).  Even in void
regions, cell-to-cell comparisons shows less than a percent difference
-- thus the effect of artificial collisionality is small.  The fact
that the scatter goes negative shows that some cells are cooler in the
run with coarser dark matter particle mass. It is likely an indication
that other discretization effects (the deposition of particles and the
addition of gravitational force in the Euler equations) introduce at
least the same amount of error as collisional heating.

As a result, we observe that the difference in flux statistics is
small, less than 0.5 per cent in the flux PDF, and less than 1 per cent for the 1D
power spectrum up to $k = 10$ $h$ Mpc$^{-1}$.  This is below the level
we are presently concerned with in making theoretical \Lyaf
predictions. Note that the small effect particle sampling plays is
not completely unexpected: simulations focusing on the IGM suffer from
coarse particle sampling in the relevant regions, but the mass of dark
matter particles is still small enough in today's
simulations. Therefore the (real) radiative cooling still dominates
over the (artificial) collisional heating process.

%% file: sec10.tex
\section{Conclusions}
\label{sec:conclusion}

We have investigated simulated \Lyaf statistics over the redshift
range $2 \leq z \leq 4$. A large suite of simulations covering box
sizes of 10-80 $\hinv$Mpc, and resolutions 10-78 $\hinv$kpc have
enabled us to understand the numerical requirements for future sets of
simulations aimed at constraining cosmological parameters using the \Lyaf.
While we model gas dynamics using a very accurate finite-volume
numerical method, the additional physics which enters as a source
(heating) term in the energy equations as well as those used to
calculate the ionization state of a primordial chemistry gas is
accounted for in a more approximative way. In our optically thin
simulations, the gas is rapidly ionized by the assumed UVB at high
redshift, and in a short time changes its ionization fraction by order
unity -- a model of sudden reionization. We neglect the effects of
self-shielding in high over-density regions as those do not contribute
to the \Lyaf signal. The temperature boost during the sudden
reionization depends only on the shape of the spectrum of ionizing
radiation, and in fact one expects a range of spectral shapes
responsible for ionizing different gas elements. Modeling the details
of temperature evolution during and immediately after reionization
requires full radiative transfer simulations and is beyond the scope
of our needs here, as the thermal memory of the IGM gradually fades
after the epoch of reionization (mostly due to Compton cooling, see
\citealt{hui_and_gnedin_1997}).

As the IGM fills the simulation box, it is fruitless to try to resolve
it with adaptive refinement; similarly, as a large portion of the
\Lyaf signal arises from near mean and under-dense regions (especially
at at higher redshifts), Lagrangian methods do not offer any advantage
over a fixed grid PDE solver. Needless to say the fixed grid approach
is computationally expensive, especially in the 3D case presented
here, and thus it is important to determine the minimal resolution
requirements needed to bring our simulations to 1 per cent accuracy.  We have
explored this in section \ref{sec:resolution} arriving at the
conclusion that $\sim20$ $\hinv$kpc resolution is good enough
over the relevant redshift ranges we wish to consider for the \Lyaf.
While in places --- for example the flux PDF --- a coarser resolution
would suffice, the study of 1D power spectra brings with it a more
stringent requirement. While we explored resolution convergence, we
were also able to show that Nyx behaves well on this
multi-dimensional, multi-physics problem, exhibiting the expected
second order convergence. As shown in section \ref{sec:resolution},
this opens the possibility of achieving a desired accuracy at reduced
cost, via extrapolation of lower-resolution runs.

We also explored other numerical artifacts which can easily mask a
physical process in the IGM and/or spoil the quality of cosmological
predictions. After finding an appropriate resolution, we have
explored the effects of finite box-size, i.e.~missing modes in \Lyaf
simulations. By running simulations with all cosmological and
numerical parameters but the box-size fixed, we were able to show that
40-80 $\hinv$Mpc boxes are large enough for all relevant statistics
including the 3D power spectrum in redshift space, i.e.~$P(k, \mu)$.
For the first time, we were able to perform a \Lyaf simulation fulfilling
both the resolution requirement set by the Jeans / filtering scale and
the box-size requirement set by large-scale flows.
That enabled us to examine the accuracy of splicing 1D power spectra,
a common approach in the case when a full range of simulations are not
feasible \citep{mcdonald2003, Borde2014}.
We show that accuracy of splicing is only 5-10 per cent, and that the error
has a clear scale-dependence.

\begin{figure}
  \centering
  \includegraphics[width=\columnwidth]{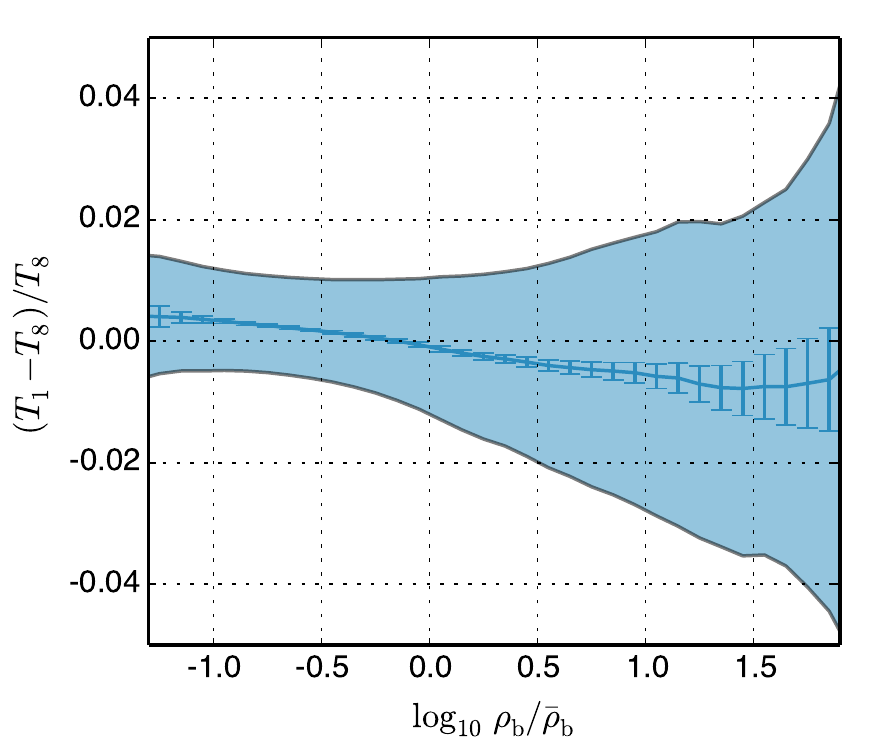}
  \caption{Change in the gas temperature in a standard run
    with 1 particle per cell ($T_1$), and an ``overloaded'' run with with 8
    particles per cell on average ($T_8$). The line shows the difference in the
    median temperatures and the fill shows the normalized median absolute
    deviation in temperatures. The median deviation is roughly symmetric and the
    resulting effect on the \Lyaf statistics is negligible.  In addition, we show
    statistical (Poisson) errorbars in each density bin.}
  \label{fig:dt_overload}
\end{figure}

As the UV background is the largest uncertainty in \Lyaf simulations,
it is very common that a cosmological model is evolved with a UVB
model, with the resulting optical depth field subsequently rescaled to
match desired, observed mean flux value.
In Section \ref{sec:uvb}, we examined the effects of changing the UVB and
rescaling the mean flux.
We found that while qualitatively it is possible to change the UVB
{\it post festum} in analysis, one suffers a few percent effect in the
UVB rates propagating back into the gas evolution which is most visible
in the flux PDF.  We note that the 1D $P(k)$ appears largely insensitive
to such rescaling for $k \lesssim 3 \times 10^{-2}$ km$^{-1}$s,
the $k$-range relevant for current observational data.

We find that resolution requirements for convergence on the line statistics
are much more demanding than on the flux statistics.
A resolution of 20 $\hinv$kpc, adequate for reproducing the flux
statistics to 1 per cent accuracy as shown in section \ref{sec:resolution},
recovers the distribution of neutral hydrogen column densities only at
$\sim$10 per cent accuracy in the range
$12.5 \lesssim \log_{10}(\NHI \, \mathrm{cm}^2) \lesssim 15$,
and even worse at higher column densities.
Similarly, the Doppler parameter ($b$) distribution is converged at the
same level, with the peak value decreasing with increasing resolution.
This is not directly relevant to modern cosmological studies which do not
rely on individual line fitting, but it is important for certain studies
of the IGM.

Finally, we have explored the effects that finite sampling of the dark
matter particles has on the statistics of the \Lyaf. The expectation
is that artificial gravitational collisionality between dark matter
particles and gas increases the gas temperature, an effect that should
be strongest in void regions. While we indeed notice, on average,
temperature increases in void regions, the effect is minor in today's
simulations even when using 1 particle per cell and CIC particle
deposition. The reason for this is the small particle mass in \Lyaf
simulations and the presence of radiative cooling, which efficiently
removes excess heat.

The advent of high performance computing power and scalable numerical
algorithms as employed in Nyx allows us to make accurate predictions
for the \Lya flux statistics, one of the most promising tools for
precision cosmology measurements in the redshift range $2 \lesssim z
\lesssim 4$. The direct simulation approach, using no {\it ad hoc}
physical assumptions, is possible for this problem.  We have made a
concentrated effort here in understanding the \Lyaf signal in
optically-thin hydrodynamical simulations, and quantifying the
accuracy of such simulations with respect to numerous numerical
effects. We are by no means the first to attempt this (indeed they go
back at least to \citealt{1992ApJS...78..341C}), however we have been
able to consolidate and improve upon earlier studies using modern
simulations with the goal of percent-level numerical precision, a
level of accuracy required for carrying out precision cosmology over
the next decade.

We note that full-range, $4096^3$, hydrodynamical simulations like the one
presented in this paper are still computationally demanding today, but will be a
fairly typical in coming years. Before one commits to running many such
simulations, it is imperative that the precision which can be obtained be
understood. Convergence testing is an invaluable tool here, as analytically soluble
problems are highly artificial in nature, and experience with them does not
necessarily translate to real cosmological runs. In our paper we have used the
Nyx code. While Nyx is focused on \Lyaf simulations, the results presented here
should be directly applicable to IGM simulations performed by the Enzo, Flash
and Ramses codes. In addition, there are several lessons applicable to a large
extent to the Gadget and Arepo codes as well. In a future work
\citep{stark_et_al_2014a} we will compare SPH and Eulerian methods for
simulating the \Lyaf.

%% file: appen_a.tex
\section{Atomic Rates}
\label{app:rates}

\begin{figure}
  \begin{center}
    \includegraphics[width=\columnwidth]{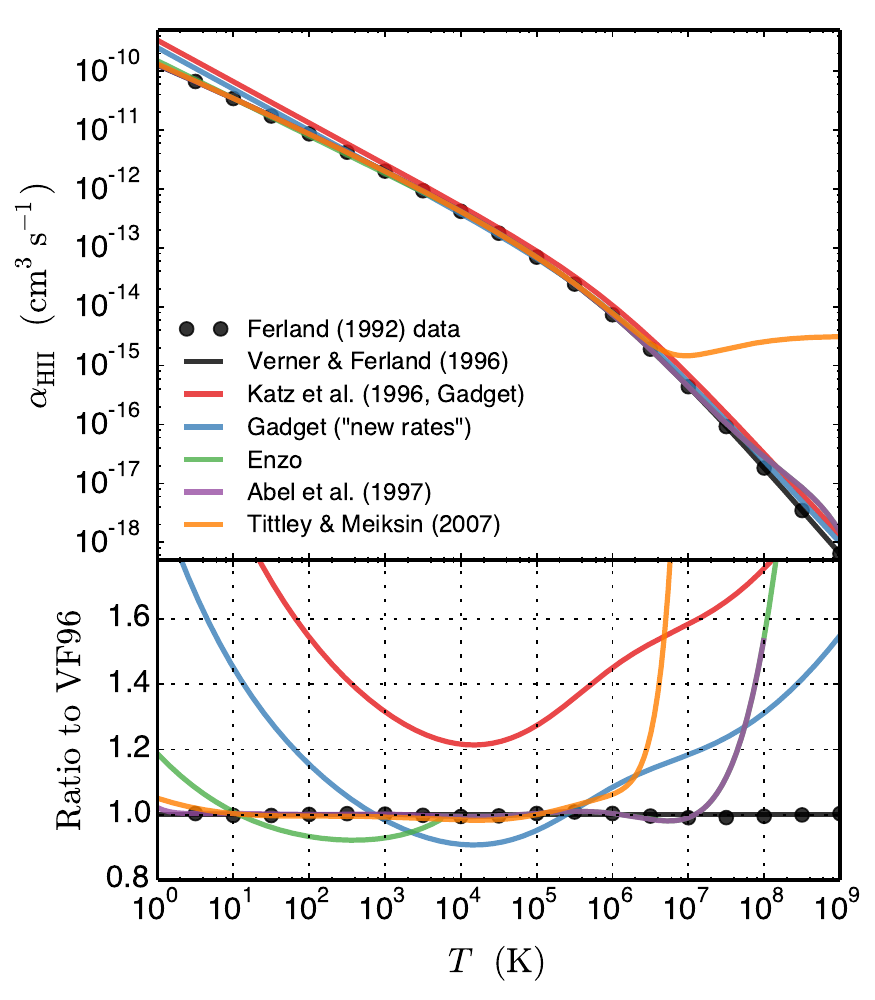}
    \includegraphics[width=\columnwidth]{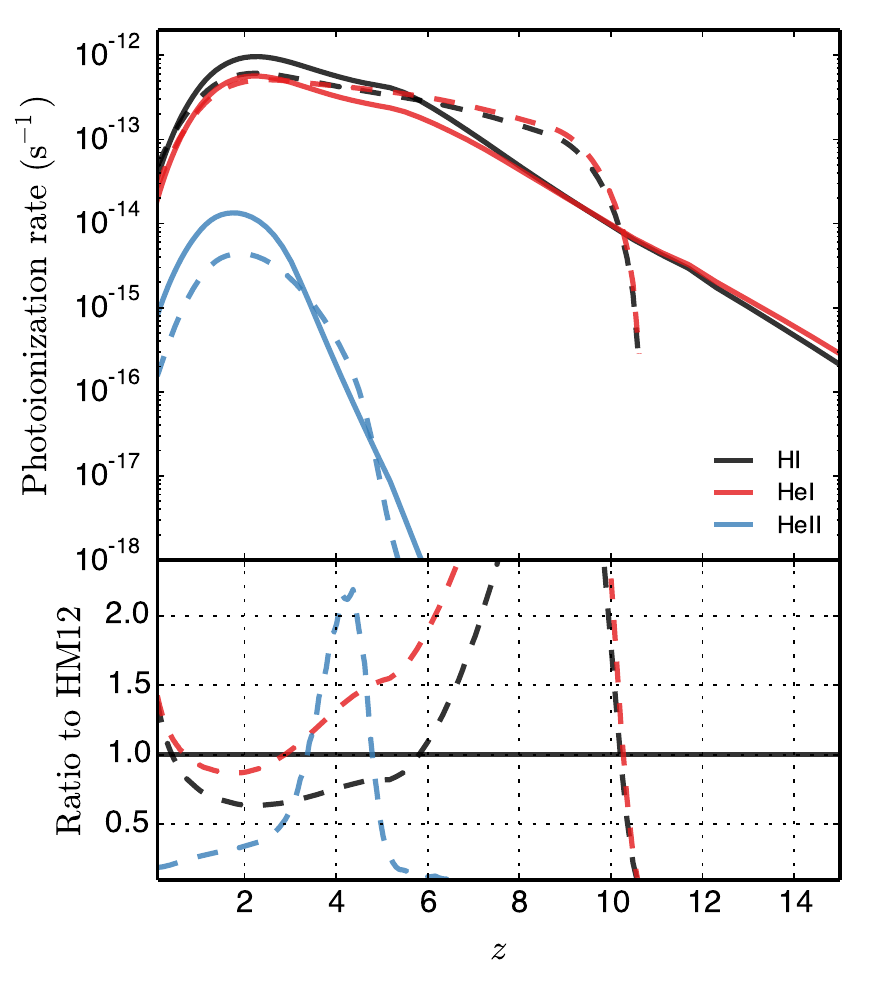}
  \end{center}
  \caption{
    \textit{Top}:
    A comparison of different hydrogen recombination rates used in recent
    simulations in the literature and \citet{1996ApJS..103..467V} calculated
    data.
    \textit{Bottom}:
    Comparison of photoionization rates published in recent works:
    solid lines are \citet{haardt_and_madau_2012}, and
    dashed lines are \citet{faucher_giguere_et_al_2009}
    (revised in \citealt{2011MNRAS.417.2982F}).}
  \label{fig:alpha_H}
\end{figure}

Here we provide some details on the reaction rates implemented in the
Nyx code. In order to provide an easy comparison to the Gadget code,
we have also implemented the atomic rates from
\citet{katz_et_al_1996}. Some of the runs presented in this paper use
those rates, as well as a companion comparison study between Nyx and
Gadget \citep{stark_et_al_2014a}. However, we also implement more
accurate rates which will be used in future Nyx studies exploring
cosmological effects in \Lyaf statistics. These are also used here for
our biggest 4096$^3$ run and the accompanying smaller runs used in the
box-size study in Section \ref{sec:box_size}. These rates are shown
in Table \ref{table:rates}, along with their references.

We explicitly keep track of the net loss of thermal energy resulting
from atomic collisional processes. Those rates are shown in
Table \ref{table:cooling}. In addition to the tabulated cooling rates,
Nyx includes cooling from inverse-Compton scattering off CMB photons
as in \citet{1968ApJ...153....1P}
\begin{equation}
  L_C = \frac{4 \sigma_T a k_B}{m_e c} n_e T^4_{\rm CMB}(z)
    \left[T - T_{\rm CMB}(z)\right] \, ,
\end{equation}
where $\sigma_T$ is the Thomson cross section, $a$ is the radiation
density constant, $k_B$ is the Boltzmann constant, $m_e$ is the
electron mass, $c$ is the speed of light, and $T_{\rm CMB}$ is the
temperature of the microwave background, which we take to be
$T_{\rm CMB} = 2.725$.

\begin{table*}
  \begin{center}
  \caption{Atomic rates in Nyx}
  \begin{tabular}{r l l}
    \toprule
    \multicolumn{1}{c}{Coefficient} & \multicolumn{1}{c}{Fitting formula [${\rm cm^3 s^{-1}}$]} &
    \multicolumn{1}{c}{Comment} \\
    \midrule
                   & \multicolumn{1}{c}{$a \left[ \sqrt{T/T_0} \left(1+\sqrt{T/T_0} \right)^{1-b}
                      \left( 1 + \sqrt{T/T_1} \right)^{1+b} \right]^{-1}$}       & \citet{1996ApJS..103..467V}\\
    \cmidrule(r){2-2}
  $\alpha_{\rm r, \HII}$  & $a=7.982\expd{-11}$, $b=0.7480$, $T_0=3.148$, $T_1=7.036\expd{5}$
                   & \\
  $\alpha_{\rm r, \HeII}$ & $a=3.294\expd{-11}$, $b=0.6910$, $T_0=15.54$, $T_1=3.676\expd{7}$
                   & $(T \leq 10^6)$\\
  $\alpha_{\rm r, \HeII}$ & $a=9.356\expd{-10}$, $b=0.7892$, $T_0=4.266\expd{-2}$, $T_1=4.677\expd{6}$
                   & $(T > 10^6)$\\
  $\alpha_{\rm r, \HeIII}$ & $a=1.891\expd{-10}$, $b=0.7524$, $T_0=9.370$, $T_1=2.774\expd{6}$
                   & \\
 %$\alpha_{\rm r, \HII}$   & $7.982\expd{-11} \left[\sqrt{T/3.148}
 %                    \left(1+\sqrt{T/3.148}\right)^{0.252} \left(1+\sqrt{T/7.036\expd{5}}
 %                    \right)^{1.748}\right]^{-1}$                                      & VF96\\
 %$\alpha_{\rm r, \HeII}(T \leq 10^6)$   & $3.294\expd{-11} \left[\sqrt{T/15.54}
 %                    \left(1+\sqrt{T/15.54}\right)^{0.309} \left(1+\sqrt{T/3.676\expd{7}}
 %                    \right)^{1.691}\right]^{-1}$                                      & VF96\\
 %$\alpha_{\rm r, \HeII}(T>10^6)$   & $9.356\expd{-10} \left[\sqrt{T/4.266\expd{-2}}
 %                    \left(1+\sqrt{T/4.266\expd{-2}}\right)^{0.2108} \left(1+\sqrt{T/4.676\expd{6}}
 %                    \right)^{1.7892}\right]^{-1}$                                     & VF96\\
 %$\alpha_{\rm d, \HeII}$  & $1.9\expd{-3}
 %                    \left( 1 + 0.3 e^{\frac{-9.4\expd{4}}{T}} \right)
 %                    e^{\frac{-4.7\expd{5}}{T}}  T^{-\frac{3}{2}}$                     & AP73\\
 %$\alpha_{\rm r, \HeIII}$ & $1.891\expd{-10} \left[\sqrt{T/9.37}
 %                    \left(1+\sqrt{T/9.37}\right)^{0.2476} \left(1+\sqrt{T/2.774\expd{6}}
 %                    \right)^{1.7524}\right]^{-1}$                                     & VF96\\
     \midrule
 $\alpha_{\rm d, \HeII}$  & \multicolumn{1}{c}{$1.9\expd{-3}
                     \left( 1 + 0.3 e^{\frac{-9.4\expd{4}}{T}} \right)
                     e^{\frac{-4.7\expd{5}}{T}}  T^{-\frac{3}{2}}$ }
                     & \citet{1973AnA....25..137A}\\
     \midrule
                   & \multicolumn{1}{c}{$A \dfrac{\left( 1 + P\ U^{1/2}\right)}{(X+U)} U^m e^{-U}$ ;
                     $U = \dfrac{11604.5 E}{T}$ }                                      & \citet{1997ADNDT..65....1V}\\
    \cmidrule(r){2-2}
    $\Gamma_{e, \HI}$    & $A = 2.91\expd{-8}$, $E = 13.6$, $P=0$, $X=0.232$, $m=0.39$    & \\
    $\Gamma_{e, \HeI}$   & $A = 1.75\expd{-8}$, $E = 24.6$, $P=0$, $X=0.180$, $m=0.35$    & \\
    $\Gamma_{e, \HeII}$  & $A = 2.05\expd{-9}$, $E = 54.4$, $P=1$, $X=0.265$, $m=0.25$    & \\
    \bottomrule
  \end{tabular}
  \label{table:rates}
  \medskip

  Recombination ($\alpha_i$) and collisional ionization ($\Gamma_{e\, i}$) rates
  in the Nyx code. $\alpha_{\rm d, \HeII}$ is the dielectronic recombination rate
  of singly ionized helium. Temperature is in K, and rates are tabulated in the
  code in the temperature range $1 \leq T \leq 10^9$ K.

\end{center}
\end{table*}

\begin{table*}
  \begin{center}
  \caption{Cooling rates in Nyx}
  \begin{tabular}{r l l}
    \toprule
    \multicolumn{1}{c}{Type} & \multicolumn{1}{c}{Fitting formula [${\rm erg \, cm^{3} s^{-1}}$]} &
    \multicolumn{1}{c}{Comment} \\
    \midrule
 Bremsstrahlung & $1.426\expd{-27} T^{\nicefrac{1}{2}} Z_i^2 \langle g_{ff} \rangle$ ;
                  $\langle g_{ff} \rangle = \begin{cases}0.79464+0.1243\log(T/Z^2); & T/Z^2\leq3.2\expd{5}K\\
                                                         2.13164-0.1240\log(T/Z^2); & T/Z^2>3.2\expd{5}K
                                                         \end{cases}$
                & \citet{1987ApJ...318...32S}\\
    \midrule
 Neutral Hydrogen & & \citet{1991ApJ...380..302S}\\
                  & $10^{-20} {\rm exp} (213.7913-113.9492y+25.06062y^2-
                                         2.762755y^3+0.1515352y^4$ &\\
                  &  \hspace{1.4cm} $-3.290382\expd{-3}y^5 -
                       1.18415\expd{5}T^{-1})$  & $2\expd{3} \leq T \leq 10^5$\\
                  & $10^{-20} {\rm exp} ( 271.25446-98.019455y+14.00728y^2-
                                          0.9780842y^3+3.356289\expd{-2}y^4-$ &\\
                  &  \hspace{1.4cm} $-4.553323\expd{-4}y^5 -
                       1.18415\expd{5}T^{-1})$  & $T > 10^5$\\
                  & \multicolumn{1}{c}{$y \equiv \ln (T)$} &\\
    \midrule
        Helium &  & \citet{1981MNRAS.197..553B}\\
    \cmidrule(r){1-1}
        $\HeI$ & $9.38\expd{-22} T^{\nicefrac{1}{2}} e^{-285335.4/T} \left(1+\sqrt{\dfrac{T}{5\expd{7}}}
                   \right)^{-1}$ & \\
       $\HeII$ & $\left( 5.54\expd{-17} T^{-0.397} e^{-473638/T}
                  + 4.85\expd{-22} T^{\nicefrac{1}{2}} e^{-631515/T} \right) \left(1+\sqrt{\dfrac{T}{5\expd{7}}}
                  \right)^{-1}$  & \\
    \midrule
 Recombinations   & & \citet{1981MNRAS.197..553B}\\
    \cmidrule(r){1-1}
        $\HII$ & $2.851\expd{-27} T^{\nicefrac{1}{2}} \left(5.914 -\frac{1}{2}\ln\, T
                    + 0.01184 T^{\nicefrac{1}{3}} \right)$ & \\
       $\HeII$ & $1.55\expd{-26} T^{0.3647} + 1.24\expd{-13}
        \left( 1 + 0.3 e^{\frac{-9.4\expd{4}}{T}} \right)
                    e^{\frac{-4.7\expd{5}}{T}}  T^{-\frac{3}{2}}$ & \\
      $\HeIII$ & $1.140\expd{-26} T^{\nicefrac{1}{2}} \left( 6.607 - \frac{1}{2}\ln\, T
                    + 7.459\expd{-3} T^{\nicefrac{1}{3}} \right)$ & \\
    \bottomrule
  \end{tabular}
  \label{table:cooling}
  \medskip

The Cooling rates used in Nyx. Note that the helium rates are from
\citet{1981MNRAS.197..553B} but were modified by a different temperature factor
than in \citet{1992ApJS...78..341C}.
In the Bremsstrahlung expression, $Z = 1$ for $\HII$ and $\HeII$, and $Z = 2$
for $\HeIII$. Temperature is in K, and the rates are tabulated in the code in
temperature range $1 \leq T \leq 10^9$ K.

  \end{center}
\end{table*}

The atomic rates are a compilation of observed laboratory data, and as such,
the fitting functions are used to interpolate and extrapolate between and
beyond these data points. In the literature many different fits to the
atomic rates have been used in IGM simulations. To build intuition
for the differences they make on the \Lyaf flux, we present the
different hydrogen recombination rates found in several works,
including the Enzo and Gadget codes used for most of the \Lyaf
simulations in recent years. The optical depth $\tau$ of neutral
hydrogen absorption is proportional to the number density of neutral
hydrogen, which is itself proportional to $\alpha_{\HII}(T)
\Gamma^{-1} n_H n_e$, for a case when hydrogen is highly 
ionized and in photoionization equilibrium (see eq.~\ref{eq:equil_species}). 
Therefore, an error in the hydrogen
recombination rate directly propagates to the same error in $\tau$. As
Figure \ref{fig:alpha_H} demonstrates, some of the fits are inaccurate
by $\sim$20 per cent at $T=10^4$ K, although the case of hydrogen 
recombination can actually be calculated from first principles, as done e.g.~in
\citet{Ferland1992}. We show a large temperature range for
completeness; at low temperatures three-body recombination is dominant
at most densities, whereas at very high temperatures the neutral
hydrogen fraction is vanishingly small.

Another essential ingredient in modeling the thermal and ionization
history of the IGM is the ultra-violet background, and
especially the hydrogen photoionization rate, $\Gamma_{HI}$. Due to
its low surface brightness this is not a directly measurable quantity
in observations, however, indirectly it can be seen via such
astrophysical phenomena as the quasar proximity effect. These kinds
of measurements are quite uncertain, and instead one often tries to
calculate the UVB intensity and spectral shape by combining all
possible sources of ionizing flux \citep{haardt_and_madau_2012}.
These calculations are also quite uncertain and have a large number of
input assumptions.

It is beyond the scope of this work to examine all the potential
physical processes and simulations used to create these different
models of the UVB, as well as their accuracy; we refer interested
readers a recent work by \citet{kollmeier2014}. Instead, in
Figure \ref{fig:alpha_H} we simply show the differences in
photoionization rates of the most recent works on this topic by
\citet{haardt_and_madau_2012} and
\citet{faucher_giguere_et_al_2009}. Note that the latter rates were
updated in 2011 \citep{2011MNRAS.417.2982F} The right panel of
Figure \ref{fig:alpha_H} clearly demonstrates that differences between
the two works --- and therefore our understanding of the UVB --- are
rather significant. The effect of different rates on the temperature
of the IGM we show in Figure \ref{fig:T0_evolution}, confirming that
the ``feedback'' onto the dynamical evolution of the gas is much
smaller.

%% file: appen_b.tex
\section{Optical Depth Calculation Details}
\label{app:tau}

The cross section for a resonant line scattering is:
\begin{equation}
  \sigma_\nu = \frac{\pi e^2}{m_e c} f_{lu} \frac{1}{\Delta \nu_{\rm D}}
    \phi_\nu
\end{equation}
where $e$ is the fundamental electric charge, $f_{lu}$ is the
oscillator strength, $m_e$ is the electron mass, $\Delta \nu_{\rm D}$
is the Doppler width, and $\phi_\nu$ is the line profile. In addition,
$\Delta \nu_{\rm D} = b / c \nu_0$, where $b$ is the Doppler parameter
(the velocity broadening scale), and $\nu_0$ is the line center
frequency. We assume that there are no extra kinematic components to
the broadening in this work, so $b = \sqrt{2 k_{\rm B} T / m_{\rm H}}$.
In general, the line profile for this process is given by
the Voigt profile:
\begin{equation}
  \phi_{\nu, \mathrm{V}} = \frac{1}{\pi^{1/2}} H(a, x)
    = \frac{a}{\pi^{3/2}} \int_{-\infty}^{\infty}
      \frac{ \exp(-y^2) }{ (x - y)^2 + a^2 } dy
\end{equation}
where $a = \Gamma_{ul} / 4 \pi \Delta \nu_{\rm D}$ is the ratio of the
damping width to the Doppler width, and $x = (\nu - \nu_0) / \Delta
\nu_{\rm D}$ is the shift from line center. However, for densities and
temperatures typical in the IGM we may use the Doppler profile
instead, which is just the Gaussian core of the Voigt profile:
\begin{equation}
  \phi_{\nu, \mathrm{D}} = \frac{1}{\pi^{1/2}} \exp(-x^2).
\end{equation}

\begin{figure*}
  \begin{center}
    \includegraphics[width=6in]{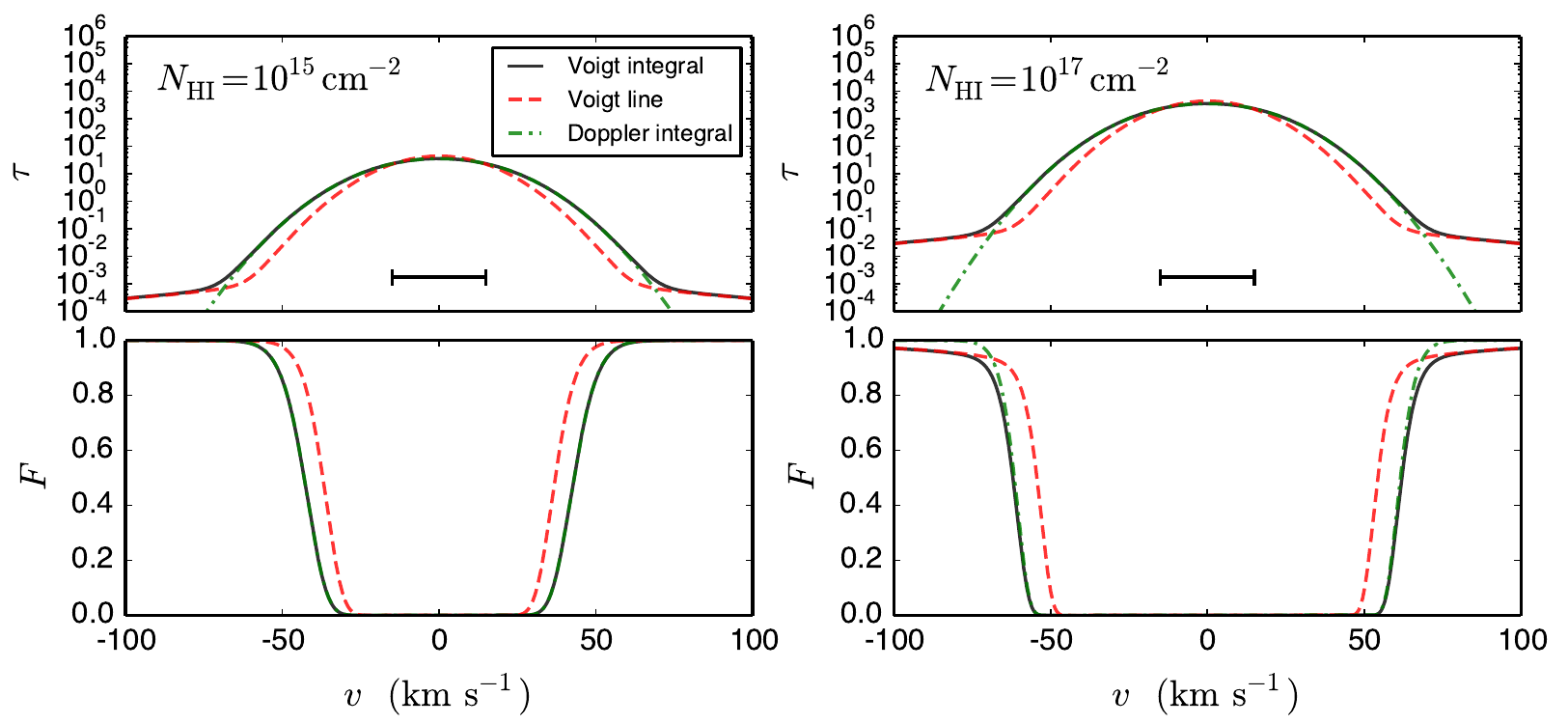}
  \end{center}
  \caption{A comparison of the optical depth across a uniform
    absorber computed with three methods, two using the full
    sightline integral with either a Voigt profile (`Voigt
    integral') or a Doppler profile (`Doppler integral'), and one
    approximating the absorber as a static system (`Voigt
    line'). See the text for more details.}
  \label{fig:voigt_doppler}
\end{figure*}

The difference in optical depth computed with a Voigt versus a Doppler
profile is very small in the regime we are interested in ($\tau_0 <
100$). Figure \ref{fig:voigt_doppler} shows the optical depth to a
single absorber with uniform \HI\ density and temperature,
computed three different ways. The system spans a comoving scale of
roughly 300 $\hinv$kpc, and corresponds to $\Delta v = 30$
km s$^{-1}$ for our cosmology's $H(z=2)$. The difference
between the left and right panels is the column density, where a
system with a typical \Lyaf column density is shown on the left and a
weak Lyman Limit System is on the right. The Voigt integral and
Doppler integral versions are the full sightline integral, and only
differ by the line profile assumed. A third computation approximates
the feature as a single line with line center at the center of the
absorber (``Voigt line''). The Voigt line version follows the damping
wings, but has the wrong shape near line center. It does not account
for the change in the line center of the gas across the system and is
therefore too narrow. The Doppler integral version correctly traces
the Gaussian shape near line center, but misses the damping
wings. However, for the low column density lines that make up the
\Lyaf, the damping wings add optical depth at the level of $\sim
10^{-3}$. After the transformation to $F$, such a small $\tau$ is far
from detectable. For column densities of Lyman Limit Systems though,
the damping wings contribute $\tau \sim 0.1$, which clearly shows up
in $F$. In \Lyaf observations, contamination from LLSs and DLAs is
masked out or taken into account in error estimates, so we actually
want to avoid modeling their contamination here. Additionally, our
simulations do not properly model the \HI\ density and
temperature in the high density regions that give rise to LLSs and
DLAs since they do not include radiative transfer. Because the
difference in the resulting optical depth is very small, and the
Doppler profile is simpler and faster to compute, we use the Doppler
line profile in this work.

In order to produce statistics at a single redshift, we also compute
the optical depth at a fixed redshift. That is, we do not account for
the speed of light when we cast rays in the simulation; we use the gas
state at a single cosmic time. The simulated spectra are not meant to
look like full \Lyaf spectra, but just recover the statistics of the
flux in a small redshift window. The path length in the sightline
integration is then $dr = a \, dx = dv / H$, where $r$ is the proper
distance, $x$ is the comoving distance, $v$ is the Hubble flow
velocity, and $H$ is the Hubble expansion rate at that redshift. In
velocity coordinates, the optical depth is
\begin{equation}
  \tau_v = \frac{\pi e^2 f_{lu} \lambda_0}{m_e c H}
      \int n_{\rm X} \frac{1}{\pi^{1/2} b}
      \exp \left[ - \left( \frac{v - v_0}{b} \right)^2 \right] dv.
\end{equation}
Although the gas data is fixed at the grid resolution, we can choose
an arbitrary spectral resolution $N_{\rm pix}$ along the LOS. We also
take the gas values as constant across each cell. With $i$ as the cell
index, and $j$ as the pixel index, the discretized version of the
optical depth is
\begin{equation}
  \tau_j = \frac{\pi e^2 f_{lu} \lambda_0}{m_e c H}
    \sum_i n_{{\rm X},i}
    \left[ \mathrm{erf}( y_{i-1/2} ) - \mathrm{erf}( y_{i+1/2} ) \right]
\end{equation}
where $y = (v_j - v_{\parallel, i} - v) / b$ is the line center shift
from the pixel velocity in terms of the broadening scale,
$v_{\parallel, i}$ is the component of the gas peculiar velocity
parallel to the sightline, and $v$ is the Hubble velocity. The
velocity coordinates are also periodic on the domain scale $[0,
\dot{a} L)$. It is fortunate the optical depth integration reduces to
an analytic expression, as this makes the calculation more robust and
straightforward. Previous studies have used the midpoint expression
for this integral, but we found that this created too large an error
when the sampling scale $\Delta v$ was $\sim 2$ km s$^{-1}$ or
larger, whereas the analytic version explicitly
conserves the optical depth for any $\Delta v$.

%% file: appen_c.tex
\section{The Effect of Spectral Resolution}
\label{app:specres}

As noted in the previous section, we have the choice to evaluate the
spectra at an arbitrary resolution. Given a vector of the simulated
values $s_i = \left(n_{\rm X, i}, v_{\parallel, i}, b_i \right)$, at
position $v_i$ along a skewer, we can evaluate the optical depth at
any $v_j$. The resolution requirement here is essentially set by the
linewidths, so that we capture all fluctuations that should be present
in the spectra. Given that the most narrow lines have $b \sim 5$ km/s,
the required spectral sampling should be similar.

We test that we have adequate spectral resolution by taking the
L10\_N512 snapshot a redshift $z=2.5$, recomputing the optical depth
at lower and higher spectral resolution and checking the effect on
various flux statistics. In the rest of the paper, we have taken the
spectral grid to be the same as the simulation grid ($512^3$ in this
case). Here we also try the grids 512x512x128, 256, 1024, and 2048,
which spans a factor of 4 worse to 4 better than the default. We see
essentially no change in the mean flux (less than $10^{-6}$).  We can
actually see the effect in the flux PDF, but the difference is still
very small, and does not appear to be a systematic difference, but
behaves more like noise from the slightly different sampling. Over all
bins, the RMS difference compared to the $N_{\rm pix} = 2048$ PDF is
less than $10^{-3}$ in all cases.  For the 1D flux power, the RMS
difference is also less than $10^{-3}$ including points up to $k = 30$
$\hinv$Mpc. On very small scales ($k > 50$ $\hinv$Mpc),
the results do depend heavily on spectral resolution,
with high spectral resolution results having tens of percent larger
power. However, this result is of no concern since by this scale, the
dimensionless 1D power is already $< 10^{-6}$. It seems that even at
the worst resolution, all relevant lines are resolved.